\documentclass[apj]{emulateapj}
\usepackage[colorlinks,linkcolor={blue},citecolor={blue},urlcolor={red}]{hyperref}
\usepackage{mathtools}
\usepackage{epsfig,graphicx,natbib,amsmath,amsfonts,amssymb}
\usepackage{color}
\usepackage{multirow}
\usepackage{booktabs}
\usepackage{amssymb}
\usepackage{threeparttablex} 
\usepackage{booktabs} 

\usepackage{longtable}
\usepackage{array} 
\usepackage[dvipsnames]{xcolor}  

\newcommand{\zgas}{$Z_{\rm gas}$}
\newcommand{\mgas}{$M_{\rm gas}$}
\newcommand{\mz}{$M_{\rm Z}$}

\newcommand{\re}{\Reff}
\newcommand{\msolar}{${\rm M}_\odot$}

\newcommand{\lgmstar}{$\log (M_\ast/$\msolar)}

\newcommand{\Reff}{{$R_{\rm e}$}}

\newcommand{\myemail}{\email{enci.wang@phys.ethz.ch}}

\defcitealias{Wang-19}{W19}
\defcitealias{Dopita-16}{DOP16}
\defcitealias{Pilyugin-16}{PG16}

\shorttitle{Gas-phase metallicity and variations in SFR}
\shortauthors{Wang \& Lilly}

\graphicspath{{fig/}}

\begin{document}
\title {Gas-phase metallicity as a diagnostic of the drivers of star-formation on different spatial scales}

\author {Enci Wang\altaffilmark{1},
Simon J. Lilly\altaffilmark{1}
} \myemail

\altaffiltext{1}{Department of Physics, ETH Zurich, Wolfgang-Pauli-Strasse 27, CH-8093 Zurich, Switzerland}

\begin{abstract}  

We examine the correlations of star formation rate (SFR) and gas-phase metallicity $Z$. We first predict how the SFR, cold gas mass and $Z$ will change with variations in inflow rate or in star-formation efficiency (SFE) in a simple gas-regulator framework. 
The changes $\Delta {\rm log}$SFR and $\Delta {\rm log} Z$, are found to be negatively (positively) correlated when driving the gas-regulator with time-varying inflow rate (SFE). 
We then study the correlation of $\Delta {\rm log}$sSFR (specific SFR) and $\Delta {\rm log}$(O/H) 
from observations, at both $\sim$100 pc  and galactic scales, based on two 2-dimensional spectroscopic surveys with different spatial resolutions, MAD and MaNGA.  
After taking out the overall mass and radial dependences, which may reflect changes in inflow gas metallicity and/or outflow mass-loading, we find that $\Delta {\rm log}$sSFR and $\Delta {\rm log}$(O/H) on galactic are found to be negatively correlated, but $\Delta {\rm log}$sSFR and $\Delta {\rm log}$(O/H) are positively correlated on $\sim$100 pc scales within galaxies. If we assume that the variations across the population reflect temporal variations in individual objects, we conclude that variations in the star formation rate are primarily driven by time-varying inflow at galactic scales, and driven by time-varying SFE at $\sim$100 pc scales. 
We build a theoretical framework to understand the correlation between SFR, gas mass and metallicity, as well as their variability, which potentially uncovers the relevant physical processes of star formation at different scales.

\end{abstract}


\section{Introduction}
\label{sec:introduction}
 
 
 Heavy elements, "metals", are produced in the universe through nuclear synthesis in massive stars, and are partly returned to the interstellar medium via the explosive collapse of supernovae. 
 The gas-phase metallicity of galaxies is therefore a powerful diagnostic of the different processes in galaxy evolution, including gas inflow, star formation in cold gas clouds, and wind-driven outflows of gas from galaxies.  The Oxygen abundance, mostly produced on short time-scales (a few Myr) by the rapid collapse and violent explosion of massive stars, i.e. the Type II supernovae, is widely used as an observationally accessible proxy of the metallicity of the gas in galaxies \citep[e.g.][]{Lequeux-79, Wheeler-89, Kewley-02, Pettini-04, Tremonti-04}. 
  
  Observationally, the gas-phase metallicity is usually measured based on the flux ratios of emission lines in star-forming HII regions \citep[e.g.][]{Kobulnicky-04,  Tremonti-04, Pettini-04,  Maiolino-08,  Kewley-08, Perez-Montero-09, Pilyugin-10, Rosales-Ortega-12,  Marino-13,  Dopita-13, Vogt-15, Dopita-16, Pilyugin-16}, such as [OII]$\lambda$3227, [OIII]$\lambda$4363, H$\beta$, [OIII]$\lambda$5007, H$\alpha$, [NII]$\lambda$6584, and [SII]$\lambda\lambda$6717,6731.   These emission lines are mostly excited by O and B stars, which must have formed recently, within the last $<$10 Myr, and the gas-phase metallicity measured in this way can therefore be treated as the current ``instantaneous" metallicity of the gas out of which the stars have formed.  This timescale is much shorter than the $\sim 10^{10}$ year lifetime of the galaxies or even the $\sim 10^9$ year gas depletion timescale \citep{Leroy-08, Bigiel-08, Shi-11}.
  
  In the literature, there is a number of empirical relations to derive the Oxygen abundance based on the combination of some of these emission lines \citep[see][and references therein]{Kewley-08, Sanchez-17}. As a whole, measurements of Oxygen abundance based on these different approaches are certainly positive correlated, but the absolute values and the ranges of these measurements are not consistent with each other \citep[e.g.][]{Kewley-08, Blanc-15, Sanchez-19}, due to the different methods, different samples and different emission lines used in the calibration. 
  
Even given the inconsistency of estimations of the gas-phase metallicity by different approaches, there is no dispute that the galaxy integrated gas-phase metallicity is strongly correlated with the stellar mass \citep[e.g.][]{Lequeux-79, Tremonti-04}. Based on fibre spectra of the central regions of large numbers of galaxies from SDSS \citep[the Sloan Digital Sky Survey;][]{Stoughton-02}, \cite{Tremonti-04} established a tight stellar-mass/gas metallicity relation (MZR) for star-forming (SF) galaxies spanning over three orders of magnitude in mass and one order of magnitude in gas-phase metallicity. The relation is relatively steep at low mass end, but flattens at stellar masses above $10^{10.5}$\msolar. Furthermore, the gas-phase metallicity appears to have a larger dispersion towards the lower stellar masses. The MZR is found to be evolving with redshift in the sense that galaxies at higher redshift tend to be more metal-poor with respect to the galaxies of the similar stellar mass in the local universe \citep[e.g.][]{Savaglio-05, Maier-06, Maiolino-08}.  The existence of the MZR relation can be explained by  one or the combination of these factors: the outflow by the supernova-driven winds \citep{Larson-74,Tremonti-04, Finlator-08}, the different star formation efficiencies in galaxies \citep{Brooks-07, Mouhcine-07, Calura-09},  and the variations in the initial mass function across galaxy population \citep{Koppen-07}. 

The gas-phase metallicity is also found to be correlated with other global properties of galaxies, such as the SFR \citep[e.g.][]{Ellison-08, Mannucci-10, Andrews-13} and half-light radius \citep[e.g.][]{Ellison-08, Wang-18b}. Based on the large sample of galaxies from SDSS, \cite{Mannucci-10} found that the negative correlation with star formation rate (SFR) is strong at low stellar mass end, and becomes less significant with increasing stellar mass. 
 Furthermore, they claimed that there was a universal epoch-independent  mass-metallicity-SFR relation $Z(M_*,{\rm SFR})$, i.e. that the apparent evolution in the MZR could be accounted for, phenomenologically, by the higher SFR encountered in high redshift galaxies.  This universal $Z(M_*,{\rm SFR})$ is therefore known as the ``fundamental metallicity relation'' \citep[FMR;][]{Mannucci-10, Lara-Lopez-10, Richard-11, Nakajima-12, Cresci-12, Dayal-13, Salim-14, Cresci-19, Huang-19, Curti-20}. 
 \cite{Cresci-19} finds that an anti-correlation between specific SFR (sSFR, defined as the SFR divided by the stellar mass) and gas-phase metallicity at given stellar mass, regardless of what the metallicity and SFR indicators are used. 
  
 
Recently, the emergence of widespread integral field spectroscopy (IFS) for galaxy surveys, such as MaNGA \citep{Bundy-15}, CALIFA \citep{Sanchez-12} and SAMI \citep{Croom-12}, has produced many spatially resolved spectroscopic data of relatively nearby galaxies. This enables the investigation of the relations of metallicity with mass (surface density) and star-formation rates  within galaxies. A strong correlation between local stellar surface density and local gas-phase metallicity, an apparent analog to the global MZR, is found based on the spatially-resolved spectroscopic data by many authors \citep[e.g.][]{Moran-12, Rosales-Ortega-12, Barrera-Ballesteros-16, Zhu-17, Gao-18}. 

However, whether the SFR or sSFR is a second parameter to the sub-galactic resolved MZR has been debated.  By using 38 nearby galaxies from the PINGS survey, \cite{Rosales-Ortega-12} found a negative correlation of gas-phase metallicity and the local specific SFR, indicated by the H$\alpha$ equivalent width \citep[also see][]{Zhu-17, Sanchez-Almeida-18, Hwang-19}. 
However, \cite{Moran-12} and \cite{Barrera-Ballesteros-17} argued that there is no evidence for the local sSFR (or SFR surface density) to be a robust second parameter in the resolved MZR.     
More recently, based on analysis of MaNGA galaxies (with a spatial resolution of 1-2 kpc), \cite{Berhane-Teklu-20} found a negative correlation between local metallicity and local sSFR when using {\tt N2} and {\tt O3N2} metallicity indicators, but the correlation is nearly disappeared for the {\tt N2O2} and {\tt N2S2} metallicity indicators \citep[also see][]{Sanchez-Menguiano-19}. 
Furthermore, by using the HII regions of eight nearby galaxies mapped by the Multi-Unit Spectroscopic Explore (MUSE), \cite{Kreckel-19} found that the regions with highest H$\alpha$ luminosity tended to have higher gas-phase metallicity at a $\sim$100 pc scale, indicating a {\it positive} correlation between metallicity and sSFR. Similarly, \cite{Ho-18} found that the oxygen abundances are higher in the spiral arms than in the inter-arm regions for NGC 2997, at a similar spatial resolution.  A clear picture to understand these seemingly contradictory findings is still lacking. 
 
Efforts have been made to understand the global star formation rates and metal content of galaxies, by looking at the balance between inflow, outflow and star formation \citep[e.g.][]{Schaye-10, Bouche-10, Dave-11, Lilly-13, Belfiore-19}. In particular, \cite{Dave-12} gave an analytic formalism that describes the evolution of the stellar mass, gas mass and metallicity of galaxies, assuming an equilibrium state in which the mass of the gas reservoir is assumed to be constant, i.e. $\dot{M}_{\rm gas}\sim$0. This scenario is also known as the ``reservoir'' or ``bathtube'' model \citep{Bouche-10, Dave-12}. Because the gas mass does not change, this model is not able to regulate the SFR. \cite{Lilly-13} released this restriction and allowed the mass of gas reservoir to change, so that the SFR is regulated by the changing gas mass adjusting to the inflow rate. This is known as the ``gas-regulator'' model. The gas-regulator model of \cite{Lilly-13} produces an analytic form of the mass metallicity relation that has the SFR naturally as a second parameter, i.e. $Z(M_*,{\rm SFR})$.  Further, the form of this is set by the basic parameters of the regulator model, specifically the star-formation efficiency and the mass-loading of the wind $\lambda$, both of which may vary with the overall stellar mass.  However, if these mass-dependent parameters are independent of epoch, as is quite plausible, then the form of $Z(M_*,{\rm SFR})$ will also not evolve.  The gas-regulator model is therefore naturally able to produce an epoch-independent FMR.  
     
The whole point of the \cite{Lilly-13} gas-regulator model is that the mass of gas in the galaxy can change.  In previous papers, we have explored the dynamical behaviour of the gas regulator model as it adjusts to variations in the inflow rate or other parameters and find that it can well explain several features of the galaxy population, and especially observations of the variation of SFR within galaxies and across the galaxy population. 

Based on a well defined sample of galaxies on the Star Formation Main Sequence \citep[SFMS; e.g.][]{Brinchmann-04, Daddi-07, Salim-07} from MaNGA, \citet[][hereafter \citetalias{Wang-19}]{Wang-19} investigated their SFR surface density ($\Sigma_{\rm SFR}$),  and found that the dispersion in the relative $\Sigma_{\rm SFR}$ (correcting for different effective radii) at a given relative radius in galaxies with similar stellar mass, decreases with increasing gas depletion time. The gas depletion timescale ($\tau_{\rm dep}$) is defined as the total cold gas mass divided by the current SFR for individual galaxies, which is exactly the inverse of star formation efficiency (SFE).
By driving a gas-regulator system with a periodic time-varying inflow rate, \citetalias{Wang-19} found that
regions with shorter gas depletion times are better able to follow variations in inflow rate, and therefore produce a larger dispersion in $\Sigma_{\rm SFR}$ at a given driving period and amplitude. It was suggested that this feature of the gas regulator model could produce the observed relation between the scatter of $\Sigma_{\rm SFR}$ with the inferred gas depletion time (see more details in \citetalias{Wang-19}).  Similarly, the dynamical gas regulator model can also qualitatively explain the observed dependence of the dispersion of the overall SFMS on stellar mass and stellar surface density \citep[][]{Wang-18b, Davies-19}. 

Consistent with, but quite independent of, our \citetalias{Wang-19} analysis, \cite{Wang-20a} found that regions with shorter gas depletion times also exhibit larger dispersions in the temporal changes in the SFR, as parameterized by SFR$_{\rm 5Myr}$/SFR$_{\rm 800Myr}$, the ratio of SFR averaged within the last 5 Myr to the SFR averaged within the last 800 Myr. 
The SFR$_{\rm 5Myr}$/SFR$_{\rm 800Myr}$ was estimated by the equivalent widths of H$\alpha$ emission and H$\delta$ absorption. The results in \cite{Wang-20a} therefore confirm that that the variation in the $\Sigma_{\rm SFR}$ profiles in \citetalias{Wang-19} are indeed apparently due to real {\it temporal} variations in the SFR within galaxies, rather than any intrinsic differences between galaxies in the population.  

Furthermore, based on the same dataset in \cite{Wang-20a}, \cite{Wang-20b} constrained the power spectral distribution (PSD) of the star formation histories of galaxies, i.e. the contribution of the variations in SFR at different timescales.  This too showed highly consistent results with our earlier results in \citetalias{Wang-19} and \cite{Wang-20a}. All these results strongly support the importance of the dynamical response of the simple gas-regulator system to a time-varying inflow in producing the variations of SFR or $\Sigma_{\rm SFR}$ at galactic scale.


Since the dynamical gas-regulator model has gained success in reproducing the un-evolving FMR \citep{Lilly-13}, and interpreting the dispersion of $\Sigma_{\rm SFR}$ across the galaxy population \citetalias{Wang-19}, it is interesting to further explore the behaviour of this system, and in particular to look again at its response to variations also in star-formation efficiency, and to explore further the gas-phase metallicity as a diagnostic tool.  This is the focus of the current paper.


In this work, we extend the work of \cite{Lilly-13} and \citetalias{Wang-19} and look at the metal-enrichment process in the dynamical gas-regulator framework. We will present the basic assumptions and continuity equations of the dynamical gas-regulator model in Section \ref{sec:2.1} and examine how the SFR, and the total mass, metal mass and gas-phase metallicity of the gas reservoir, vary in response to time-variations in the inflow rate of gas into the system and/or time-variations in the SFE (Section \ref{sec:2.2} and \ref{sec:2.3}). In addition, we will also explore how the wind mass-loading factor, the metallicity of the inflowing gas, and the yield (defined as the mass of metals returned to the interstellar medium per unit mass that is locked up into long-lived stars), can all modify these responses (Section \ref{sec:2.4}). 

We then turn to look for evidence of the predicted responses of the dynamic gas regulator in observational data. In Section \ref{sec:3}, we introduce the data used in this work, including the IFS data from the MaNGA survey and from the MUSE Atlas of Disks \citep[MAD;][]{Erroz-Ferrer-19}. The MaNGA sample is taken from \cite{Wang-18a} and \citetalias{Wang-19}, and includes nearly 1000 SF galaxies with typical spatial resolutions of 1-2 kpc, while the MAD sample has only 38 SF galaxies but with the spatial resolution down to 100 pc or even less. Therefore, the MaNGA sample is suitable to study the global effects at galactic and sub-galactic scale, while MAD galaxies can be used to study the effects on the scale of HII regions or individual molecular clouds. In Section \ref{sec:4} and \ref{sec:5}, we present the main observational results and compare them with the model predictions of the dynamical gas-regulator model. In Section \ref{sec:6}, we discuss our results compared with previous findings, and present the implications from the perspective of our understanding of relationship between SFR, cold gas mass and gas-phase metallicity at different physical scales. We summarize the main results of this work in Section \ref{sec:7}.   
 
Throughout this paper, we assume a flat cold dark matter cosmology model with $\Omega_m=0.27$, $\Omega_\Lambda=0.73$ and $h=0.7$ when computing distance-dependent parameters. 
The metallicity in this work is throughout the gas-phase metallicity. In constructing  the dynamical gas-regulator model (see Section \ref{sec:2}), we denote the gas-phase metallicity as Z, which is the mass ratio of metals to cold gas. However, observationally, the gas-phase metallicity is usually indicated by the Oxygen abundance, 12+$\log$(O/H), where the $\log$(O/H) is the ratio of the element number density of Oxygen to Hydrogen in logarithmic space. Therefore, we use 12+$\log$(O/H) to indicate the gas-phase metallicity from observations, which is usually used in Section \ref{sec:3}, \ref{sec:4}, and \ref{sec:5}. 

\section{The dynamic response of the gas-regulator model}
\label{sec:2}

\subsection{Basic continuity equations} \label{sec:2.1}

The basic idea of gas-regulator model is that the formation of stars is instantaneously determined by the mass of a cold gas reservoir, which is regulated by the interplay between inflow, outflow and star formation \citep{Lilly-13}. The instantaneous SFR can be written as: 
\begin{equation} \label{eq:1}
    {\rm SFR}(t) = M_{\rm gas}(t) \cdot {\rm SFE}(t), 
\end{equation}
where the SFE$(t)$ is the instantaneous star formation efficiency. We note that the SFE is the inverse of the gas depletion time ($\tau_{\rm dep}$) by definition, i.e. SFE$\equiv 1/\tau_{\rm dep}$.  Following the work of \cite{Lilly-13} and \citetalias{Wang-19}, we assume that the mass loss due to outflow is scaled by the instantaneous SFR$(t)$ with a mass-loading factor $\lambda$, i.e. $\lambda$SFR$(t)$. The fraction of stellar mass that is returned to the interstellar medium through winds and supernova explosions is denoted as $R$. We utilize the instantaneous return assumption and take $R = 0.4$ from stellar population models \citep[e.g.][]{Bruzual-03}. We define the effective gas depletion timescale ($\tau_{\rm def,eff}$) as the gas depletion timescale not only due to star formation but also the wind-loading outflow, i.e. $\tau_{\rm dep,eff}= \tau_{\rm def}/(1-R+\lambda)$. We denote the inflow rate as $\Phi(t)$, and the metallicity of infalling gas as $Z_0$. The metal mass of the gas reservoir is denoted as $M_{\rm Z}(t)$. The yield, i.e. the mass of metals returned to the interstellar medium per unit mass of instantaneously formed stars, is denoted as $y$. 

The basic continuity equations for gas and metals are \citep[see equations 9 and 20 in][]{Lilly-13}:  
\begin{equation} \label{eq:2}
\begin{split}
\frac{dM_{\rm gas(t)}}{dt} = & \ \Phi(t) - (1-R)\cdot{\rm SFR}(t) - \lambda \cdot {\rm SFR}(t) \\
   = & \ \Phi(t) - (1-R+\lambda) \cdot {\rm SFE}(t)\cdot M_{\rm gas}(t)
\end{split}
\end{equation}

\begin{equation} \label{eq:3}
\begin{split}
\frac{dM_{\rm Z}(t)}{dt} = &y {\rm SFR}(t) - Z(t) \cdot (1-R+\lambda) \cdot {\rm SFR}(t) + \Phi(t) Z_{\rm 0} \\
                   = & y {\rm SFE}(t)\cdot M_{\rm gas}(t) - (1-R+\lambda) {\rm SFE}(t) \cdot M_{\rm Z}(t) \\
                    & + \Phi(t) \cdot Z_{\rm 0}
\end{split}
\end{equation}
where $Z(t) = M_{\rm Z}(t)/M_{\rm gas}(t)$ by definition.

We apply the instantaneous re-cycling approximation in this work, i.e. we assume that the metal-enhancement by star formation is instantaneous.
We are ignoring the timescale for supernova ejecta to mix with the interstellar medium and start forming a new generation of stars (see more discussion in Section \ref{sec:6.4}), which can be a few tens of Myr \citep{Roy-95}.

In Equation \ref{eq:2} and \ref{eq:3}, there are a total of five quantities driving the solution: the possibly time-varying $\Phi(t)$, SFE$(t)$, and the (assumed constant) $\lambda$, $Z_0$ and $y$. The $M_{\rm gas}(t)$, SFR$(t)$, $M_{\rm Z}(t)$ and $Z(t)$ are then the response of the regulator system to these five quantities. 
In this work we assume the $\lambda$, $Z_0$ and $y$ are time-independent. 
The mass-loading factor is tightly correlated with the stellar mass of galaxies \citep{Hayward-17}, and is not likely to change significantly on timescales of Gyr given the current relatively low sSFR$\sim$0.1 Gyr$^{-1}$ of local SFMS galaxies.  
The yield $y$ is a physical parameter reflecting the nuclear synthesis activity and the relative number of massive stars, which is expected to be tightly correlated with the initial mass function (IMF) but which is not likely to change significantly on Gyr timescale. However, we note that $\lambda$, $Z_{\rm 0}$ and even $y$ may well change from galaxy-to-galaxy and even from region-to-region within individual galaxies. 

Equations \ref{eq:2} and \ref{eq:3} are the two basic continuity equations in the gas-regulator model. The following analysis in Section \ref{sec:2} will focus on the solution of these two equations by driving the regulator system with sinusoidal $\Phi(t)$ or sinusoidal SFE$(t)$, and investigate the correlation between the resulting instantaneous SFR$(t)$ and the gas-phase metallicity $Z(t)$. This correlation will be the main diagnostic used in our later analysis of observational data. 

In Section \ref{sec:2.2} and \ref{sec:2.3}, we will investigate the properties of the gas-regulator model using sinusoidally varying inflow rate or SFE.  Although this is undoubtedly artificial in isolation, we can argue that any given inflow rate or SFE can be expressed as a linear combination of sinusoidal functions with different frequencies via the Fourier transform. 
The gas-regulator system will respond to these individual sinusoidal components independently. For any given time-varying inflow rate, the resulting  $M_{\rm gas}$ (or $M_{\rm Z}$) should be the superposition of the resulting $M_{\rm gas}$ (or $M_{\rm Z}$) obtained from these sinusoidal variations of inflow rate. This statement is also true for SFE, when the variation of SFE is small (see details in Section \ref{sec:2.3}).   

\subsection{Driving the gas-regulator system with a time-varying inflow rate} \label{sec:2.2}

\begin{figure*}
  \begin{center}
    \epsfig{figure=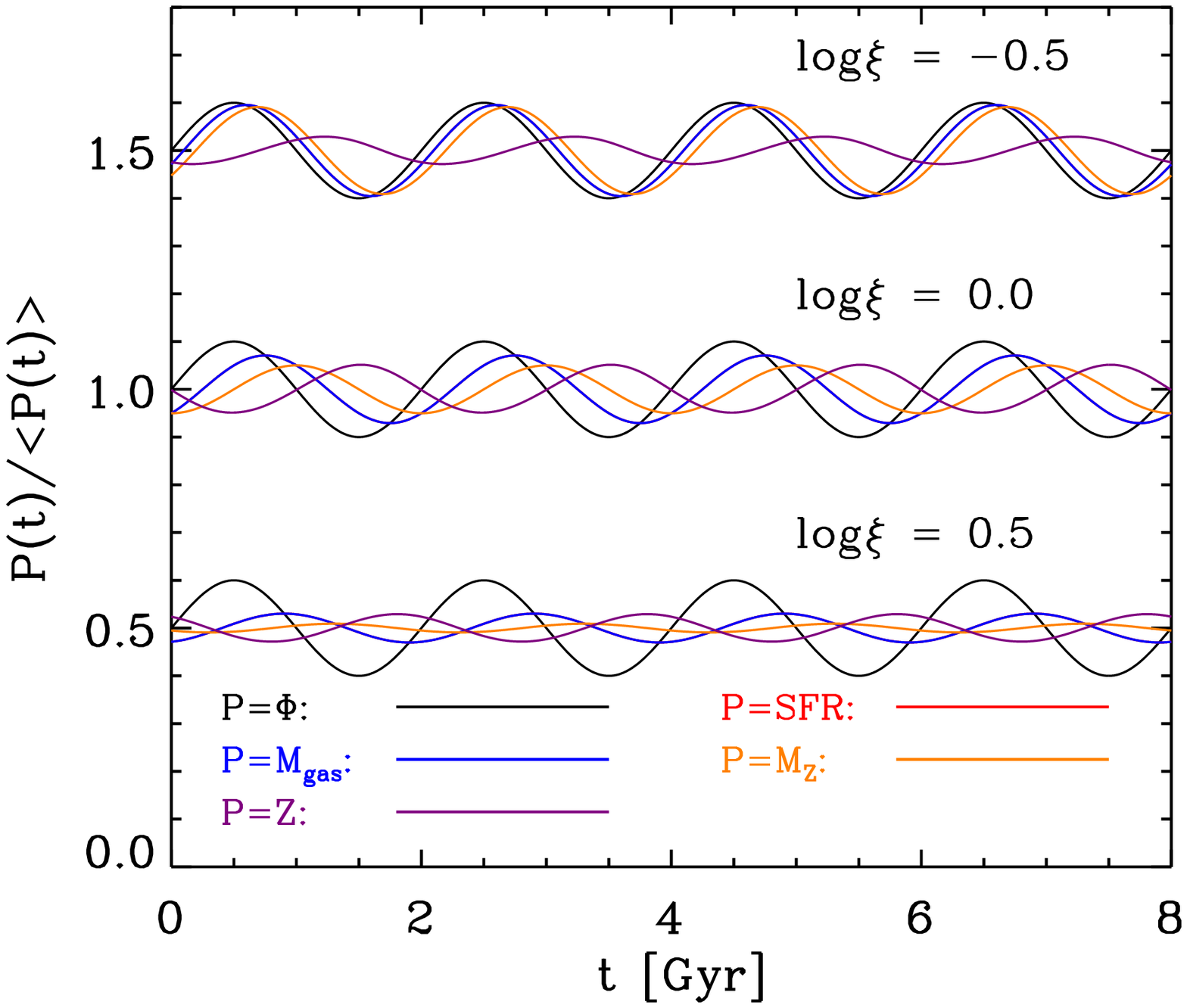,clip=true,width=0.32\textwidth}
    \epsfig{figure=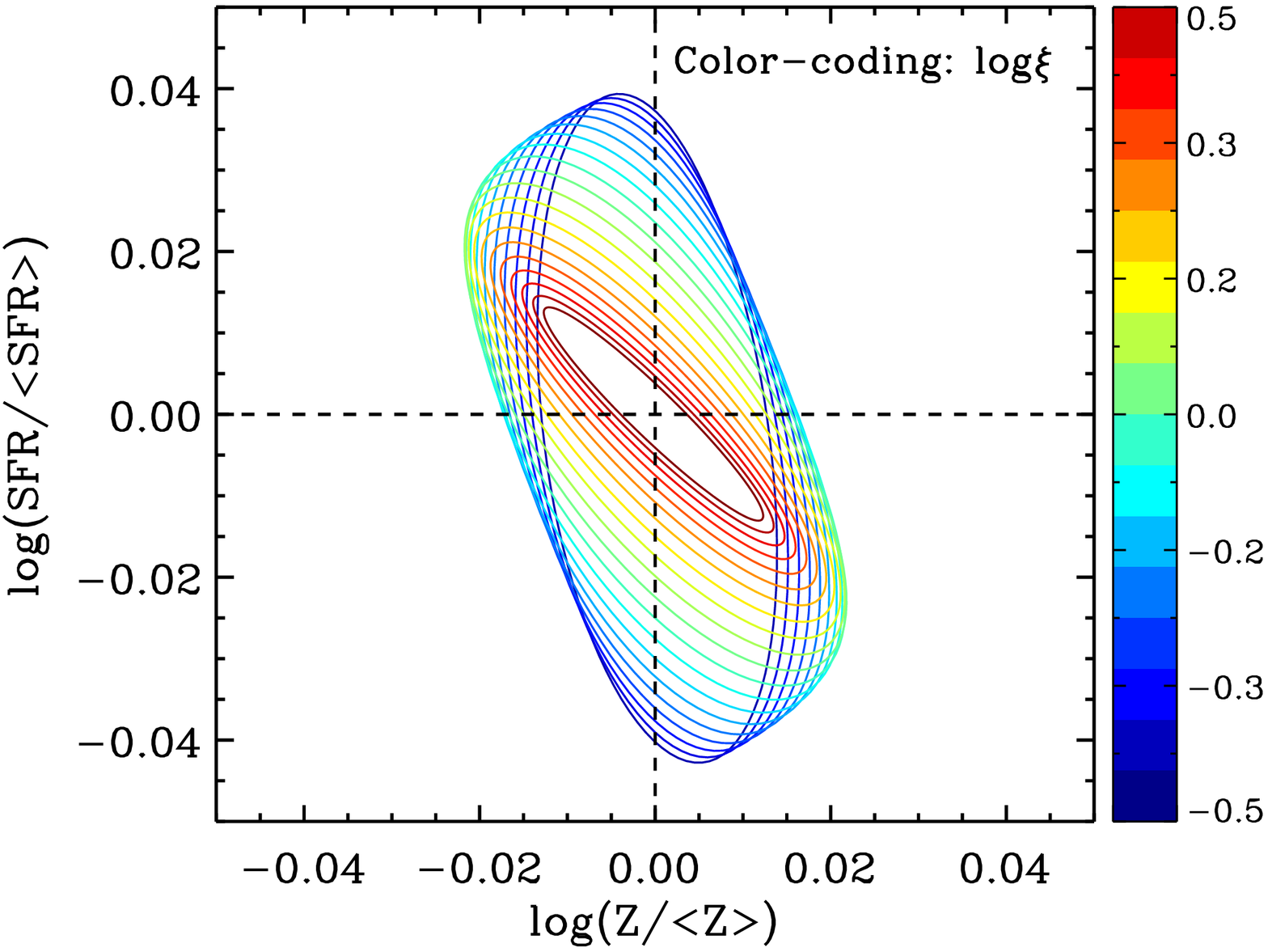,clip=true,width=0.33\textwidth}
    \epsfig{figure=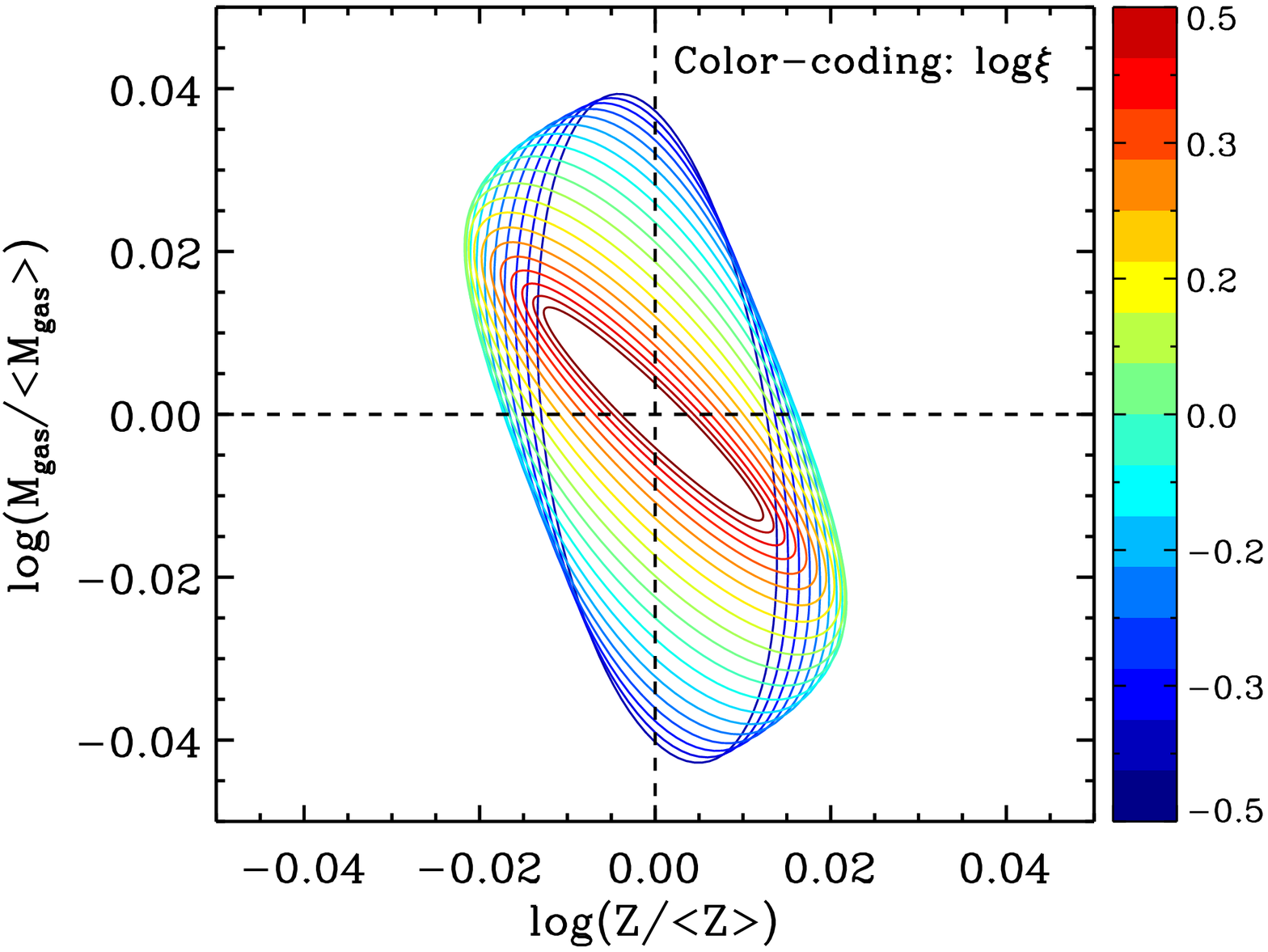,clip=true,width=0.33\textwidth}
    
    \epsfig{figure=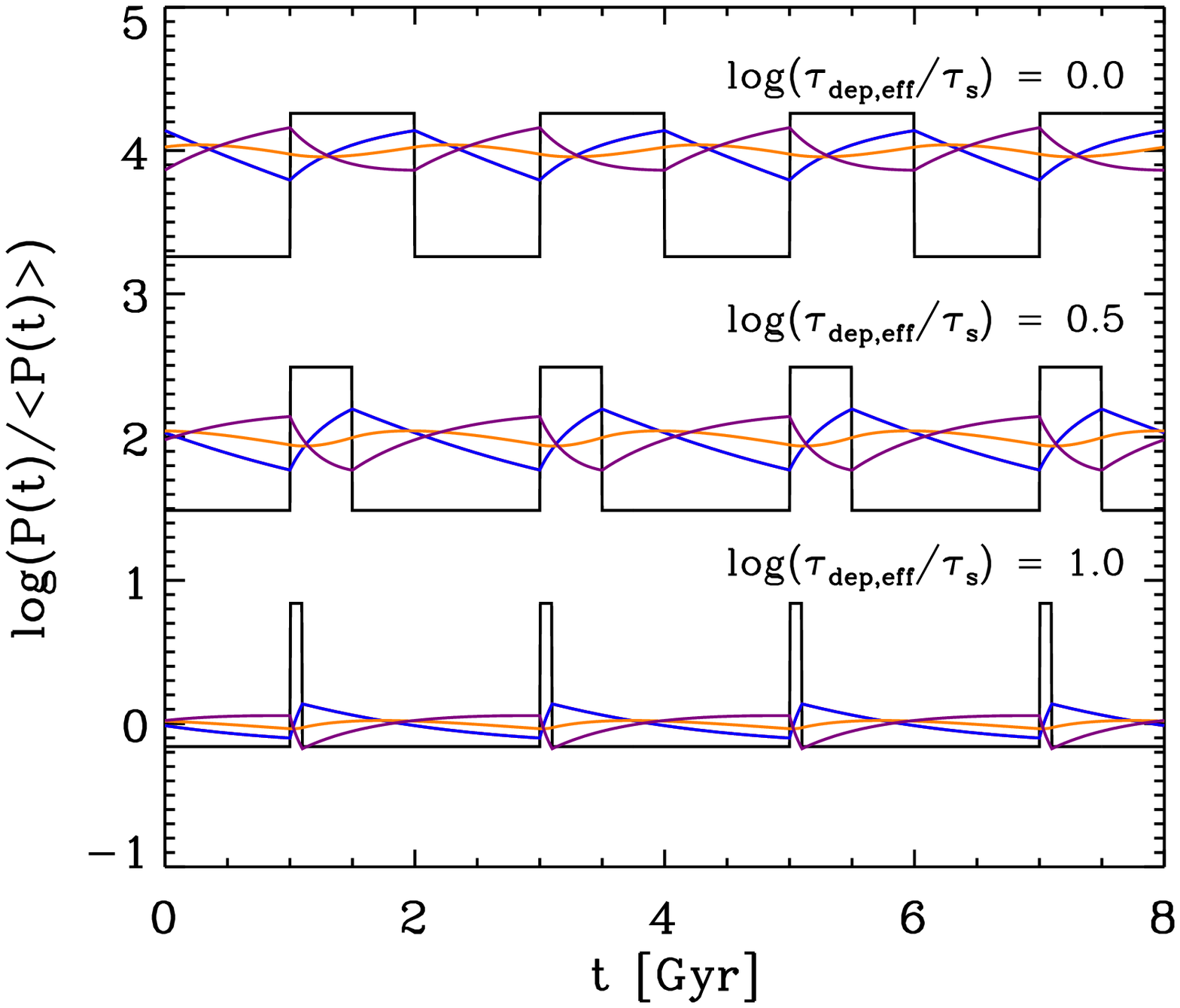,clip=true,width=0.32\textwidth}
    \epsfig{figure=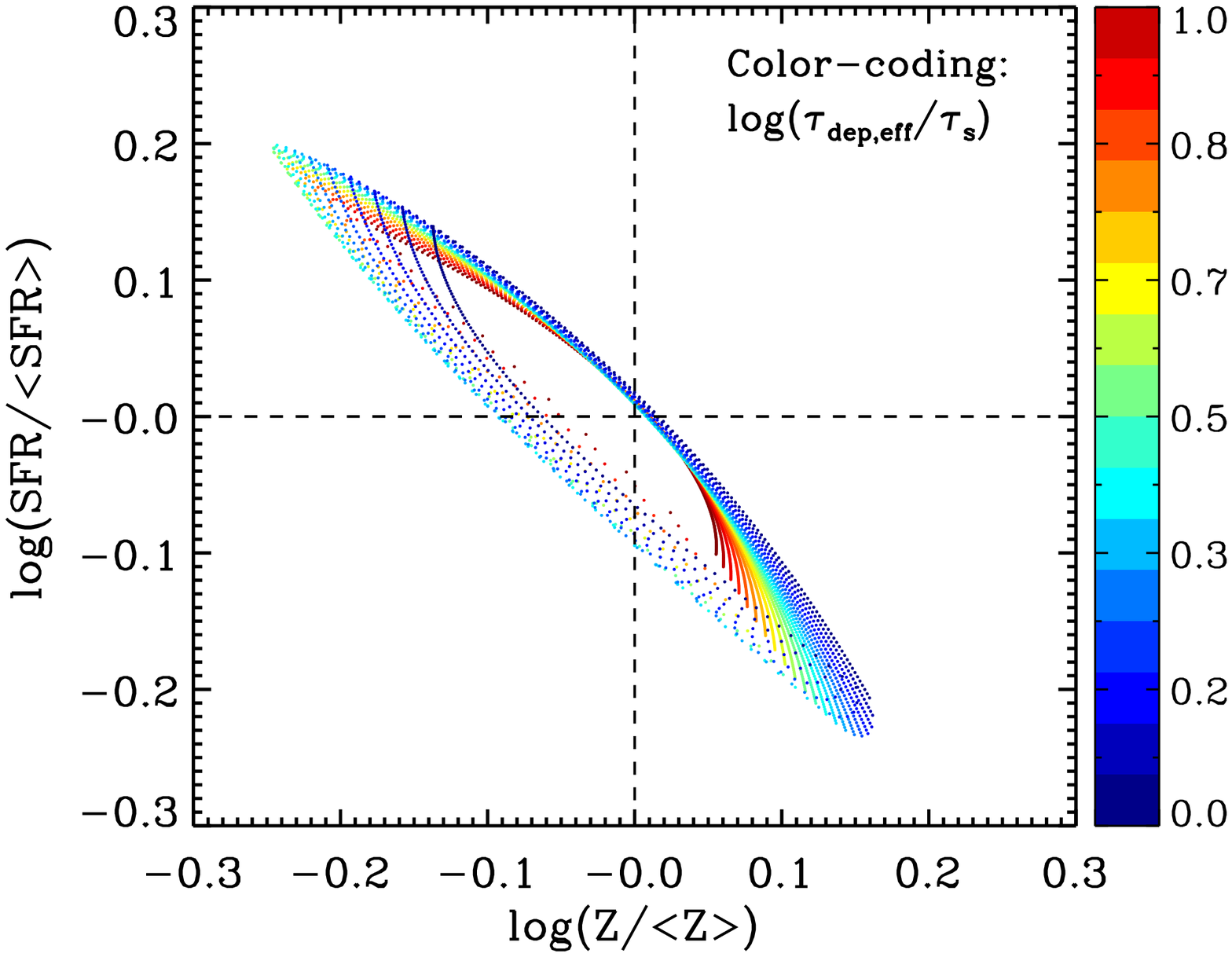,clip=true,width=0.33\textwidth}
    \epsfig{figure=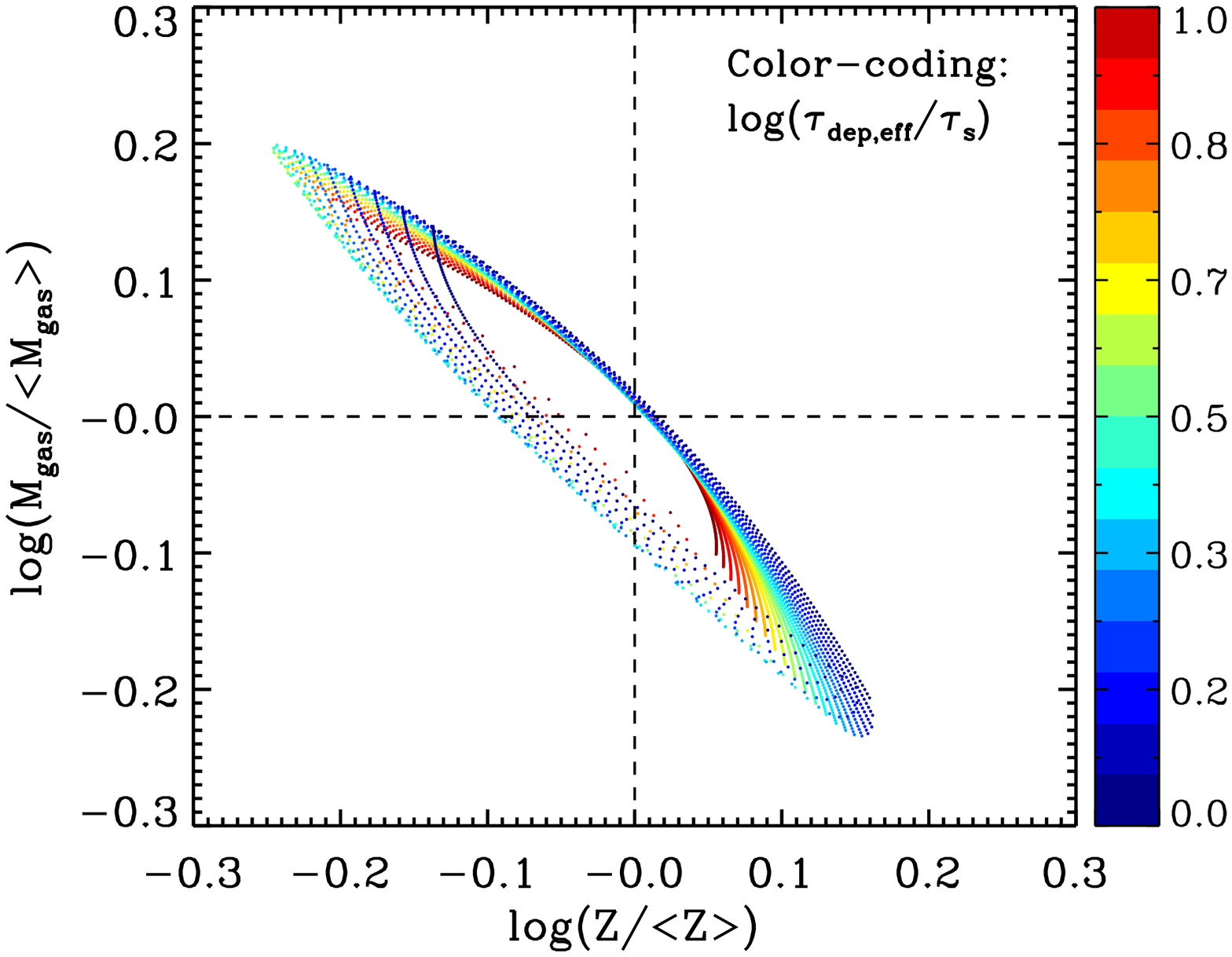,clip=true,width=0.33\textwidth}
    \end{center}
  \caption{Illustration of the SFR, \mgas\ and $Z$ in response to a sinusoidally varying inflow rate (upper panels) and a periodic step function inflow rate (lower panels), both with a constant SFE, in the gas regulator framework. 
  Upper left panel: Examples of the SFR$(t)$, \mgas$(t)$, \mz$(t)$, and $Z(t)$ (scaled to their mean values) in response to the sinusoidal inflow rate at three different $\xi$ (see text).  The cases of different $\xi$ are separated with arbitrary offsets ($-0.5$, 0.0, $+$0.5) in the y-axis for display purposes.
  Upper middle panel: The correlation of SFR and $Z$ in logarithmic space for different $\xi$. 
 Upper right panel: The correlation of SFR and \mgas\ in logarithmic space for different $\xi$ (see text).
  The lower panels are similar to the top panels, but for a periodic step function of the inflow rate.  For illustration, the period of the step-function is set to be 2 Gyr, and the $\tau_{\rm dep,eff}$ is set to be 1 Gyr. We denote the period that inflow rate is at its higher value as ``high-phase'', and the rest as ``low-phase''.
  The duration of the high-phase inflow rate ($\tau_{\rm s}$) within one period varies from 0.1$\tau_{\rm dep,eff}$ to $\tau_{\rm dep,eff}$.  The different colors in the lower middle and right panels are for the cases of different $\tau_{\rm dep,eff}/\tau_{\rm s}$, and the data points are equally spaced in time and so their density reflects the speed of the change of the model. 
  Since the SFE is set to be constant over time, the SFR$(t)$ is always overlapped with \mgas$(t)$ in the two left panels, and the middle panels are the same as the right-most panels.   }
  \label{fig:1}
\end{figure*}

We now first drive the gas-regulator system with time-varying inflow and time-invariant SFE. As in \citetalias{Wang-19}, we input the simple case that inflow rate is a linear sinusoidal function of time with a period of $T_{\rm p}$: 
\begin{equation} \label{eq:4}
  \Phi(t) = \Phi_{\rm 0} + \Phi_{\rm t} \cdot {\rm sin}(2\pi t/T_{\rm p}). 
\end{equation}
Then, 
we look for solutions in which $M_{\rm gas}(t)$ and $M_{\rm Z}(t)$ are sinusoidal functions phase-shifted from the inflow rate. 
\begin{equation} \label{eq:5}
\begin{split}
    M_{\rm gas}(t) = &\  M_{\rm 0} + M_{\rm t} \cdot {\rm sin}(2\pi t/T_{\rm p} - \delta) \\
    M_{\rm Z}(t)   = &\  M_{\rm Z0} + M_{\rm Zt} \cdot {\rm sin}(2\pi t/T_{\rm p} - \beta).
\end{split}
\end{equation}


In \citetalias{Wang-19}, by substituting $M_{\rm gas}(t)$ into Equation \ref{eq:2} and equalising the various time-dependent terms in the usual way, we have the solution of $M_{\rm gas}(t)$:  
\begin{equation} \label{eq:6}
\begin{split}
   M_{\rm 0} = & \  \Phi_{\rm 0}\tau_{\rm dep,eff} \\
   \delta \ \ = & \ {\rm arctan}(\xi) \\
   \frac{M_{\rm t}}{M_{\rm 0}} = & \ \frac{1}{(1+\xi^2)^{1/2}} \times \frac{\Phi_t}{\Phi_0},
\end{split}
\end{equation}
where $\tau_{\rm def,eff}$ is the effective gas depletion time, define as $\tau_{\rm dep}\cdot (1-R+\lambda)^{-1}$ or ${\rm SFE}^{-1}\cdot (1-R+\lambda)^{-1}$ in Section \ref{sec:2.1}, and $\xi$ is the ratio of the effective gas depletion timescale to $T_{\rm p}(2\pi)^{-1}$, i.e. 
\begin{equation} \label{eq:6.1}
\xi \equiv 2\pi \tau_{\rm dep,eff}/T_{\rm p}. 
\end{equation}

As we discussed in \citetalias{Wang-19}, the amplitude and phase-delay of the output $M_{\rm gas}(t)$ strongly depends on the parameter $\xi$, the relative timescale of gas depletion time to the driving period.   At fixed $T_{\rm p}$, galaxies or regions with shorter gas depletion time are more able to follow the changes of the inflow rate, leading to a smaller phase-delay and  larger amplitude, and vice versa (see more discussion in section 4 of \citetalias{Wang-19}).  

In the similar way, we substitute Equations \ref{eq:5} and \ref{eq:6} into Equation \ref{eq:3}, and equate the various time-dependent terms to find the solution of $M_{\rm Z}(t)$:
\begin{equation} \label{eq:8}
\begin{split}
   M_{\rm Z0} = &\  (y_{\rm eff}+Z_{\rm 0})\Phi_{\rm 0}\tau_{\rm dep,eff} \\
   \beta \ \  = & \ {\rm arctan}[\frac{2y_{\rm eff}\xi + Z_{\rm 0}\xi(1+\xi^2)}{y_{\rm eff}(1-\xi^2)+Z_{\rm 0}(1+\xi^2)}] \\
   \frac{M_{\rm Zt}}{M_{\rm Z0}} = &\  \frac{(1+\eta^2)^{1/2}}{1+\xi^2}\times \frac{\Phi_{\rm t}}{\Phi_{\rm 0}},
\end{split}
\end{equation}
where
\begin{equation} \label{eq:9}
    y_{\rm eff} \equiv y\cdot(1-R+\lambda)^{-1}
\end{equation}
and 
\begin{equation} \label{eq:10}
    \eta = \xi Z_0 \cdot (y_{\rm eff}+Z_0)^{-1}. 
\end{equation}
If $\beta$ is less than zero, then $\beta$ should equal to $\beta$+$\pi$. 
The shorthand $\eta$ is defined for convenience. We remind readers that the effective yield $y_{\rm eff}$ defined in this way is {\it different} from some previous papers \citep[e.g.][]{Edmunds-90, Garnett-02}. We prefer this definition because we believe it is more fundamental. 

If we assume that the inflow gas is in a pristine state, i.e. $Z_0\sim0$, then the solution of $M_{\rm Z}(t)$ can be simplified further to be 
\begin{equation} \label{eq:12}
    \begin{split}
     M_{\rm Z0} = &\  y_{\rm eff}\Phi_{\rm 0}\tau_{\rm dep,eff} \\
     \beta \ \  = &\  {\rm arctan}[\frac{2\xi}{1-\xi^2}] \\
     \frac{M_{\rm Zt}}{M_{\rm Z0}} = & \ \frac{1}{1+\xi^2}\times \frac{\Phi_{\rm t}}{\Phi_{\rm 0}}.
    \end{split} 
\end{equation}

Interestingly, in this specific case with $Z_0\sim0$, the phase-delay of $M_{\rm Z}(t)$ is twice that of $M_{\rm gas}(t)$, i.e. $\beta=2\delta$. Similar to $M_{\rm gas}(t)$, 
the phase-delay $\beta$ and relative amplitude $M_{\rm Zt}/M_{\rm Z0}$ of $M_{\rm Z}(t)$ strongly depend on the parameter $\xi$. At fixed $T_{\rm p}$, galaxies or regions with shorter (effective) gas depletion time, can more easily follow the change of inflow rate and gas mass, resulting in smaller $\beta$ and larger $M_{\rm Zt}/M_{\rm Z0}$. 
Specifically, if $\xi$ is much less than unity, then both $\delta$ and $\beta$ are close to zero, and both $M_{\rm t}/M_{\rm 0}$ and $M_{\rm Zt}/M_{\rm Z0}$ are close to $\Phi_{\rm t}/\Phi_{\rm 0}$. In other words, when the (effective) gas depletion time is much less than the driving period, i.e. $\xi \ll 1$,  the change of  mass of gas reservoir and of the mass of metals in the gas-regulator system can nearly follow the change of inflow rate, with little phase-delay and with nearly the same relative amplitude of variation. If, however, $\xi$ is much larger than 1, then $\delta$ is close to $\pi/2$, $\beta$ is close to $\pi$, and both $M_{\rm t}/M_{\rm 0}$ and $M_{\rm Zt}/M_{\rm Z0}$ are close to zero. This means that, when the (effective) gas depletion time is much longer than the driving period, i.e. $\xi \gg 1$, the gas-regulator system is unable to follow the relatively fast changes in the inflow rate, resulting in little variation in either $M_{\rm gas}(t)$ or $M_{\rm Z}(t)$. 

The dependence of $M_{\rm gas}(t)$ and $M_{\rm Z}(t)$ on $\xi$ can be clearly seen in the top left panel of Figure \ref{fig:1}, where we show examples of the evolution of $M_{\rm gas}$ (blue), SFR (red), $M_{\rm Z}$ (orange) and $Z$ (purple) when driving the gas-regulator system with periodic sinusoidal inflow. For illustrative purpose, we set $Z_{\rm 0}=0$,   $\Phi_{\rm t}/\Phi_{\rm 0}=0.1$, $T_{\rm p}=1$ Gyr, and $\log \xi=-0.5$, 0.0, and 0.5.

Given the solutions of $M_{\rm gas}(t)$ and $M_{\rm Z}(t)$ in Equation \ref{eq:6} and \ref{eq:8}, the resulting SFR$(t)$ and $Z(t)$ can be easily obtained. 
Since the SFE is assumed here (in this subsection) to be time-invariant, the change of SFR will exactly follow the change of cold gas mass. Therefore, the blue and red lines in the top left panel of Figure \ref{fig:1} are overlapped together. However, the $Z(t)$, i.e. the ratio of $M_{\rm Z}(t)$ to $M_{\rm gas}(t)$, has a more complicated behavior than SFR(t), because it is not a sinusoidal function. The variation of the metallicity depends on the amplitude of variations in $M_{\rm Z}(t)$ and $M_{\rm gas}(t)$, as well as the phase-delay between the two. 

To clarify the correlation between the instantaneous SFR$(t)$ and $Z(t)$, we plot the $\log ({\rm SFR}(t)/\langle {\rm SFR}\rangle)$ vs. $\log (Z(t)/\langle {Z}\rangle)$ and $\log (M_{\rm gas}(t)/\langle {M_{\rm gas}}\rangle)$ vs. $\log (Z(t)/\langle {Z}\rangle)$ for a set of different $\xi$ in the top middle and right panels of Figure \ref{fig:1}, where $\langle {Z}\rangle$, $\langle {\rm SFR}\rangle$ and $\langle {M_{\rm gas}}\rangle$ are the average metallicity, SFR and cold gas mass, respectively.  Since the $\log ({\rm SFR}(t)/\langle {\rm SFR}\rangle)$ is a relative quantity, i.e. $\log {\rm SFR}(t) - \log \langle {\rm SFR}\rangle$, we also denote the $\log ({\rm SFR}(t)/\langle {\rm SFR}\rangle)$ as $\Delta \log$SFR. In the same way, we denote the $\log (Z(t)/\langle {Z}\rangle)$ as $\Delta \log Z$, and $\log (M_{\rm gas}(t)/\langle {M_{\rm gas}}\rangle)$ as $\Delta \log M_{\rm gas}$. 

As shown, at all the different $\xi$ shown here, the gas-regulator model predicts that $\Delta \log$SFR and $\Delta \log Z$ are {\it negatively} correlated when the system is driven with a sinusoidal inflow rate. The slope and the tightness of the $\Delta \log$SFR-$\Delta \log Z$ correlation strongly depend on $\xi$. Generally speaking, at fixed $T_{\rm p}$, the correlation of $\Delta \log$SFR-$\Delta \log Z$ becomes weaker and steeper with increasing the effective gas depletion time. The slope of $\Delta \log$SFR-$\Delta \log$Z relation is always steeper than $-1$. This means that the gas-regulator model requires that the scatter of $\Delta \log Z$ is always less than or equal to the scatter of $\Delta \log$SFR. We will come back to this point later. 


In addition to the sinusoidal inflow rate, we also for completeness explored the effect of a periodic step function in the inflow rate. We solve the Equation \ref{eq:2} and \ref{eq:3} numerically for this case. The bottom panels of Figure \ref{fig:1} show the resulting $M_{\rm gas}(t)$, SFR$(t)$, $M_{\rm Z}(t)$ and $Z(t)$, as well as the resulting correlation between $\Delta \log$SFR (and $\Delta \log M_{\rm gas}$) and $\Delta \log Z$. In generating the plots, we set the period of step function as 2 Gyr, and change the upper-state duration of inflow rate ($\tau_{\rm s}$). We allow the $\tau_s$ varying from $0.1\tau_{\rm dep,eff}$ to $\tau_{\rm dep,eff}$. 

As shown in the bottom-left panel of Figure \ref{fig:1}, a sudden increase of inflow rate causes an increase of the SFR (or cold gas mass) and a decrease of gas-phase metallicity, and vice versa. This therefore also leads to a {\it negative} correlation between SFR (or cold gas mass) and metallicity, i.e. between $\Delta \log$SFR (or $\Delta \log M_{\rm gas}$) and $\Delta \log Z$, consistent with the result for the sinusoidal variation in the top panels of Figure \ref{fig:1}. 



Of course, observationally we cannot follow the temporal evolution of a single galaxy.  In \citetalias{Wang-19} we therefore explored the scatter of the instantaneous SFR in a population of gas regulators, $\sigma({\rm SFR})$, when they are driven with simple sinusoidal $\Phi(t)$, and showed that within this population $\sigma({\rm SFR})$/$\sigma(\Phi)$ is a monotonically decreasing function of $\xi$:

\begin{equation} \label{eq:13}
\frac{\sigma({\rm SFR})}{\sigma(\Phi)} = \frac{1}{(1+\xi^2)^{1/2}} \cdot (1-R+\lambda)^{-1}.
\end{equation}

In Equation \ref{eq:13}, the scatter of SFR and $\Phi$ are calculated in linear space, while in the observations, the scatter of SFR is usually measured in logarithmic space. Here we present an approximate analytical solution of $\sigma({\rm \log SFR})/\sigma(\log \Phi)$, which can be written as: 
\begin{equation} \label{eq:14}
\frac{\sigma({\rm \log SFR})}{\sigma(\log \Phi)} \approx \frac{1}{(1+\xi^2)^{1/2}}.
\end{equation}
As can be seen in Equation \ref{eq:14}, if the scatter is measured in logarithmic space, the factor $1-R+\lambda$ vanishes, since this can be viewed as a constant ``inefficiency'' in the star-formation \citep[see][]{Lilly-13}. The details to derive the Equation \ref{eq:14} are given in the Appendix \ref{sec:A}. 

The left panel of Figure \ref{fig:3} shows the numerical solution (black solid curve) and the approximate analytical solution (gray dashed curve) of $\sigma({\rm \log  SFR})/\sigma(\log \Phi)$ as a function $\log\xi$, which are in excellent agreement. Specifically, in obtaining the numerical solution, we first solve the Equation \ref{eq:2} to obtain the SFR($t$) at a set of different $\xi$. Then we calculate the $\sigma(\log {\rm SFR})$ within a single period for each $\xi$. The $\sigma(\log Z)$ is calculated in the similar way. 
The analytic solution provides the physical insight to how $\xi$ determines the response of gas-regulator model, while the numerical solution provides a double-check for the analytic solution, and provides the judgment for the validation of the {\it approximate} analytic solution.  

As pointed out in \citetalias{Wang-19} and \cite{Wang-20b}, when driving the gas-regulator with time-varying inflow rate, the amplitude of the variations in SFR is reduced from the amplitude of variations in the inflow rate by a frequency-dependent ($\nu=1/T_{\rm p}$) response curve, i.e. Equation \ref{eq:14} or equation 9 in \citetalias{Wang-19}. In other words, for a given inflow rate with any given PSD$_{\log\Phi}$($\nu$), the power spectral distribution of the resulting logSFR, PSD$_{\rm  \log SFR}$($\nu$), can be written as: 
\begin{equation} \label{eq:28}
  \begin{split}
   {\rm PSD}_{\rm \log SFR}(\nu) \approx & \frac{\sigma^2(\log {\rm SFR})}{\sigma^2(\log \Phi)}\cdot {\rm PSD}_{\log \Phi}(\nu) \\
       \approx & \frac{1}{1+\xi^2} \cdot {\rm PSD}_{\log \Phi}(\nu) \\
       = & \frac{1}{1+(2\pi\tau_{\rm dep,eff}\nu)^2} \cdot {\rm PSD}_{\log \Phi}(\nu). 
  \end{split}
\end{equation}
The Equation \ref{eq:28} is established under the condition that the variation of inflow rate is small, and therefore the logSFR(t) and log$\Phi(t)$ are close to sinusoidal functions. However, as examined in Appendix \ref{sec:C}, we directly input the sinusoidal $\Phi(t)$ in logarithmic space with a large amplitude (0.5 dex), and find that the Equation \ref{eq:28} is still valid.  Therefore, we conclude that the Equation \ref{eq:28} theoretically predicts the connection between the PSDs of $\log$SFR and $\log \Phi$, when driving the gas-regulator with time-varying inflow rate \citep[also see \citetalias{Wang-19};][]{ Wang-20a,Tacchella-20}. However, from the observations, the inflow rate history of galaxies is not of course a directly observable quantity. 

Finally, we come to the scatter of the gas-phase metallicities in the population. As shown in top middle panel of Figure \ref{fig:1}, the ratio of the scatter in $Z(t)$ to the scatter of SFR$(t)$ (i.e. what would be observed in a given population of regulators at fixed time) is predicted to be strongly dependent on $\xi$. Here we present the approximate analytical solution of $\sigma({\rm \log Z})/\sigma(\log {\rm SFR})$, which can be written as: 
\begin{equation} \label{eq:15}
\frac{\sigma(\log Z)}{\sigma(\log {\rm SFR})} \approx \frac{\xi}{(1+\xi^2)^{1/2}} \cdot \frac{1}{1+Z_0/y_{\rm eff}}.
\end{equation}
The detailed derivation of Equation \ref{eq:15} is shown in Appendix \ref{sec:A}. Similar to Equation \ref{eq:14}, Equation \ref{eq:15} provides the link between the variability of $\log$SFR and $\log$Z. However, one can not write the correlation of the PSD$_{\rm \log SFR}$ and PSD$_{\rm \log Z}$ in a similar way as Equation \ref{eq:28}, because the resulting metallicity is not a sinusoidal function. 
Since the $\sigma({\rm \log Z})/\sigma(\log {\rm SFR})$ only depends on $\xi$ at fixed $Z_{\rm 0}/y_{\rm eff}$, we therefore present both the numerical (the blue, green and red curves) and the analytic solutions (the gray dashed curves) of the $\sigma(\log Z)/\sigma(\log {\rm SFR})$ in the left panel of Figure \ref{fig:3} at three different $Z_{\rm 0}/y_{\rm eff}$, which are again in excellent agreement. For different $Z_{\rm 0}/y_{\rm eff}$, the $\sigma({\rm \log Z})/\sigma(\log {\rm SFR})$ monotonically increases with $\log \xi$, which is opposite to $\sigma({\rm \log SFR})/\sigma(\log \Phi)$.  Intriguingly, if $Z_0=0$, the Equation \ref{eq:14} and \ref{eq:15} are strictly symmetrical across the axis of $\log \xi=0$. 
Unlike Equation \ref{eq:14}, the Equation \ref{eq:15} prediction of the gas-regulator model relates two readily observable quantities, the instantaneous SFR and the instantaneous gas-phase metallicity. 

\subsection{Driving the gas-regulator system with time-varying SFE} \label{sec:2.3}

\begin{figure*}
  \begin{center}
    \epsfig{figure=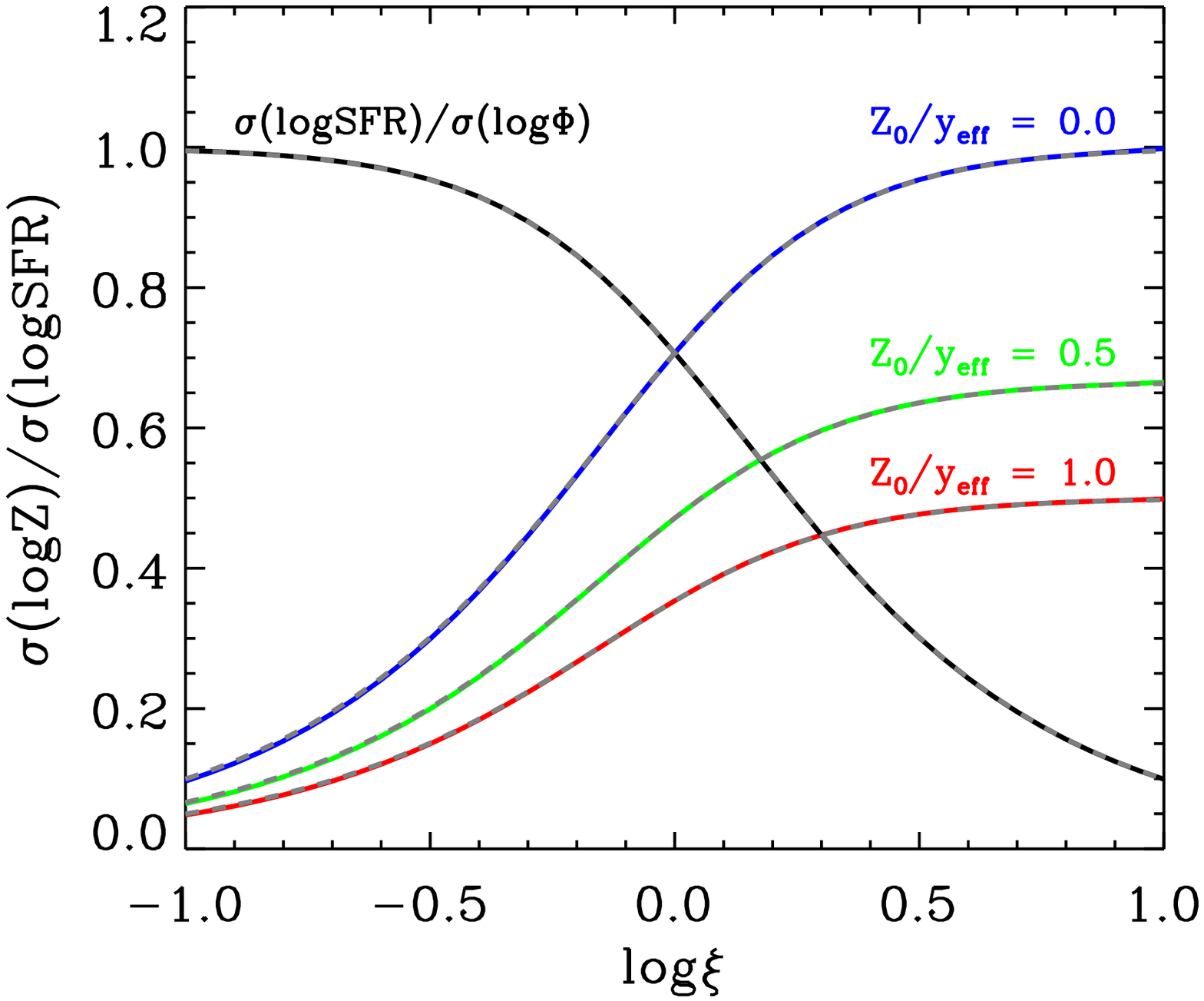,clip=true,width=0.42\textwidth}
    \epsfig{figure=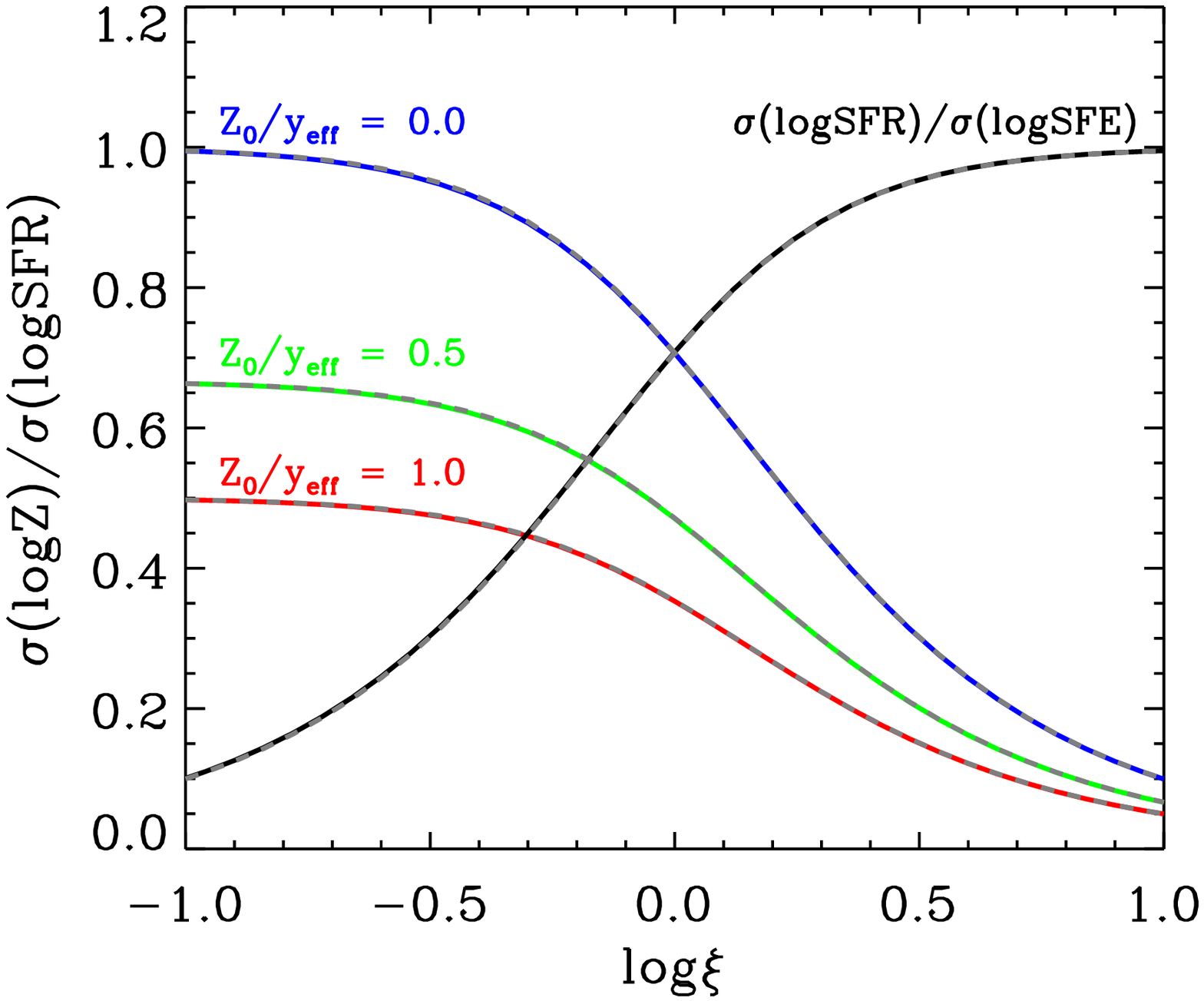,clip=true,width=0.42\textwidth}
    \end{center}
  \caption{Left panel: The ratio of $\sigma$($\log Z$) to $\sigma$($\log$SFR) as a function of $\xi$ determined by the numerical calculation, when the gas regulator is driven with a sinusoidal inflow rate ($\Phi_{\rm t}/\Phi_{\rm 0}=0.1$) and constant SFE. The colored lines show the relations for three different $Z_{\rm 0}/y_{\rm eff}$. In the meanwhile, we also display the ratio of $\sigma$($\log$SFR) to $\sigma$($\log \Phi$) as a function of $\xi$ as a black line. Each line is followed by a gray dashed line, which shows the approximate analytic solution (see Equation \ref{eq:14} and \ref{eq:15}).
  Right panel: The same as the left panel but for the case in which the gas regulator system is driven with a sinusoidal SFE and constant inflow rate (see Equation \ref{eq:19} and \ref{eq:20}). }
  \label{fig:3}
\end{figure*}

\begin{figure*}
  \begin{center}
    \epsfig{figure=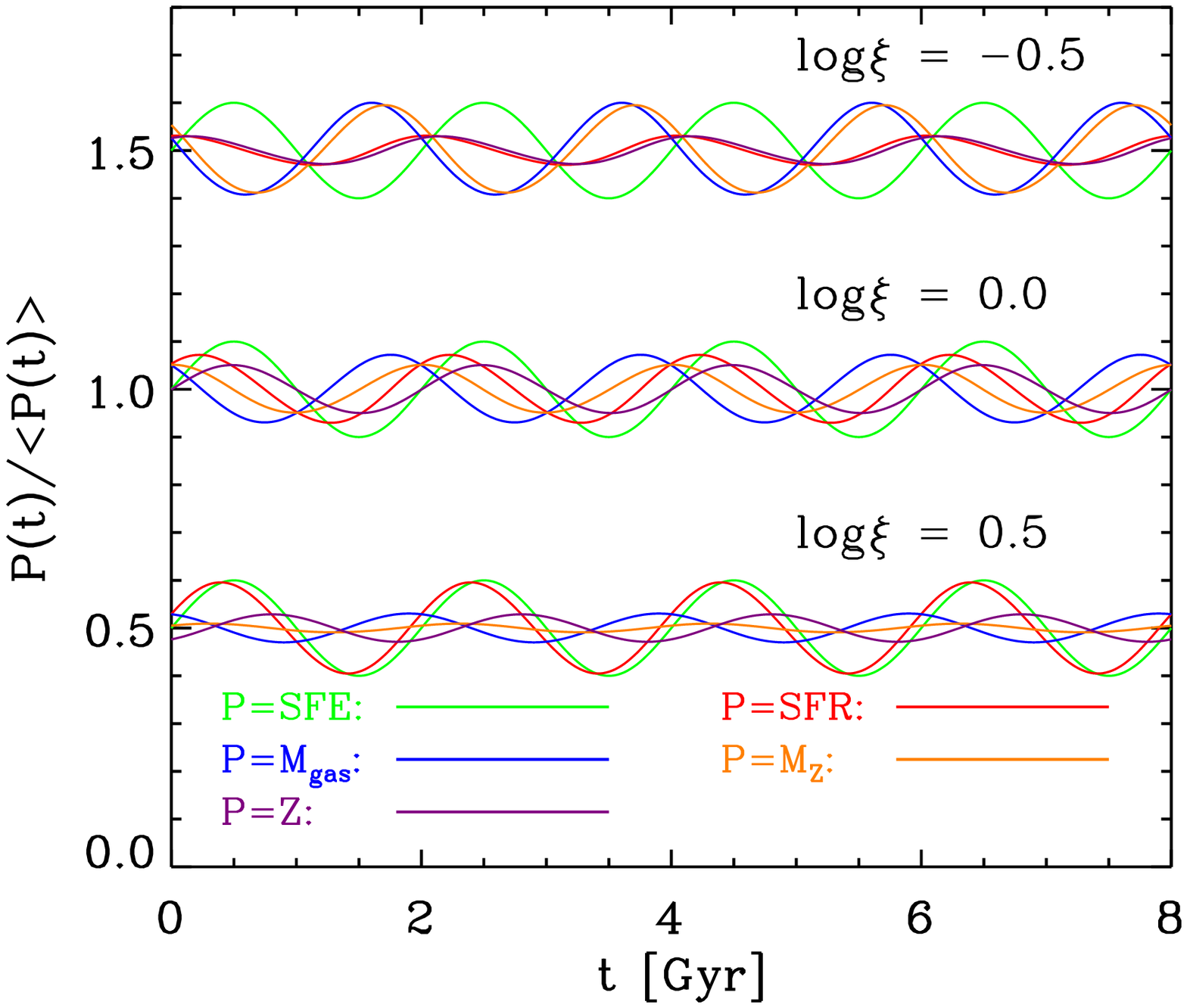,clip=true,width=0.32\textwidth}
    \epsfig{figure=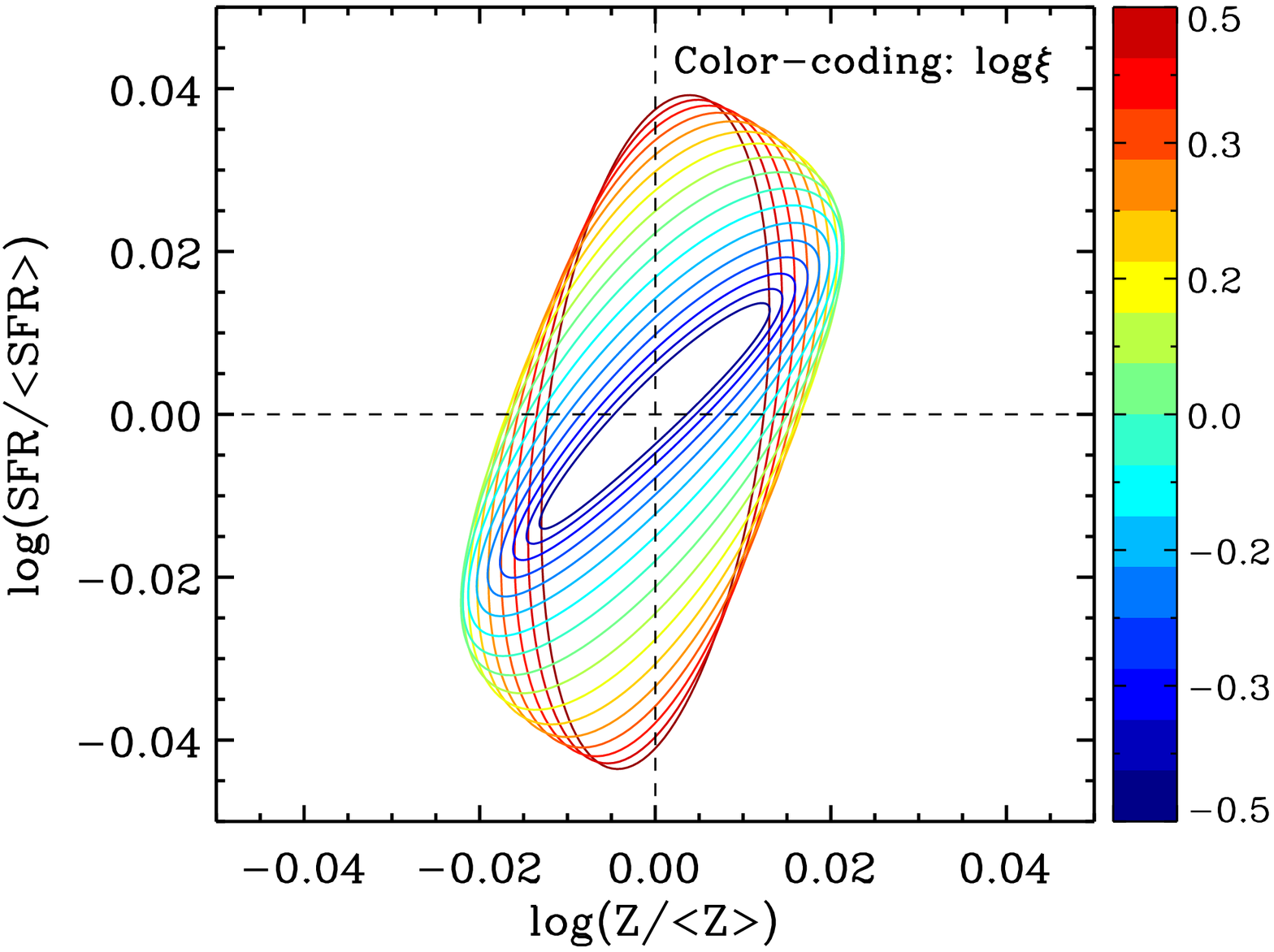,clip=true,width=0.33\textwidth}
    \epsfig{figure=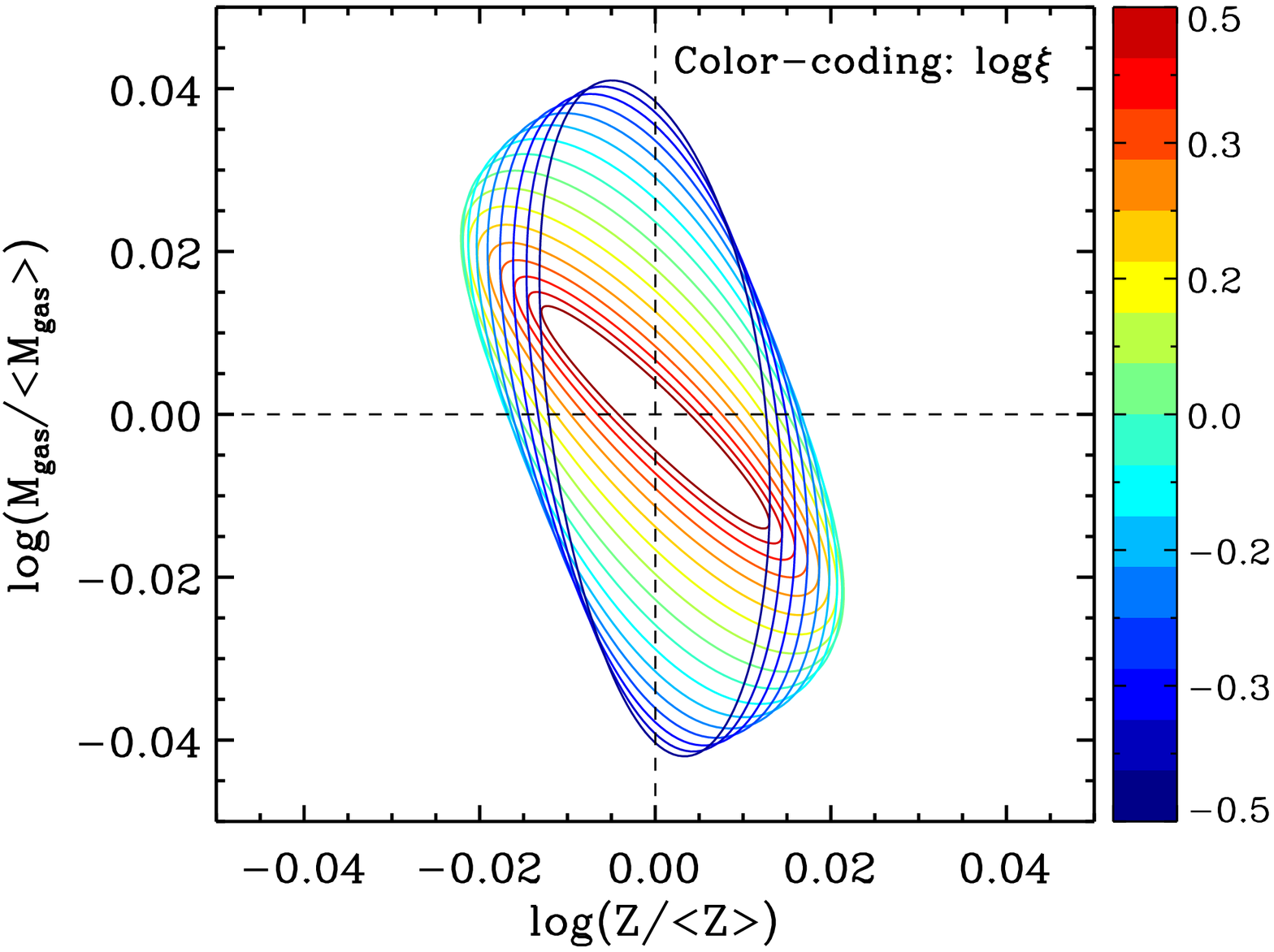,clip=true,width=0.33\textwidth}
    
    \epsfig{figure=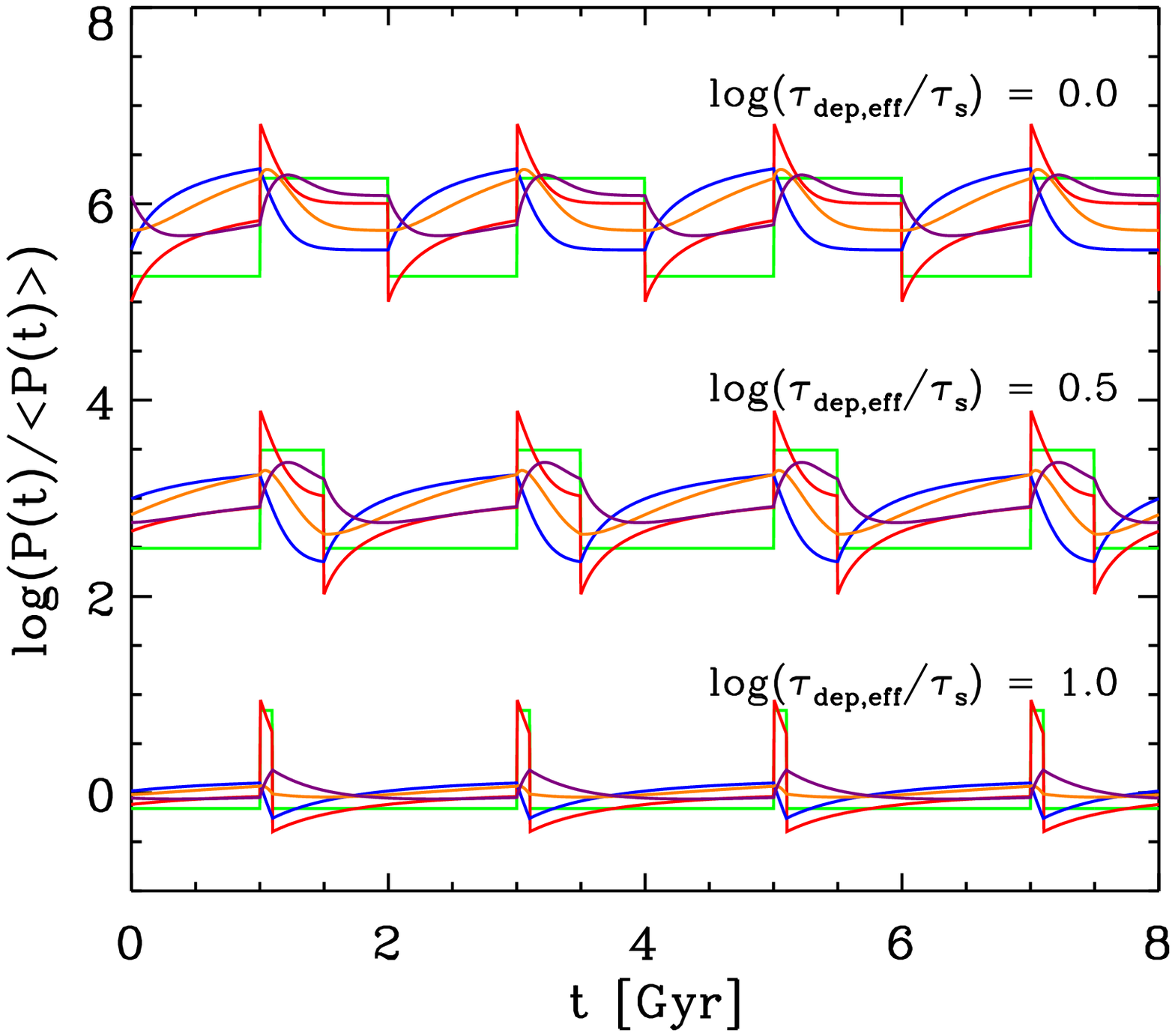,clip=true,width=0.32\textwidth}
    \epsfig{figure=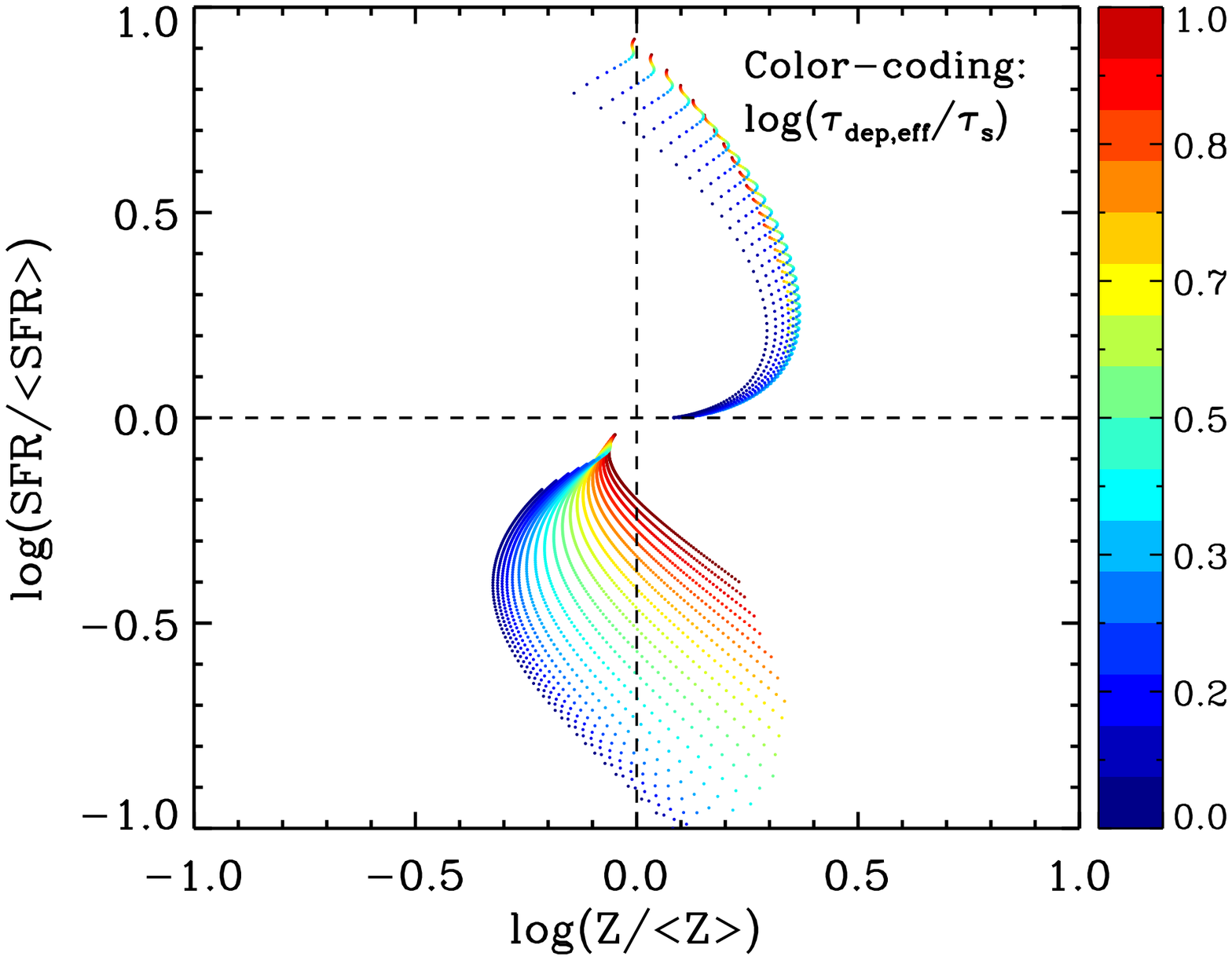,clip=true,width=0.33\textwidth}
    \epsfig{figure=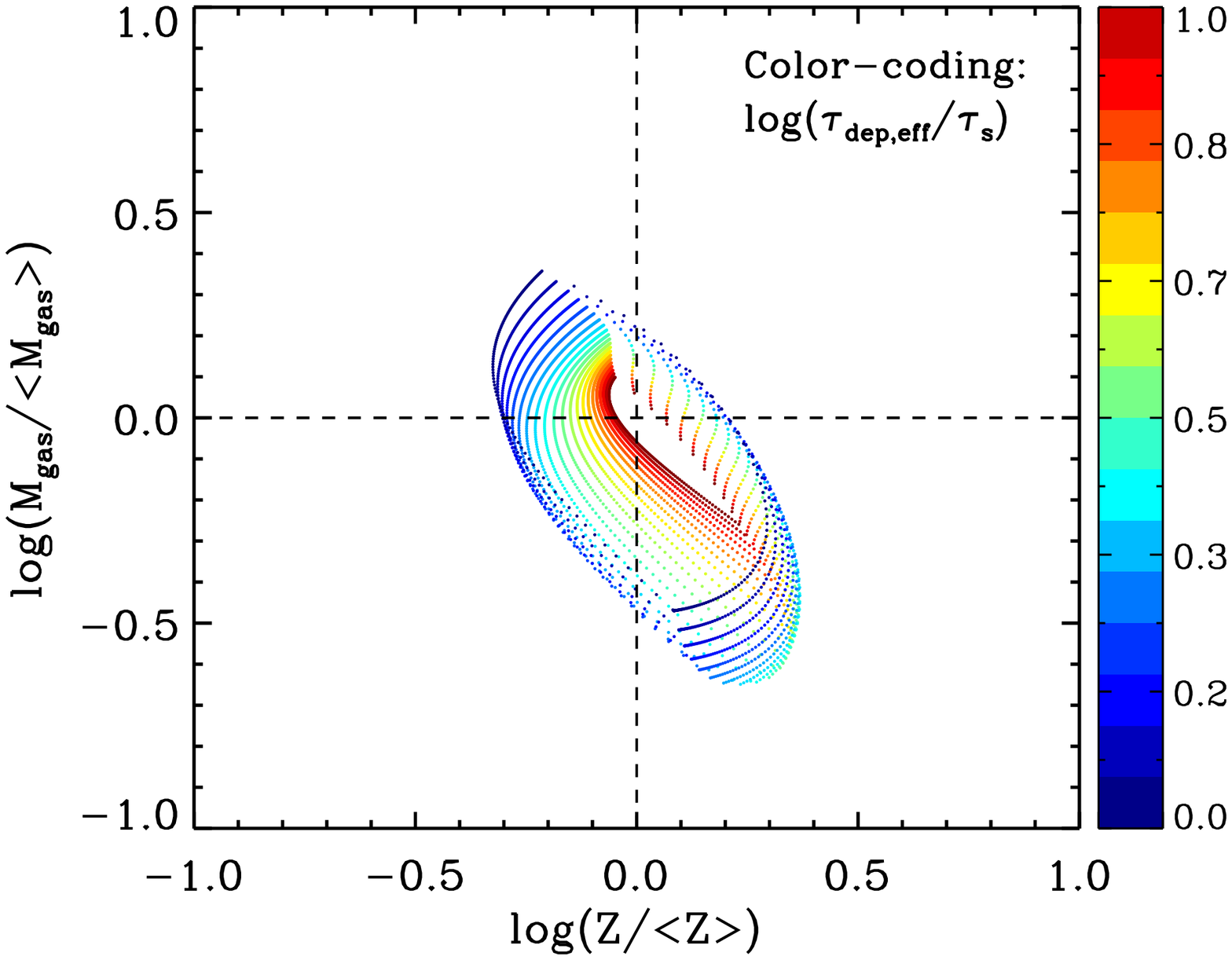,clip=true,width=0.33\textwidth}
    \end{center}
  \caption{Illustration of the SFR, \mgas\ and $Z$ in response to a time-varying SFE$(t)$ of the sinusoidal function (upper panels) and the periodic step function (lower panels) with constant inflow rate, in the gas regulator framework. The panels, lines and colors are the same as those in Figure \ref{fig:1}, except the gas regulator system is now driven with periodic SFE$(t)$ and constant inflow rate. }
  \label{fig:2}
\end{figure*}

In the previous subsection, we looked at the behavior of the gas-regulator system with time-varying inflow and time-invariant SFE. In this subsection, we will explore the behaviour when the regulator is driven with a constant inflow rate but experiences a time-varying SFE.  

Similar to Section \ref{sec:2.2}, we first input a time-invariant inflow, i.e. $\Phi(t)=\Phi_0$, and a sinusoidally varying SFE(t): 
\begin{equation} \label{eq:16}
      {\rm SFE}(t) = {\rm SFE_0} + {\rm SFE_t} \cdot {\rm sin}(2\pi t/T_{\rm p}). 
\end{equation} 
The driving period of SFE$(t)$ is again denoted as $T_{\rm p}$. As before, we look for the solution of $M_{\rm gas}(t)$ and $M_{\rm Z}(t)$ in terms of Equation \ref{eq:5}. However, we note that, unlike in Section \ref{sec:2.2}, the solutions of $M_{\rm gas}(t)$ and $M_{\rm Z}(t)$ are not exact sinusoidal functions, but can be represented as approximations to sinusoids. We assume the variation of the input SFE(t) is small, i.e. SFE$_{\rm t}\ll $SFE$_{\rm 0}$. Therefore, the variations of the resulting $M_{\rm gas}(t)$ and $M_{\rm Z}(t)$ are also small, i.e. $M_{\rm t}\ll M_{\rm 0}$ and $M_{\rm Zt}\ll M_{\rm Z0}$.

Actually, based on Equation \ref{eq:2} and \ref{eq:3}, varying the SFE is mathematically (but of course not physically) equivalent to varying inflow rate as ${\rm SFE_{\rm t}} = -\Phi_{\rm t}/\Phi_{\rm 0}\times {\rm SFE}_{\rm 0}$ and ${\rm SFE}_{\rm t}\ll{\rm SFE}_{\rm 0}$. Therefore, the solution of Equation \ref{eq:2} and \ref{eq:3} can be directly written as: 
\begin{equation} \label{eq:17}
\begin{split}
   M_{\rm 0} = &\  \Phi_{\rm 0}\tau_{\rm dep,eff} \\
   \delta \ \  = &\  {\rm arctan}(\xi) \\
   \frac{M_{\rm t}}{M_{\rm 0}} = &\  - \frac{1}{(1+\xi^2)^{1/2}}\times \frac{{\rm SFE}_{\rm t}}{{\rm SFE}_{\rm 0}}, 
\end{split}
\end{equation}
and 
\begin{equation} \label{eq:18}
\begin{split}
   M_{\rm Z0} = &\  (y_{\rm eff}+Z_{\rm 0})\Phi_{\rm 0}\tau_{\rm dep,eff} \\
   \beta \ \  = & \ {\rm arctan}[\frac{2y_{\rm eff}\xi + Z_{\rm 0}\xi(1+\xi^2)}{y_{\rm eff}(1-\xi^2)+Z_{\rm 0}(1+\xi^2)}] \\
   \frac{M_{\rm Zt}}{M_{\rm Z0}} = &\  -\frac{(1+\eta^2)^{1/2}}{1+\xi^2}\times \frac{{\rm SFE}_{\rm t}}{{\rm SFE}_{\rm 0}}.
\end{split}
\end{equation}
We emphasize that the definitions of $\xi$ and $\eta$ are the same as in Section \ref{sec:2.2}. 

Comparing Equation \ref{eq:17} and \ref{eq:18} with Equation \ref{eq:6} and \ref{eq:8}, the difference in sign indicates the fact that, when driving the regulator with time-varying $\Phi(t)$, the increase of $\Phi(t)$ produces an increase of both $M_{\rm gas}(t)$ and $M_{\rm Z}(t)$ with some time-delays, while when driving the regulator with time-varying SFE$(t)$, the increase of SFE$(t)$ leads to decreases of both $M_{\rm gas}(t)$ and $M_{\rm Z}(t)$, again with some time-delay. We note that the difference in sign cannot be simply treated as an additional phase-delay of $\pi$, i.e. a time lag of $T_{\rm p}/2$, because
an inverse connection between the mass of cold gas and the metallicity is physically expected.


The solutions in Equation \ref{eq:17} and \ref{eq:18} are illustrated in the top-left panel of Figure \ref{eq:2}, where we show examples of the evolution of $M_{\rm gas}$ (blue), SFR (red), $M_{\rm Z}$ (orange) and $Z$ (purple). 
As previously, we also investigate the correlation between $\Delta\log$SFR (and $\Delta\log M_{\rm gas}$) and $\Delta\log Z$ at a set of different $\xi$, as shown in the top-middle (and top-right) panel of Figure \ref{fig:2}.  

In contrast with the result in Section \ref{sec:2.2}, $\Delta\log$SFR and $\Delta\log Z$ now show a strong {\it positive} correlation. However, we note that $\Delta\log M_{\rm gas}$ and $\Delta\log Z$ remain negatively correlated, as would be expected, independent of whether the gas-regulator is driven with time-varying inflow or time-varying SFE. 
In this sense, metallicity variations fundamentally reflect variations in total gas mass in the gas regulator reservoir \citep[see][]{Lilly-13, Bothwell-13, Bothwell-16}. 

As in Section \ref{sec:2.2}, we also look at periodic step functions for the time-variation of SFE. Such changes in SFE may well be more realistic than sinusoidal changes. 
The bottom panels of Figure \ref{fig:2} demonstrate the resulting $M_{\rm gas}(t)$, SFR$(t)$, $M_{\rm Z}(t)$ and $Z(t)$, as well as the correlation between $\Delta\log$SFR (and $\Delta\log M_{\rm gas}$) and $\Delta\log Z$. In generating the plots, we set the period of step function to be 2 Gyr, and change the upper-state duration ($\tau_{\rm s}$). We allow the $\tau_s$ varying from $0.1\tau_{\rm dep,eff}$ to $\tau_{\rm dep,eff}$. Here the $\tau_{\rm dep,eff}$ is calculated based on the SFE in its lower-state. 

A sudden increase of SFE causes a sudden increase, but a following decrease, of the SFR, a following decrease of $M_{\rm gas}$ and a following increase of the gas-phase metallicity. 
As a whole, it is not as immediately clear what the sign of the $\Delta\log$SFR and $\Delta\log$Z correlation will be, given the fact that the lower branch has many more data points (reflecting the longer interval of time) than the upper branch on the bottom-middle panel of Figure \ref{fig:2}. 
Therefore, it appears that the sign of the correlation between SFR and metallicity depends on the detailed forms of input time-varying SFE. 
There is a strong {\it asymmetry} in the distribution of SFR through the cycle, indicated by the number density of the data points in the bottom-middle panel of Figure \ref{fig:2}. Specifically, SFR$(t)$ stays close to its median value for most of the time, but shows a strong increase for a short period. The asymmetry becomes more significant as the relative duration of the increased phase of SFE is decreased.   However, one thing is clear: the states with strongly {\it enhanced} SFR are always {\it metal-enhanced} with respect to the mean metallicity. These phases are represented in the upper locus of points in the figure which have $Z > \langle Z \rangle$.   

Consistent with the top-right panels of Figure \ref{fig:2}, the $\Delta\log M_{\rm gas}$ and $\Delta\log Z$ we conclude that there will always overall be a negative correlation that will be most clearly seen in the highest SFR points.

Similar to Section \ref{sec:2.2}, we again present the  approximate analytical solution of $\sigma(\log {\rm SFR})/\sigma(\log {\rm SFE})$, and $\sigma(\log Z)/\sigma(\log {\rm SFR})$ when driving the gas-regulator with sinusoidal SFE. These quantities can be written as:
\begin{equation} \label{eq:19}
\frac{\sigma({\rm \log SFR})}{\sigma(\log {\rm SFE})} \approx \frac{\xi}{(1+\xi^2)^{1/2}}
\end{equation}
and 
\begin{equation} \label{eq:20}
\frac{\sigma(\log Z)}{\sigma(\log {\rm SFR})} \approx \frac{1}{(1+\xi^2)^{1/2}} \cdot \frac{1}{1+Z_0/y_{\rm eff}}.
\end{equation}
In the similar way as in Section \ref{sec:2.2}, according to Equation \ref{eq:19}, we can write the correlation between the PSDs of $\log$SFR$(t)$ and $\log$SFE$(t)$ as: 
\begin{equation} \label{eq:30}
   {\rm PSD}_{\rm \log SFR}(\nu) 
       \approx \frac{1}{1+(2\pi\tau_{\rm dep,eff}\nu)^{-2}} \cdot {\rm PSD}_{\rm \log SFE}(\nu)
\end{equation}
The Equation \ref{eq:30} is established when the variation of SFE is small, i.e. SFE$_{\rm t}\ll $SFE$_{\rm 0}$. 
The right panel of Figure \ref{fig:3} shows the numerical solution (solid curves) and the approximate analytical solution (gray dashed curves) of $\sigma(\log {\rm SFR})/\sigma(\log {\rm SFE})$ and $\sigma(\log Z)/\sigma(\log {\rm SFR})$ as a function $\log\xi$. As shown, the numerical solution is well matched by the analytical solution. 
Intriguingly, the Equation \ref{eq:19} and \ref{eq:20} are strictly symmetrical to the Equation \ref{eq:14} and \ref{eq:15}, respectively, at the axis of $\log \xi=0$.

When driving the gas-regulator with a time-varying SFE, the $\sigma(\log {\rm SFR})/\sigma(\log {\rm SFE})$ is predicted to increase with $\xi$, while $\sigma(\log Z)/\sigma(\log {\rm SFR})$ is predicted to decrease with $\xi$. Although the Equation \ref{eq:14}, \ref{eq:15}, \ref{eq:19} and \ref{eq:20} are only approximate solutions obtained in the limit of small variations of inflow rate or SFE, we have verified numerically that these equations remain reasonable approximations even when the variations of inflow rate or SFE are quite significant. In Appendix \ref{sec:C}, we examine the correlation of $\Delta \log$SFR vs. $\Delta \log$Z, when there is a large variation (0.5 dex) of inflow rate or SFE. We confirm that the model prediction of $\Delta \log$SFR vs. $\Delta \log$Z in the top-middle panels of Figure \ref{fig:1} and \ref{fig:2}, are scalable to the case in which the variation of driving factor is quite significant. 

\subsection{The effects of mass-loading, $Z_{\rm 0}$ and the yield} \label{sec:2.4}

\begin{figure*}
  \begin{center}
    \epsfig{figure=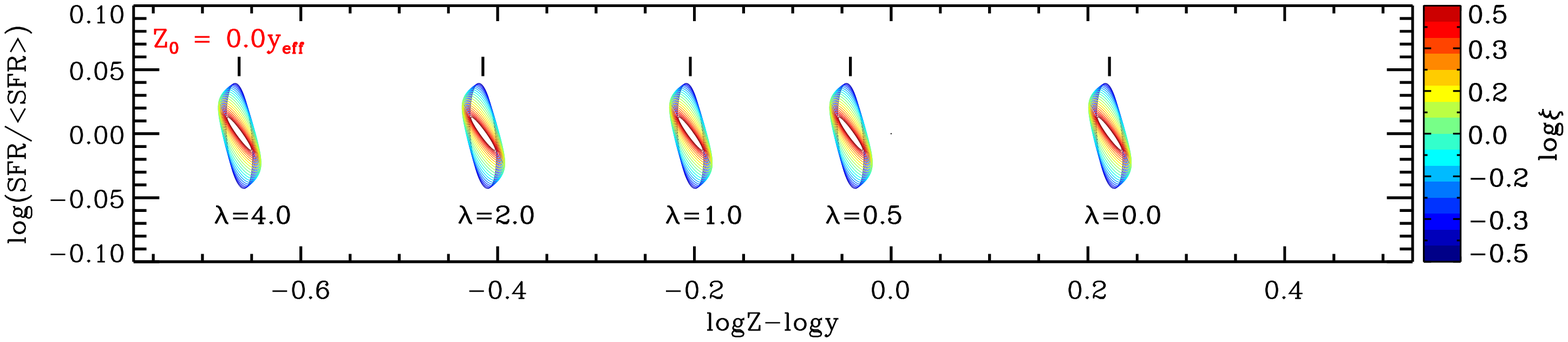,clip=true,width=0.95\textwidth}
    \epsfig{figure=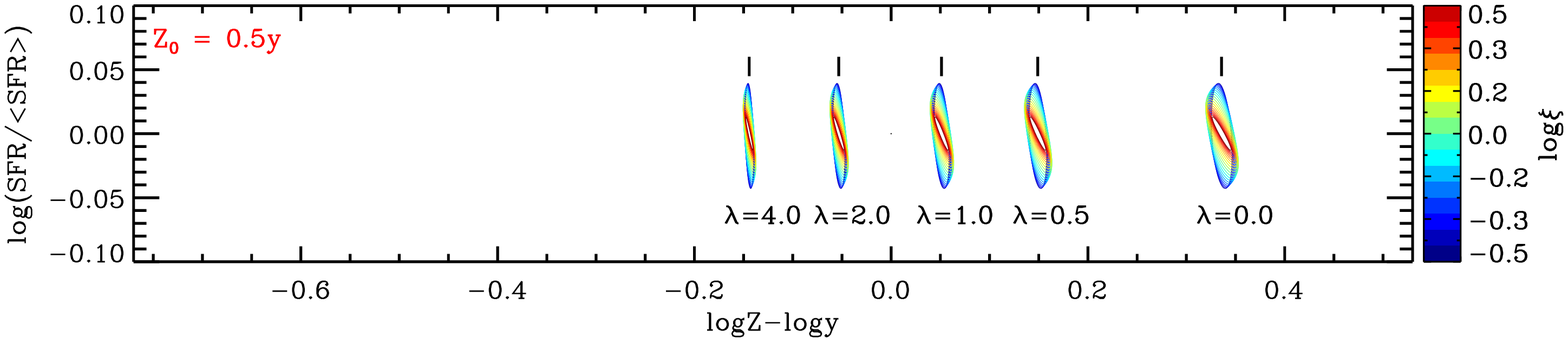,clip=true,width=0.95\textwidth}
    \epsfig{figure=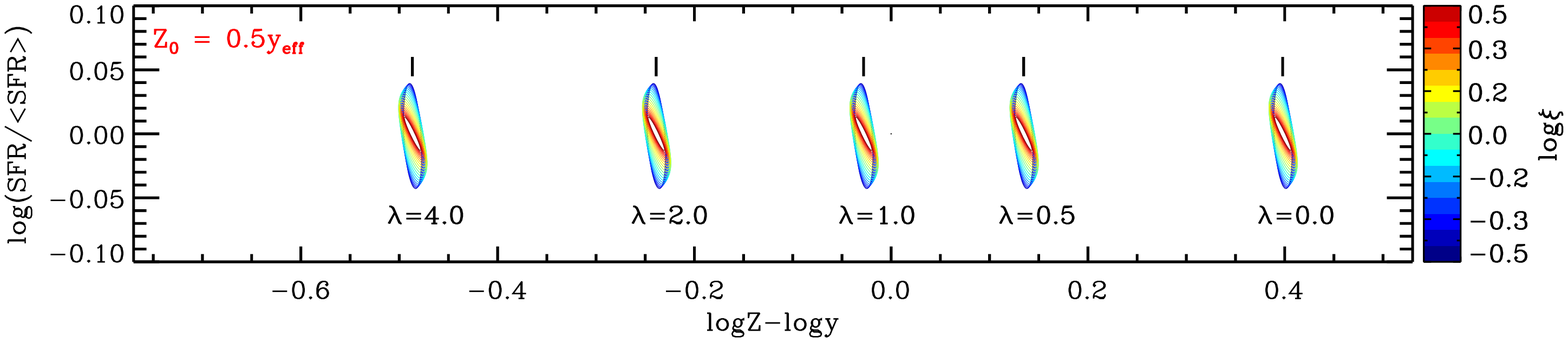,clip=true,width=0.95\textwidth}
    \end{center}
  \caption{Illustration of the role of $Z_{\rm 0}$ and the mass-loading factor $\lambda$ in shaping the correlation of SFR and $Z$, when driving the gas regulator system with a sinusoidal inflow rate and constant SFE. From top to bottom, we set the $Z_{\rm 0}$ to be 0.0, 0.5$y$ and 0.5$y_{\rm eff}$. The $y_{\rm eff}$ is defined as $y/(1-R+\lambda)$. In each panel, we explore the cases of five different mass-loading factors, i.e. $\lambda=$0.0, 0.5, 1.0, 2.0, and 4.0. The lines are color-coded with the $\xi$ as before. 
  The black line segments indicate the median gas-phase metallicity of the different five mass-loading factors for the three panels, which is exactly  $Z=y_{\rm eff}+Z_{\rm 0}$. }
  \label{fig:4}
\end{figure*}

\begin{figure*}
  \begin{center}
    \epsfig{figure=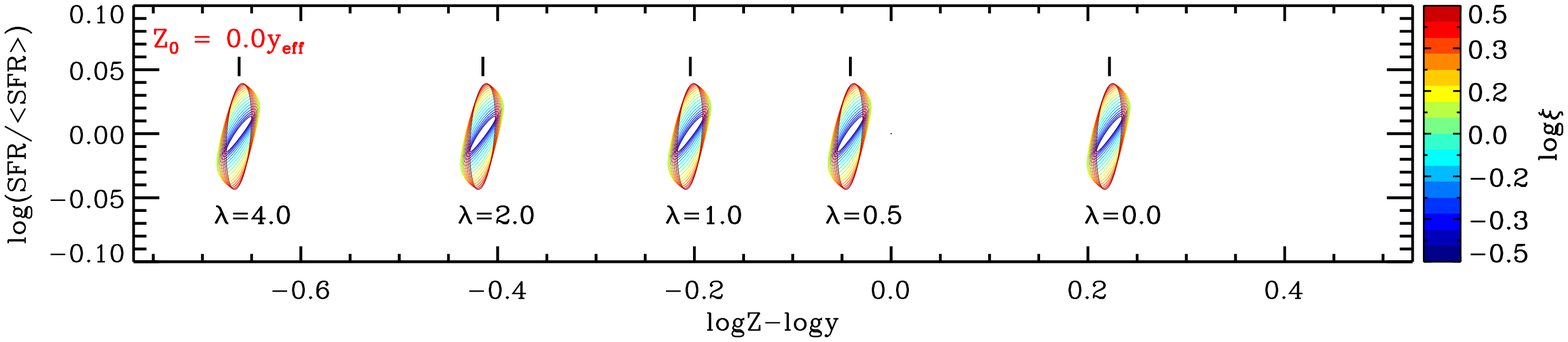,clip=true,width=0.95\textwidth}
    \epsfig{figure=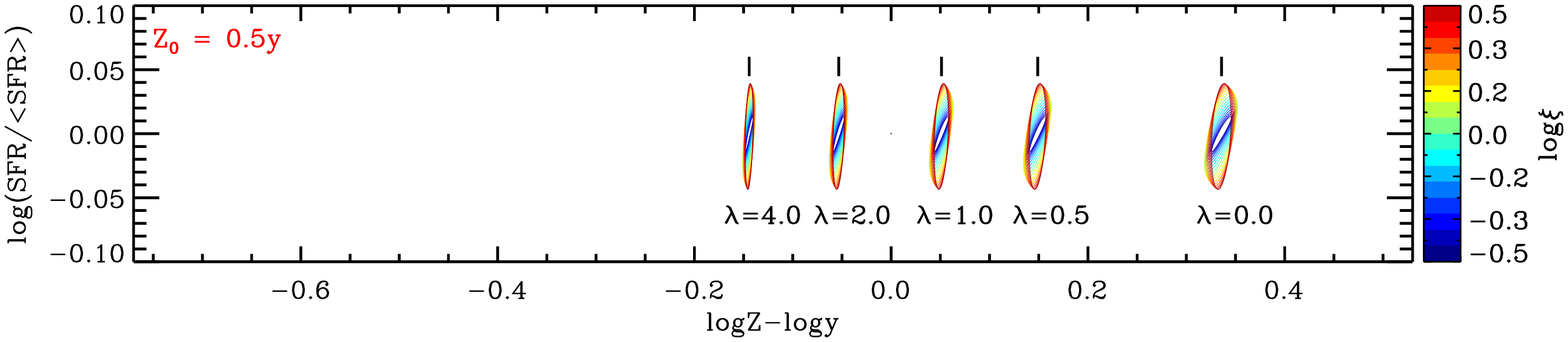,clip=true,width=0.95\textwidth}
    \epsfig{figure=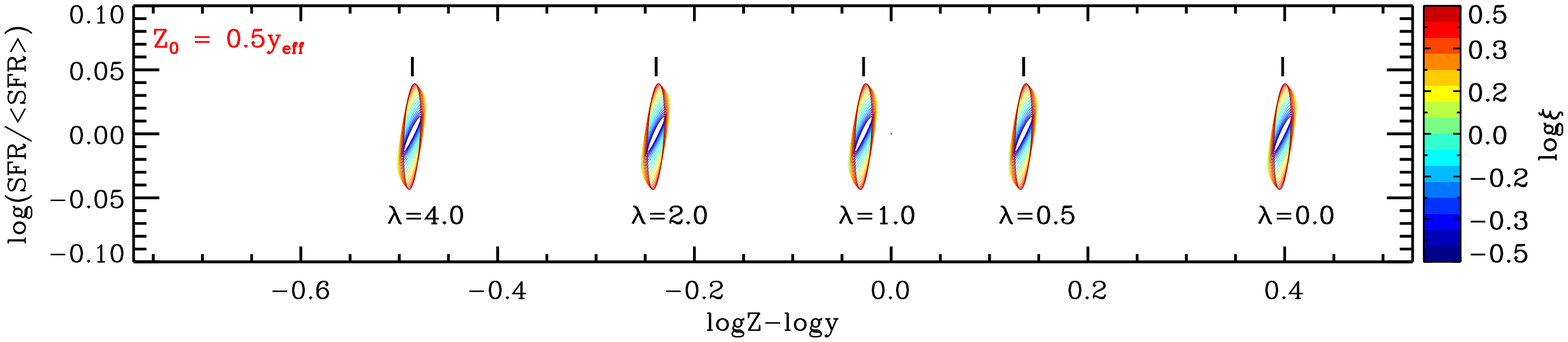,clip=true,width=0.95\textwidth}
    \end{center}
  \caption{The same as Figure \ref{fig:4}, but driving the gas regulator system with sinusoidal SFE and constant inflow. }
  \label{fig:5}
\end{figure*}

In Section \ref{sec:2.2} and \ref{sec:2.3}, we have explored the behavior of the gas-regulator system when it is driven by time-varying inflow and time-varying SFE, respectively. In this subsection, we will explore how the metallicities are modified by changes in the wind mass-loading factor and the metallicity of inflow gas.  The assumed yield enters as a simple scaling factor throughout.  

Following Section \ref{sec:2.2}, we drive the gas-regulator with a sinusoidally varying inflow rate for a set of different the mass-loading factors and $Z_{\rm 0}$.  First, we set $Z_{\rm 0} = 0$, and show the $\Delta\log$SFR vs. $\log Z - \log y$  in the top panel of Figure \ref{fig:4} for different mass-loading factors. To eliminate the dependence on the value of $y$, we set the x-axis to be $\log Z - \log y$ rather than $\log Z$. 
At $Z_{\rm 0}=0$, the relative changes in metallicity and star-formation, i.e. the $\Delta\log$SFR-$\Delta\log Z$ relation, stays the same with varying $\lambda$, while the {\it mean} metallicity decreases with increasing $\lambda$. 
We then set $Z_{\rm 0} = 0.5y$, and obtain the $\Delta\log$SFR vs. $\log Z-\log y$ shown in the middle panel of Figure \ref{fig:4}. As shown, the basic negative correlation between $\Delta\log$SFR vs. $\Delta\log Z$ still retains, but the relative change in metallicity, i.e. the scatter of metallicity in the population, decreases with increasing $Z_{\rm 0}$ and increasing $\lambda$ (if $Z_{\rm 0}\ne 0$), compared to the top panel of Figure \ref{fig:4}. Finally, we set $Z_{\rm 0}= 0.5y_{\rm eff}$, and the $\Delta\log$SFR vs. $\log Z-\log y$ is shown in the bottom panel of Figure \ref{fig:4}. 
The correlation between  $\Delta\log$SFR vs. $\Delta\log Z$ is the same when $Z_0/y_{\rm eff}$ is fixed. 

These results follow the Equation \ref{eq:15} where it is clear that the ratio of $\sigma(\log Z)$ to $\sigma(\log {\rm SFR})$ only depends on $Z_{\rm 0}/y_{\rm eff}$ once $\xi$ is fixed. At a given $\xi$, the scatter of $\log Z$ is scaled by the factor $(1+Z_{\rm 0}/y_{\rm eff})^{-1}$ with respect to $\sigma(\log Z)$ at $Z_{\rm 0}=0$. 
Another interesting point is that the mean metallicity does depend on $Z_0$, $\lambda$ and $y$. As a whole, the mean gas-phase metallicity increases with increasing $Z_0$ and/or decreasing $\lambda$. Actually, the mean SFR and metallicity can be solved analytically based on Equation \ref{eq:6} and \ref{eq:8} (or Equation \ref{eq:17} and \ref{eq:18}), which can be written as \citep[also see][]{Sanchez-Almeida-14}: 
\begin{equation} \label{eq:21}
    \langle {\rm SFR}\rangle = M_{\rm 0}\cdot {\rm SFE} = \Phi_{\rm 0} \cdot (1-R+\lambda)^{-1}.
\end{equation}
and
\begin{equation} \label{eq:22}
    \langle {Z}\rangle = \frac{M_{\rm Z0}}{M_{\rm 0}} = Z_{\rm 0}+y_{\rm eff}. 
\end{equation}

For completeness, we also look at driving the gas-regulator system with sinusoidal SFE. The results are shown in Figure \ref{fig:5}. As shown, the effects of $Z_{\rm 0}$, $\lambda$ and $y$ on the correlation between $\Delta\log$SFR and $\Delta\log Z$ follow the Equation \ref{eq:20}, and the effects of them on mean SFR and metallicity follow the Equation \ref{eq:21} and Equation \ref{eq:22}.  Although here we do not try to set a different yield, we argue that the effect of varying yield would also follow the Equation \ref{eq:15} or Equation \ref{eq:20}. 

Based on Equation \ref{eq:21} and \ref{eq:22}, we find that the mean SFR is determined by the mean inflow rate and mass-loading factor regardless of the SFE, simply because the gas reservoir adjusts to maintain long-term balance, while the mean metallicity only depends on $Z_{\rm 0}$ and $y_{\rm eff}$ regardless of how large is the inflow or SFE. 
We conclude that, under the gas-regulator framework, the metallicity is primarily determined by $Z_{\rm 0}$ and $y_{\rm eff}$ with a secondary effect on SFR (or cold gas mass) due to the time-variation of the inflow rate or SFE. 

It is important to note that the metallicity does not depend on the {\it absolute} values of inflow rate and SFE, but rather on the change in them.  Therefore, when investigating the question whether the SFR is a secondary parameter to determine the metallicity, one should look at the correlation of the relative values of SFR and $Z$ (or residuals like $\Delta\log$SFR and $\Delta\log Z$ in Figure \ref{fig:1} and Figure \ref{fig:2}), rather than the absolute values, in order to eliminate different $\langle {\rm SFR}\rangle$ and $\langle Z\rangle$ for different galaxies or regions. 

In Section \ref{sec:2.2} and \ref{sec:2.3}, we have investigated the properties of gas-regulator by driving the gas-regulator system with 1) time-varying $\Phi$ and time-invariant SFE, and 2) time-varying SFE and time-invariant $\Phi$. The most important result is that there are opposite correlations (negative and positive, respectively) between $\Delta\log$SFR and $\Delta\log Z$ for these two modes.  

However, in the real universe, the inflow rate and SFE could both be time-varying.  
If the variation of inflow rate is much more dominant than the variation of SFE, a negative correlation between $\Delta\log$SFR and $\Delta\log Z$ is expected according to the analysis of Section \ref{sec:2.2}, and vice versa. 
This implies that the correlation between $\Delta\log$SFR and $\Delta\log Z$ (at least for regions of significantly enhanced star formation) is a powerful and observationally accessible diagnostic of the {\it dominant} mechanism for the variation of SFR, changes in inflow or changes in SFE, even if both of them may vary with time.  

In the whole of Section \ref{sec:2}, we have used $\Delta\log$SFR to indicate the temporal elevation and suppression of star formation. We note that one could also use the displacement of sSFR to its {\it smooth cosmic evolution} in logarithmic space, 
$\Delta \log$sSFR. These are essentially equivalent provided that the timescales of interest (essentially a single period of oscillation) are much shorter than the inverse sSFR, so that the mass changes negligibly through the oscillation cycle.
We have examined that the difference is negligible in case of sSFR$\sim10^{-10}$yr$^{-1}$ for typical local SF galaxies when replacing $\Delta \log$SFR with $\Delta \log$sSFR. This is true for the present work, since the galaxies used in the observational part are low-redshift galaxies.   As discussed below, there are good reasons to use $\Delta \log$sSFR as the observational measure of the relative star-formation rate rather than $\Delta \log$SFR.

In addition, the metallicity we measure from the observation is the Oxygen abundance, defined as the number ratio of Oxygen to Hydrogen. We argue that in Equation \ref{eq:3}, the mass of metals can be replaced by the number of Oxygen, if the $y$ and $Z_0$ are defined in the terms of the number fraction of Oxygen.  Therefore, the model predictions in the whole Section \ref{sec:2} in terms of Z are also valid in terms of O/H, including Figure \ref{fig:1}-\ref{fig:5}, as well as the corresponding equations. 

The models predict the correlation of the temporal changes in SFR and Z within a given period for an individual gas-regulator system. Observationally, it is of course not possible to monitor a single galaxy or region for the timescales of interest. Instead, we have the SFR and metallicity for a large population of galaxies or regions. Therefore, we must make the assumption of {\it ergodicity} in order to compare the models with the observational data.  In other words, we must assume that the variations that we see of the SFR and Z across the observed galaxy population (or between different regions within a galaxy) observed at a single epoch do indeed reflect the typical temporal changes of the SFR and Z within a given system \citep[see more details in Section 3.2 of][]{Wang-20b}. 

\section{Data} \label{sec:3}

In Section \ref{sec:2}, we have established links between the variations of SFR, cold gas mass, and gas-phase metallicity when a gas-regulator system is driven by changes in inflow or SFE, and established that the sign of the correlation between changes in the easily observable SFR and changes in the metallicity is a key diagnostic in establishing whether changes are driven by variations in inflow or SFE.   While these relations have been constructed based on the {\it temporal} changes in a single system, they would of course apply also to an ensemble of such systems observed at a single epoch (assuming the phases are random), provided the assumption of ergodicity applies \citep[see][for a discussion]{Wang-20b}. The goal of the observational part of the paper is therefore to examine the correlation of $\Delta\log$SFR and $\Delta\log Z$, computed relative to suitably chosen fiducial values, at different locations.  We will wish to examine this correlation both from galaxy to galaxy, but also at different locations within galaxies, in order to try to assess the relative importance of changes in inflow and SFE on different physical scales.


In this section, we will briefly introduce the data used in the observational part of this work, namely the 38 SF galaxies from the MUSE Atlas of Disks \citep[MAD;][]{Erroz-Ferrer-19} survey, and the nearly 1000 well-defined SF galaxies from Mapping Nearby Galaxies at APO (MaNGA) survey (\citetalias{Wang-19}).  We refer the readers to \cite{Erroz-Ferrer-19} and \citetalias{Wang-19} for more details of these two galaxy samples. 

\subsection{The MAD galaxy sample} \label{sec:3.1}


The final released sample\footnote{https://www.mad.astro.ethz.ch} of the MAD survey includes 38 weakly inclined, spiral galaxies, spanning a large range of stellar mass from $10^{8.5}$ to $10^{11.2}$\msolar.  These galaxies were observed during the MUSE Guaranteed Time Observing runs on the Very Large Telescope (VLT) between 2015 April and 2017 September.  The on-source exposure time was for each galaxy 1 h and the seeing ranged between 0.4 and 0.9 arcsec. These galaxies are very nearby, with $z<0.013$, leading to an average spatial resolution of $\sim$100 pc or better. The MUSE field of view is 1 arcmin$^2$, and the wavelength coverage of the spectra is from 4650 to 9300 \AA. The coverage for MAD galaxies is in the range of  $\sim$0.5-3 effective radius (\re), with a median value of 1.3\re. The data were reduced using the MUSE pipeline \citep{Weilbacher-NG}, including bias and dark subtraction, flat fielding, wavelength calibration and so on. 

The data downloaded from the data release is not the original two-dimensional spectra, but the derived measurements from the spectra. These include the strengths of the emission lines, such as the flux map of H$\beta$, [OIII]$\lambda$4959,5007, H$\alpha$, [NII]$\lambda$6548,6583, and [SII]$\lambda$6717,6731. The emission lines are modelled by single Gaussian profiles. The fluxes of emission lines are corrected for the dust attenuation by using the Balmer decrement with case B recombination. The intrinsic flux ratio of H$\alpha$ to H$\beta$ is taken to be 2.86, assuming the electron temperature of 10$^4$ K and the electron density of $10^2$ cm$^{-3}$. The adopted attenuation curve is the CCM \citep[][]{Cardelli-89, ODonnell-94} extinction curve with $R_{\rm V}=3.1$. 
In addition to the maps of emission lines, the released MAD data also includes the maps of derived quantities, such as SFR surface density ($\Sigma_{\rm SFR}$) and stellar mass surface density ($\Sigma_*$). The SFR is derived from the H$\alpha$ luminosity \citep{Kennicutt-98, Hao-11}, assuming the \cite{Kroupa-01} IMF.  The stellar mass density is derived by fitting stellar population models to the stellar continuum spectra twice. The first continuum fitting is performed spaxel-by-spaxel, and the second fitting was performed on the stellar Voronoi tessellation (single-to-noise of 50 on the continuum) using the stellar templates from {\tt MILES} \citep{Sanchez-Blazquez-06} with {\tt pPXF} \citep[][]{Cappellari-04}. 
Then, the resulting $\Sigma_*$ map is a spaxel-by-spaxel based map, assuming that the $\Sigma_*$ is the same for all spaxels within a single Voronoi bin.

We note that in the released maps, spaxels that are located within the Seyfert and LINER regions of the BPT diagram \citep[e.g.][]{Baldwin-81, Kewley-01} are masked out. The SFR and gas-phase metallicity cannot be well measured in these masked regions.  Their exclusion should not affect our analysis and, therefore, in this work we only focus on the SF and ``composite" regions. The SF and composite regions are identified by the demarcation lines of \cite{Kauffmann-03} and \cite{Kewley-01} on the BPT diagram \citep[see figure 1 in ][]{Erroz-Ferrer-19}. 
This also means that for each galaxy, we only use some fraction of the spaxels within the MUSE field. The fraction of valid spaxels varies from galaxy-to-galaxy, with a median value of 0.33. We have checked that our result does not depend on the fraction of spaxels that are used. 

The highly-resolved MAD data enables us to investigate the correlation between star formation and metal enhancement down to the scale of giant molecular clouds (GMC). However, the MAD sample only includes 38 galaxies, which limits the statistical power when examining the galaxy population as a whole.  Therefore, we utilize complementary data on the integrated spectra of galaxies from MaNGA survey.  We do not use the individual spaxel data for the MaNGA galaxies because the resolution is so much worse than for MAD. 

\subsection{The MaNGA galaxy sample} \label{sec:3.2}

MaNGA is the largest IFS surveys of nearby galaxies up to now, and aims at mapping the 2-dimensional spectra for $\sim$10,000 galaxies with redshifts in the range of $0.01<z<0.15$. Using the two dual-channel BOSS spectrographs at the Sloan Telescope \citep{Gunn-06, Smee-13}, MaNGA covers the wavelength of 3600-10300 \AA\ at R$\sim$2000. The spatial coverage of individual galaxies is usually larger than 1.5\re\ with a resolution of 1-2 kpc.  The flux calibration, including the flux loss due to atmospheric absorption and instrument response, is accurate to better than 5\% for more than 89\% of MaNGA’s wavelength range \citep{Yan-16}. 

In this work, we use the well-defined sample of SF galaxy from \citetalias{Wang-19}. Here we only briefly describe the sample definition, and refer the reader to \citetalias{Wang-19} for further details. This galaxy sample is originally selected from the SDSS Data Release 14 \citep{Abolfathi-18}, excluding quenched galaxies, mergers, irregulars, and heavily disturbed galaxies.  The quenched galaxies are identified and excluded based on the stellar mass and SFR diagram. For each individual galaxy, the stellar mass and SFR are measured within the effective radius, i.e. $M_*(<R_{\rm e})$ and SFR($<$\re), based on the MaNGA 2-dimensional spectra. The final MaNGA sample includes 976 SF galaxies, and is a good representation of normal SFMS galaxies.

Similar to the measurements of SFR for the MAD galaxies, the map of SFR surface density of MaNGA galaxies is also determined by the extinction-corrected H$\alpha$ luminosity \citep{Kennicutt-98}. The correction of dust attenuation follows the same approach for MAD galaxies, as described in Section \ref{sec:3.1}, by using the Balmer decrement and adopting the CCM extinction curve.   
The maps of stellar mass surface density for MaNGA galaxies are obtained by fitting the stellar continuum using the public fitting code {\tt STARLIGHT} \citep{Cid-Fernandes-04}, using single stellar populations with {\tt Padova} isochrones from \cite{Bruzual-03}. 
However, we note that in determining the $\Sigma_{\rm SFR}$ and $\Sigma_*$ for MaNGA galaxies, the \cite{Chabrier-03} IMF is assumed, which is different from the one adopted for MAD galaxies. We argue that the two IMFs are quite close to each other, with only a small overall shift on SFR and $M_*$ (or $\Sigma_{\rm SFR}$ and $\Sigma_*$) which does not change any of our conclusions in this work. 

\subsection{The estimation of gas-phase metallicity} \label{sec:3.3}

The $T_{\rm e}$-based method is widely understood to represent the ``gold standard'' in determining the gas-phase metallicity \citep[e.g.][]{Skillman-89, Garnett-02, Bresolin-07, Berg-15, Bian-18}. However, the measurement of the weak [OIII]$\lambda$4363 emission line is needed, which is often not detected in the available spectra. Therefore, a set of empirical recipes have been proposed to derive the gas-phase metallicity based only on the strong emission lines \citep[e.g.][]{Kobulnicky-04, Pettini-04,  Maiolino-08,  Perez-Montero-09, Pilyugin-10, Marino-13, Vogt-15}, such as  [OII]$\lambda$3227, H$\beta$, [OIII]$\lambda$5007, H$\alpha$, [NII]$\lambda$6584, and [SII]$\lambda\lambda$6717,6731.  However, systematic offsets of 0.2 dex or more are found between these different empirical calibrations, even using the same line measurements \citep{Kewley-08, Blanc-15}. Not only are there systematic offsets, the range of the derived metallicities are also different. This is unfortunately important since we have argued that the variation of gas-phase metallicity (or the residuals of metallicity) as the SFR varies is an important diagnostic.  The dispersion in  metallicities must be considered in the context of the different ranges of the Oxygen abundance measurements using the different methods.

Unfortunately, the wavelength coverage of MUSE does not include the [OII]$\lambda$3727 and [OIII]$\lambda$4363 lines, leading to a limited number of usable strong line prescriptions. In this work, we adopt the empirical relations from \citet[][hereafter \citetalias{Dopita-16}]{Dopita-16} and \citet[][hereafter \citetalias{Pilyugin-16}]{Pilyugin-16}.  These two empirical relations are the most recently constructed and have advantages over the previous methods, although ultimately it is not known whether they are more accurate than the other methods. In the following, we briefly introduce these two calibrators. 

\subsubsection{{\tt N2S2H$\alpha$}}

The {\tt N2S2H$\alpha$} is a remarkably simple diagnostic proposed by \citetalias{Dopita-16}, which  can be written as: 
\begin{equation} \label{eq:23}
   {\tt N2S2H\alpha} = \log([{\rm NII}]/[{\rm SII}]) + 0.264\log([{\rm NII}]/{\rm H}\alpha)
\end{equation}
where [NII] is the flux of [NII]$\lambda$6584 and [SII] is the total flux of [SII]$\lambda\lambda$6717,6731. Then, the empirical relation of the metallicity can be written as: 
\begin{equation} \label{eq:24}
    12 + \log ({\rm O/H}) = 8.77 + {\tt N2S2H\alpha} + 0.45({\tt N2S2H\alpha}+ 0.3)^5. 
\end{equation}

This simple empirical relation is calibrated in the range of $7.7<12+\log({\rm O/H})<9.2$, which is valid for both MAD and MaNGA galaxies. The H$\alpha$, [NII]$\lambda$6584 and [SII]$\lambda\lambda$6717,6731 are located close together in wavelength, limiting the spectral range needed, and
making the {\tt N2S2H$\alpha$} diagnostic to be insensitive to reddening. 
The [NII]/H$\alpha$ term provides a correction for the weak dependence on the ionization parameter and gas pressure. \citetalias{Dopita-16} argued that this diagnostic should be a fairly reliable metallicity estimator with only small residual dependence on other physical parameters, and that it can be used in a wide range of environments.

However, this metallicity estimator strongly depends on the relative N/O ratio. In the calibration of {\tt N2S2H$\alpha$},  \citetalias{Dopita-16} assumed a universal correlation between N/O and O/H, which is determined based on a mixture of both stellar and nebular sources \citep{Nicholls-17}.  This means that any galaxies or regions that deviate from the adopted N/O-O/H relation would have an extra uncertainty in metallicity measurement when using {\tt N2S2H$\alpha$} as the metallicity estimator. The metallicity also depends on the relative S/O ration, while \citetalias{Dopita-16} have adopted [NII/SII] term in the metallicity indicator {\tt N2S2H$\alpha$}, which likely accounts for some of the dependence of metallicity on S/O.

\subsubsection{{\tt Scal}} \label{sec:3.3.2}

The S-calibration ({\tt Scal}) metallicity estimator has been proposed by \citetalias{Pilyugin-16}, based on three standard diagnostic line ratios: 
\begin{equation}
\begin{split}
{\tt N2} = \ & [{\rm NII}]\lambda\lambda6548,6584/{\rm H}\beta,  \\
{\tt S2} = \ & [{\rm SII}]\lambda\lambda6717,6731/{\rm H}\beta, \\
{\tt R3} = \ & [{\rm OIII}]\lambda\lambda4959,5007/{\rm H}\beta.
\end{split}
\end{equation}
The {\tt Scal} diagnostic is defined separately for the upper and lower branches of $\log {\tt N2}$. The {\tt Scal} indicator for the upper branch ($\log {\tt N2}\ge -0.6$) can be written as:
\begin{equation} \label{eq:26}
\begin{split}
    12 + \log({\rm O/H}) = \ & 8.424 + 0.030\log({\tt R3/S2}) + 0.751\log {\tt N2}  \\
                     \ &  + (-0.349 + 0.182\log({\tt R3/S2}) \\
                     \ & + 0.508\log {\tt N2})\times \log {\tt S2}, 
\end{split}
\end{equation}
and the {\tt Scal} indicator for the lower branch ($\log {\tt N2}<-0.6$) can be written as: 
\begin{equation} \label{eq:27}
\begin{split}
    12 + \log({\rm O/H}) = \ & 8.072 + 0.789\log({\tt R3/S2}) + 0.726\log {\tt N2}  \\
                     \ &  + (1.069 - 0.170\log({\tt R3/S2}) \\
                     \ & + 0.022\log {\tt N2})\times \log {\tt S2}.
\end{split}
\end{equation}
The {\tt Scal} prescription is calibrated to  metallicity measurements, from the T$_{\rm e}$-based method, of several hundred nearby HII regions. \citetalias{Pilyugin-16} found that the {\tt Scal} indicator gives metallicities in very good agreement with the $T_{\rm e}$-based methods, with a scatter of only $\sim$0.05 dex across the  metallicity range of $7.0<12+\log({\rm O/H})<8.8$. Furthermore, {\tt Scal} indicator takes advantage of three emission line ratios, which is an improvement over previous strong-line methods. 

In principle, given the wavelength coverage of MUSE, the {\tt N2} and {\tt O3N2} diagnostics, calibrated by \cite{Marino-13}, are also applicable. As pointed out by \cite{Kreckel-19}, using {\tt O3N2} (or {\tt N2}) and {\tt Scal} can result in qualitatively different results from the MUSE data. In this work, we prefer to use metallicity indicators of {\tt N2S2H$\alpha$} and {\tt Scal}, rather than {\tt N2} and {\tt O3N2}. Actually, 
\cite{Marino-13} calibrated the {\tt O3N2} and {\tt N2} diagnostics to the T$_{\rm e}$-based method, and found that the {\tt O3N2} and {\tt N2} diagnostics result in the uncertainty in Oxygen abundance of 0.18 dex and 0.16 dex, respectively. Given the similar ranges of the metallicity determined by {\tt Scal}, {\tt O3N2} and {\tt N2} for a given data set, the smaller uncertainty of {\tt Scal} diagnostic indicates a significant improvement over the previous {\tt O3N2} and {\tt N2} diagnostics. This improvement may come from the fact that 1) {\tt Scal} use more emission line ratios, and 2) these break the degeneracy between metallicity and ionization parameter \citep[also see][]{Kreckel-19}.  The later is also true for the {\tt N2S2H$\alpha$} indicator. 

\subsubsection{The contamination from diffuse ionized gas}

The fluxes of emission lines cannot be fully attributed to star formation activity alone. The diffuse ionized gas (DIG) makes a substantial contribution to the total emission-line flux from disk galaxies \citep{Walterbos-94, Ferguson-96, Greenawalt-98}, especially for regions of low H$\alpha$ surface brightness \citep{Oey-07, Zhang-17}. The line ratios for emission from HII regions and DIG are different, reflecting their different physical origins. The empirical relations for deriving metallicity and SFR from line ratios and line strengths always assume that all of the line emission is due to star formation.  This assumption is not unreasonable if the target regions are SF regions on the BPT diagram, while a significant contamination ($\sim$40\% or more) from DIG is expected in the ``composite'' and LINER regions \citep{Erroz-Ferrer-19}.  
Compared with HII regions, the DIG shows enhanced line ratios in [NII]$\lambda$6584/H$\alpha$ and [SII]$\lambda\lambda$6717,6731/H$\alpha$ \citep{Reynolds-85, Hoopes-03, Madsen-06}. The emission from DIG moves the position of SF regions towards the composite or LINER regions on the BPT diagram \citep{Sarzi-06, Yan-12, Gomes-16}.

In this work, we therefore only use the SF and composite regions of galaxies, and exclude those spaxels that are classified as Seyfert or LINERs. However, the contamination of DIG for the composite regions may still be significant.  \cite{Erroz-Ferrer-19} have identified the regions in MAD galaxies in which star formation or DIG is dominant. Following the method developed in \cite{Blanc-09}, they measured the fraction of flux coming from DIG and from HII regions, and further defined the DIG regions to be those in which the flux contribution of HII regions is less than 60\%.  They found that the HII regions show on average $\sim$0.1 dex higher metallicity than the DIG, while the metallicity radial gradient in both components is similar. Following the analysis of  \cite{Erroz-Ferrer-19}, in this work, we will use all the spaxels in the SF or composite regions to do the analysis, but ignoring whether they are classified as HII regions or DIG. However, we have recalculated our main result when only using the HII regions, and find that the basic result remains  unchanged. This indicates that the contamination of DIG is not a big concern in the present work. 

\section{Observational analysis of MAD galaxies} \label{sec:4}

\subsection{Maps and profiles of sSFR and Oxygen abundance for MAD galaxies} \label{sec:4.1}

\begin{figure*}
  \begin{center}
    \epsfig{figure=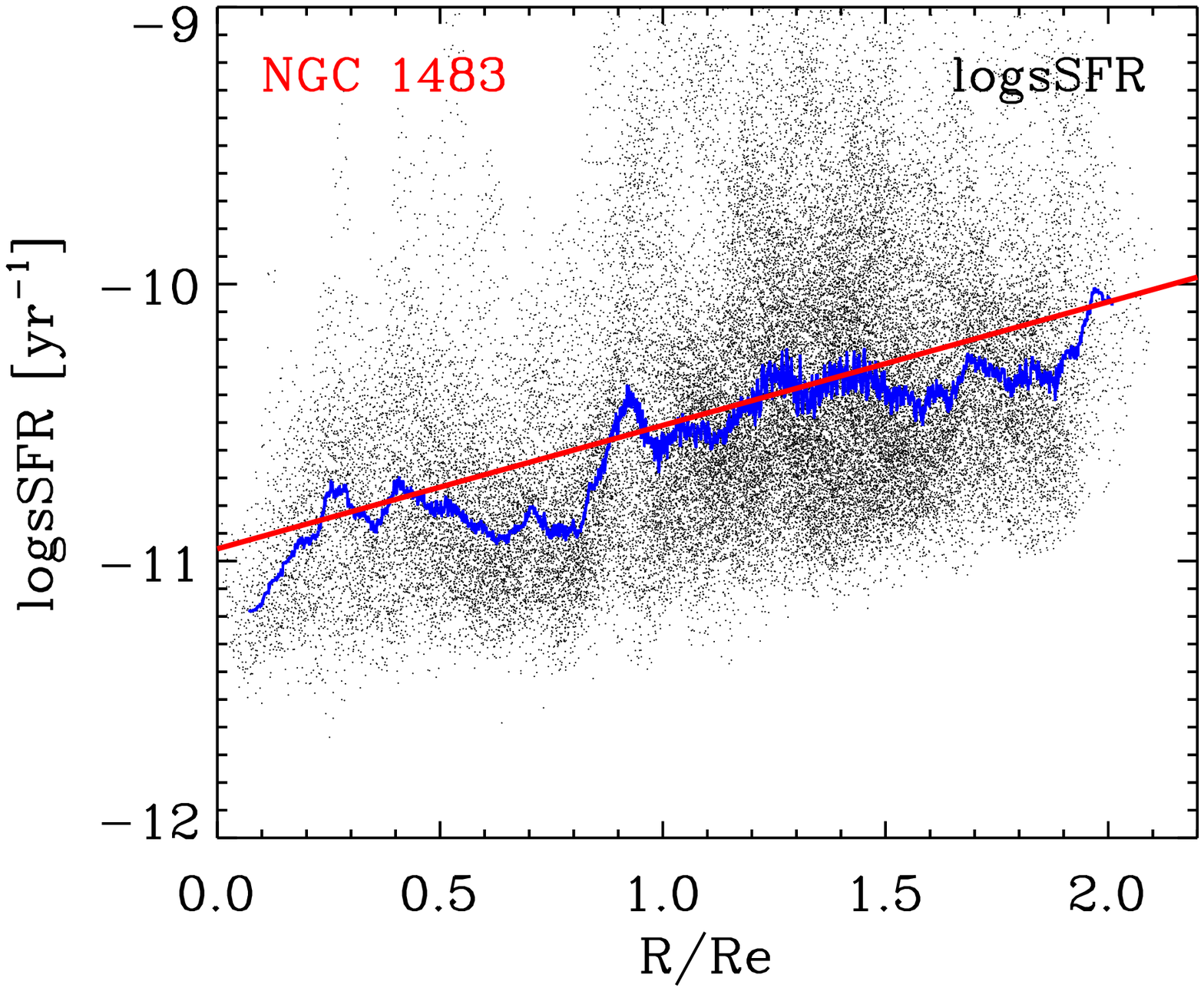,clip=true,width=0.3\textwidth}
    \epsfig{figure=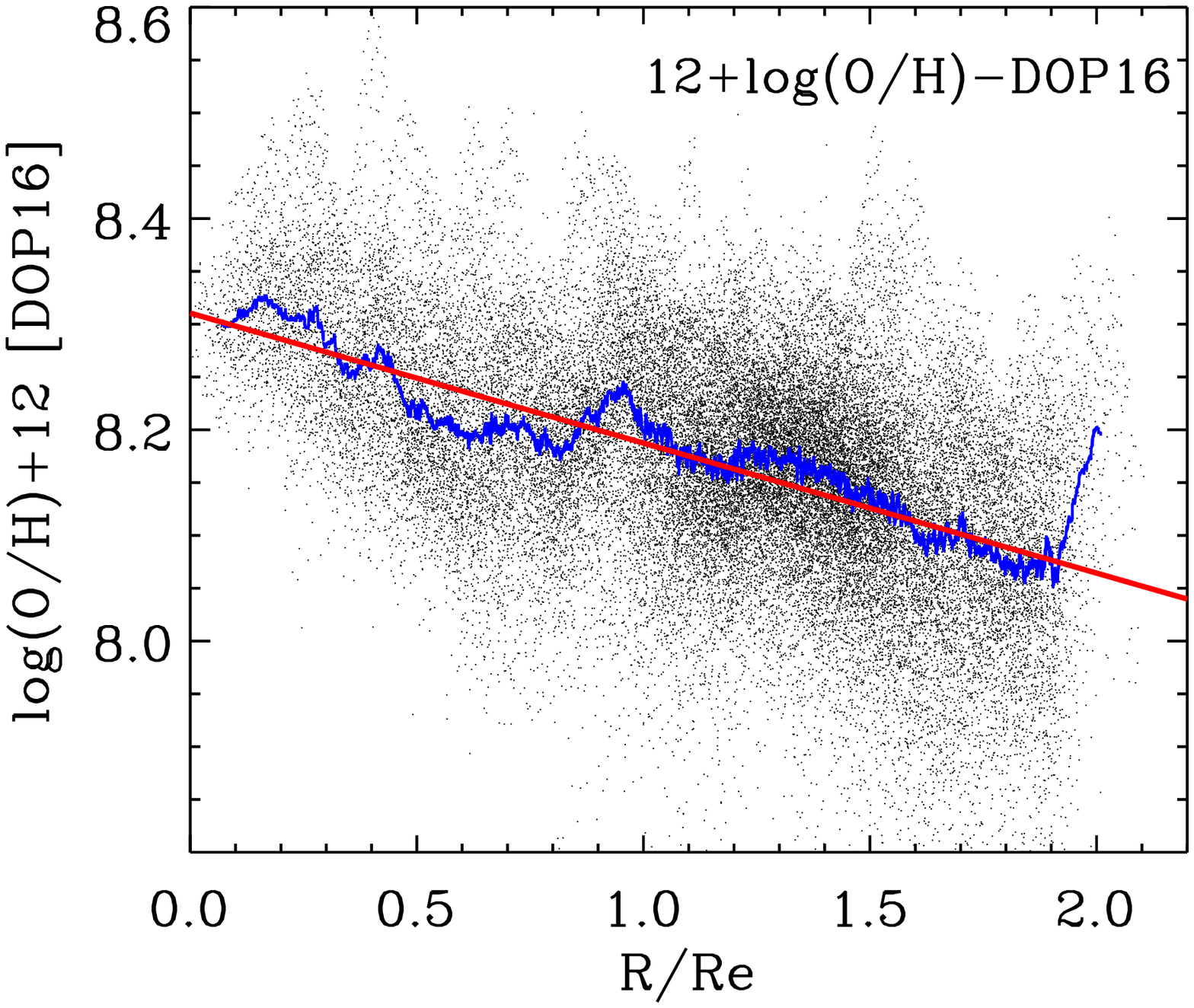,clip=true,width=0.3\textwidth}
    \epsfig{figure=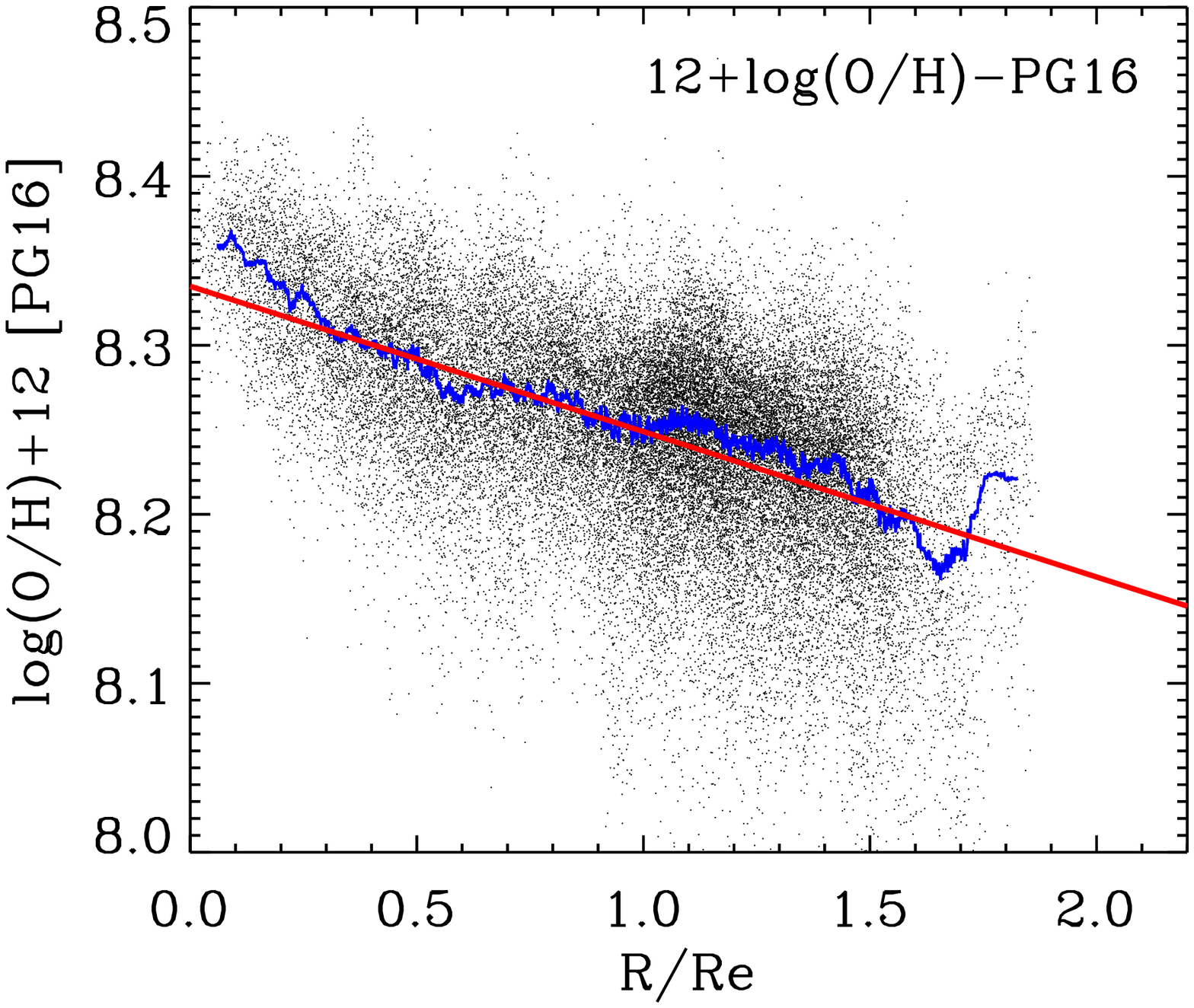,clip=true,width=0.3\textwidth}
    
    \epsfig{figure=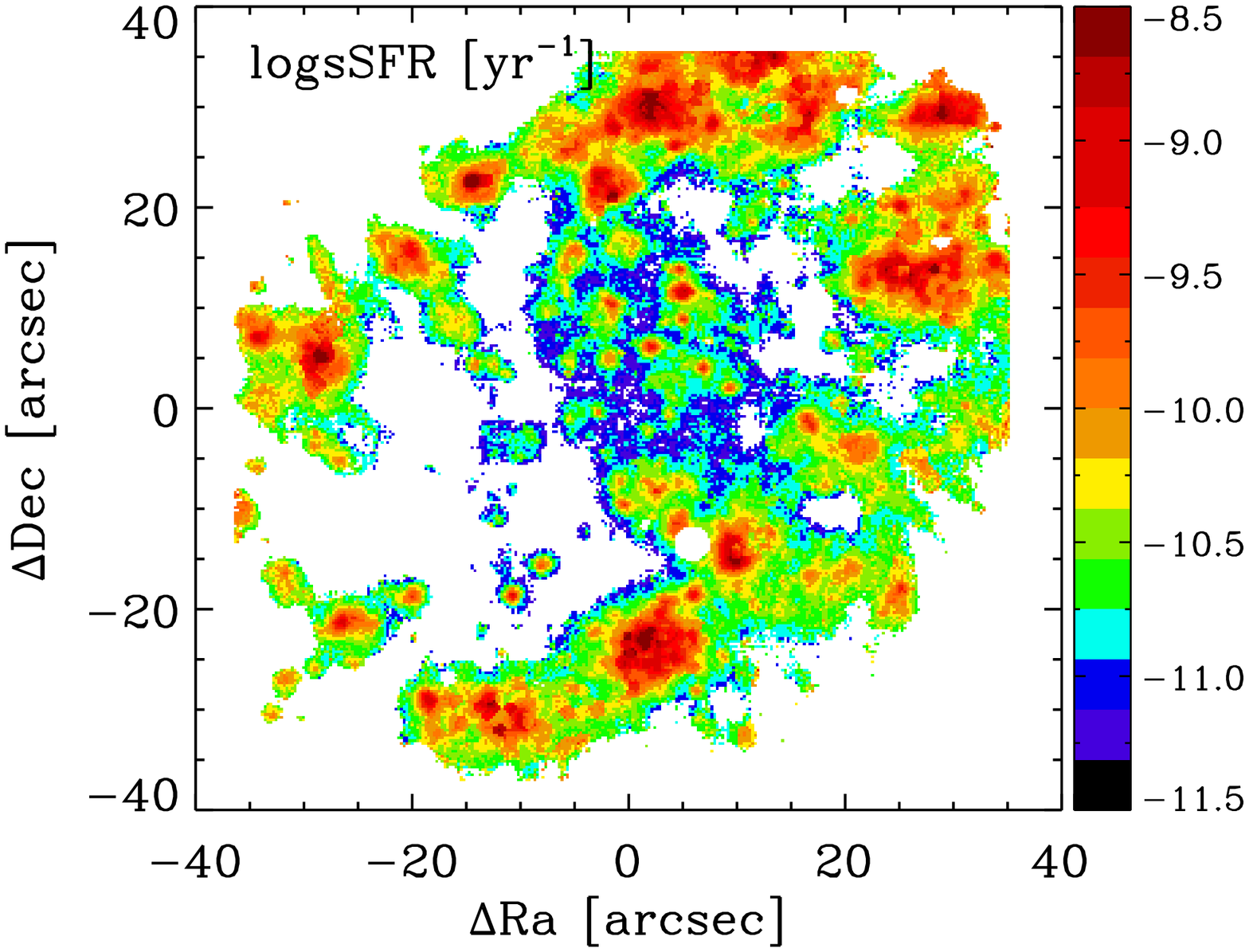,clip=true,width=0.3\textwidth}
    \epsfig{figure=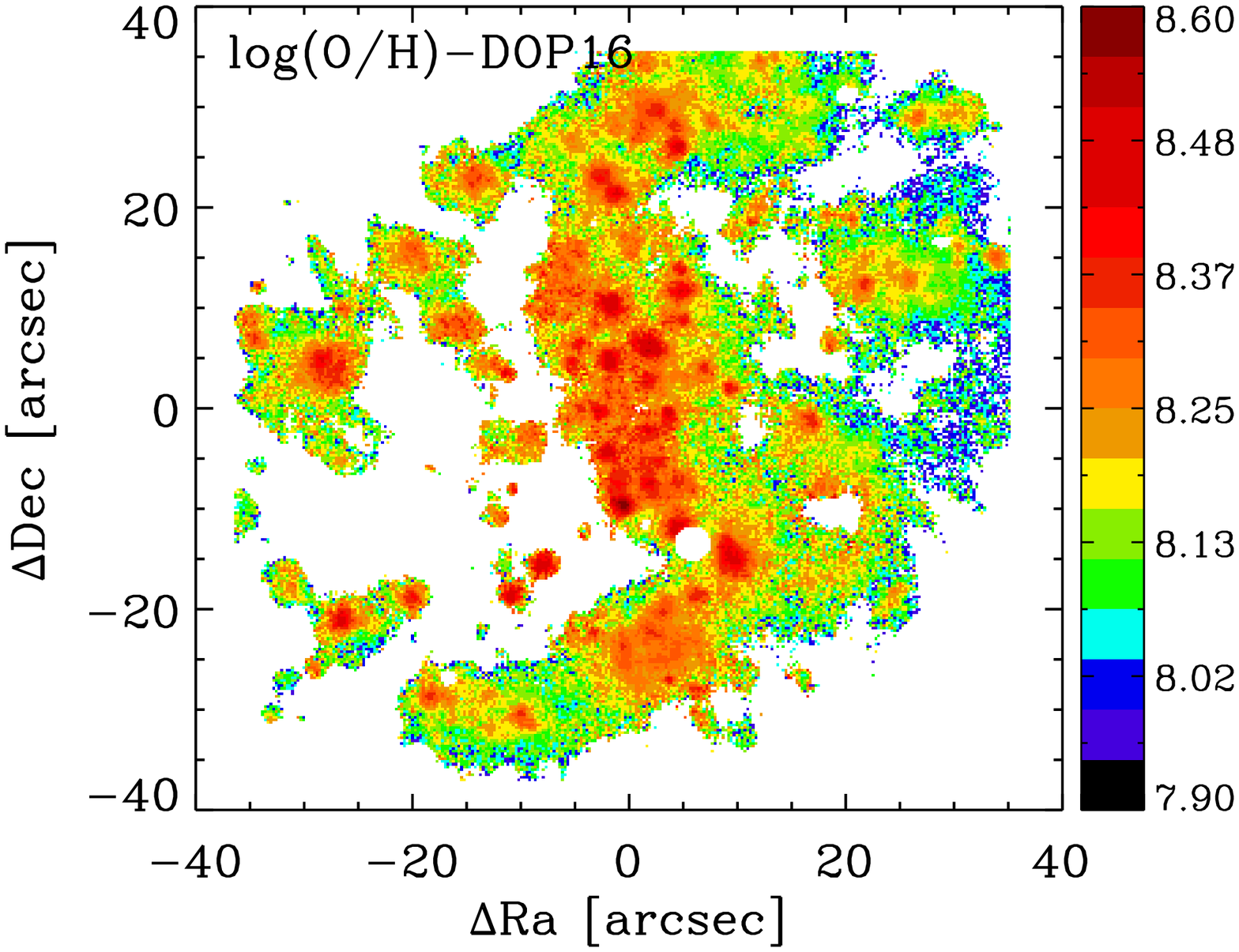,clip=true,width=0.3\textwidth}
    \epsfig{figure=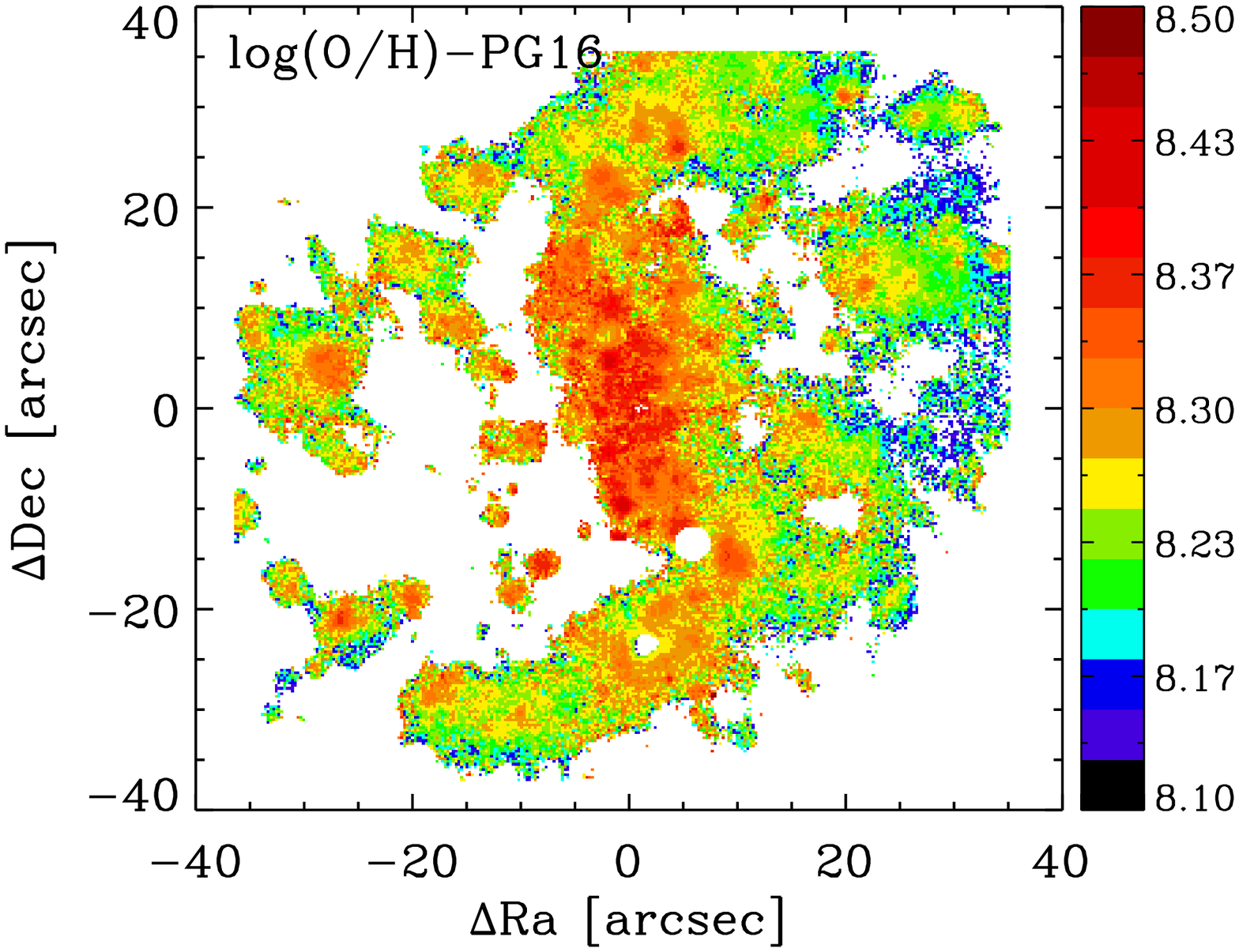,clip=true,width=0.3\textwidth}

    \epsfig{figure=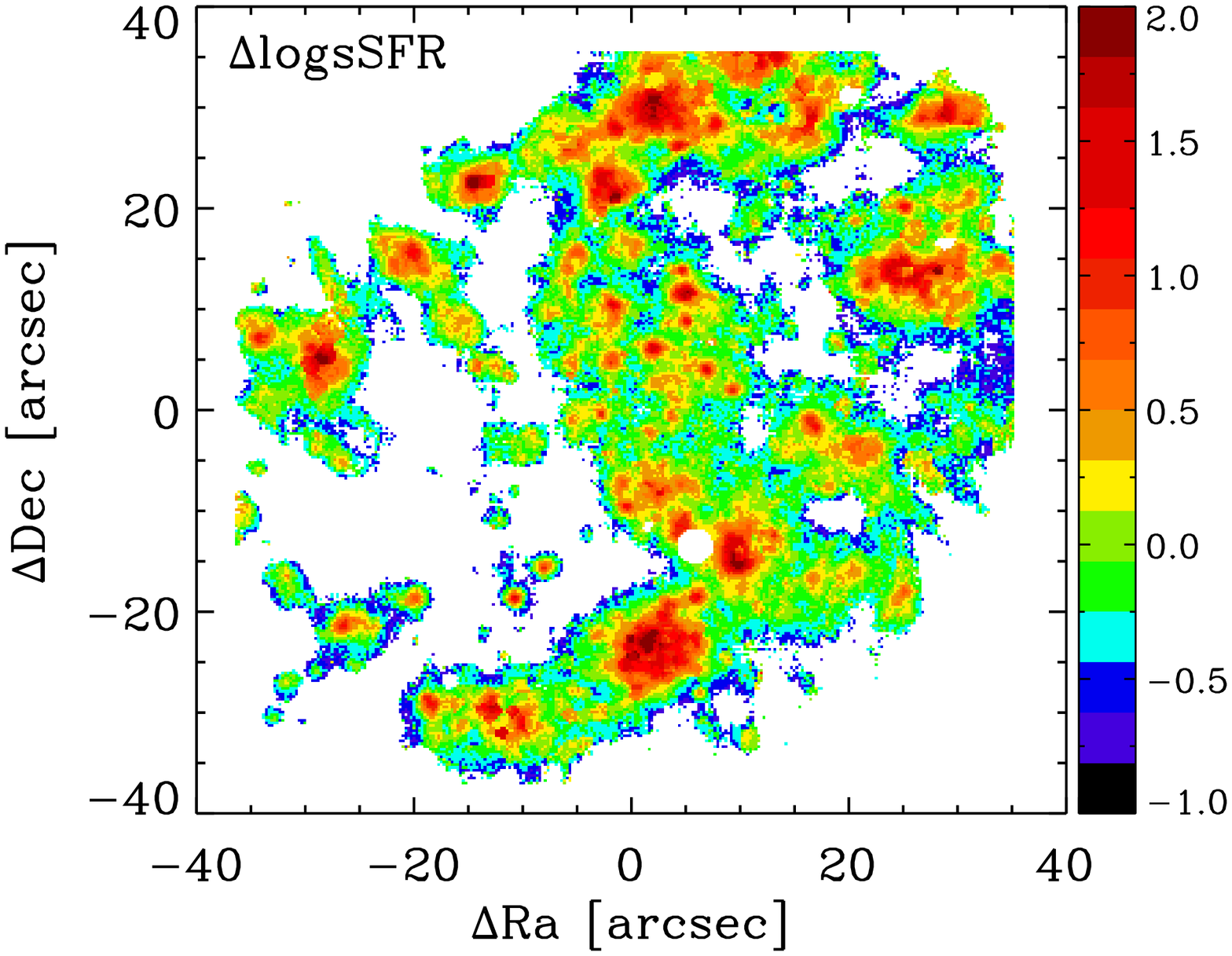,clip=true,width=0.3\textwidth}
    \epsfig{figure=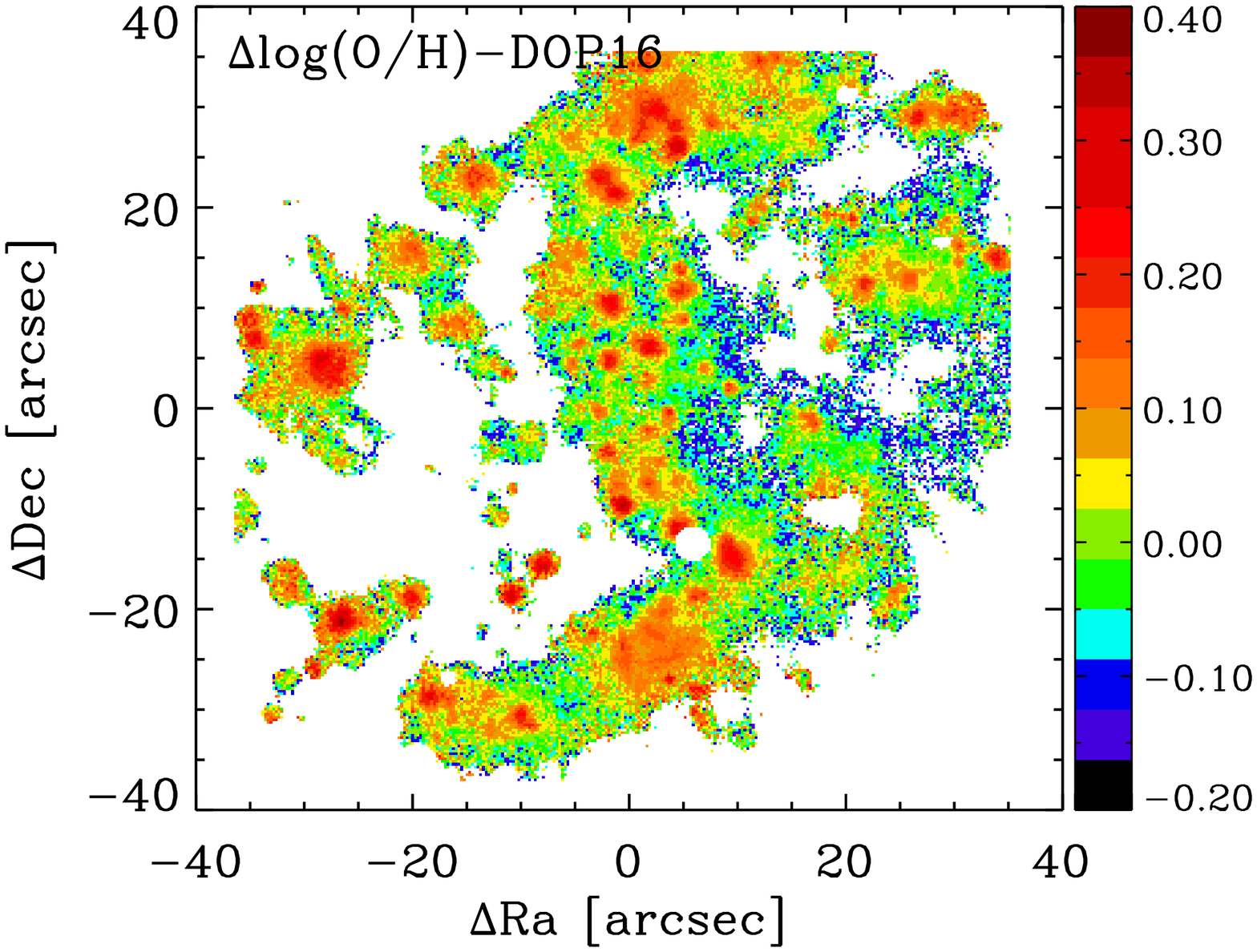,clip=true,width=0.3\textwidth}
    \epsfig{figure=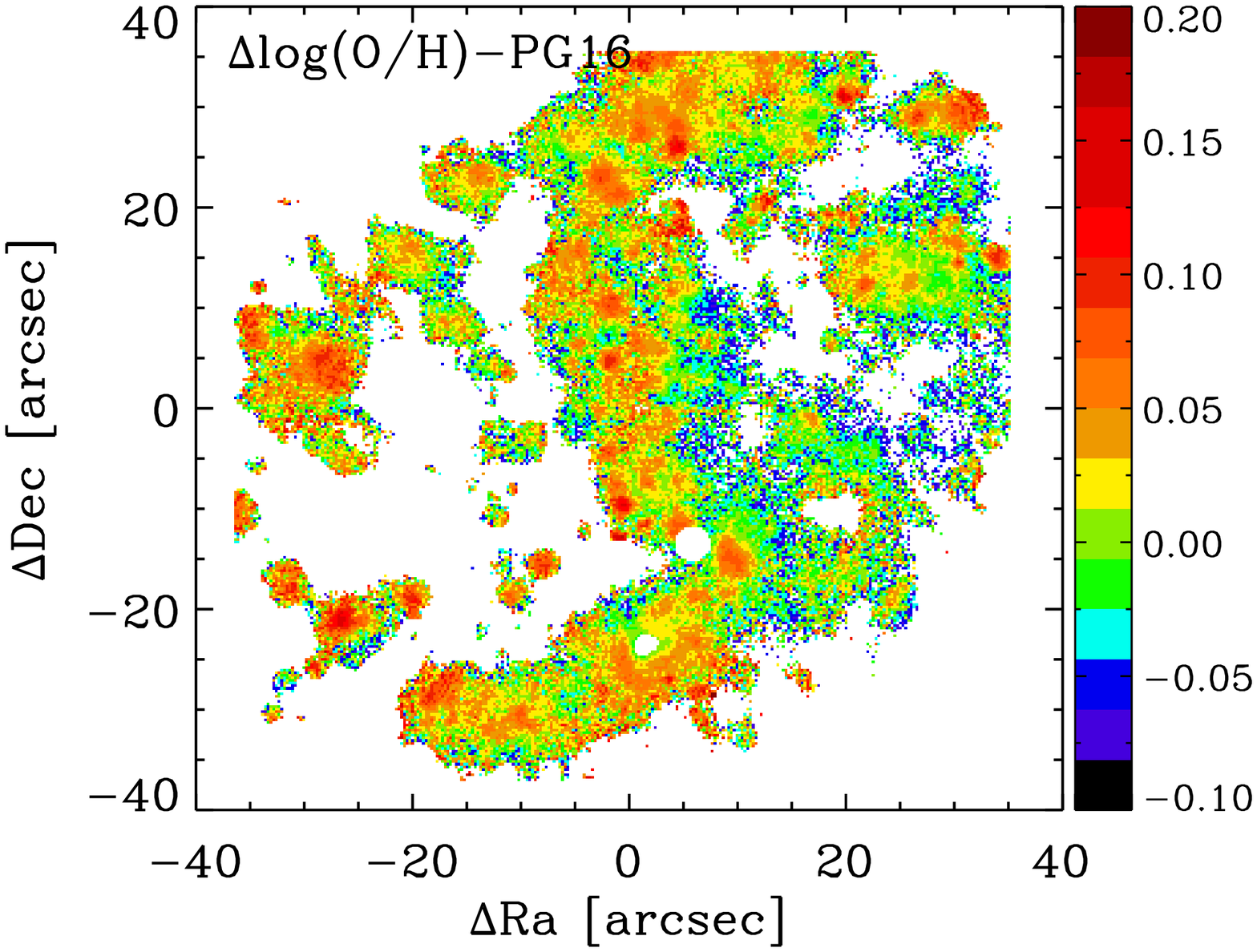,clip=true,width=0.3\textwidth}

    \end{center}
  \caption{An example of the profiles and 2-d maps of SFR and log(O/H) for a representative MAD galaxy, NGC 1483. Top three panels: The profiles of sSFR, log(O/H)-\citetalias{Dopita-16} and log(O/H)-\citetalias{Pilyugin-16} for NGC1483. In each panel, the small dots show individual spaxels from NGC 1483, the blue line is the running median profile, and the red line is a linear fit to the data points. 
  Middle three panels: The 2-d maps of sSFR, log(O/H)-\citetalias{Dopita-16} and log(O/H)-\citetalias{Pilyugin-16} for NGC1483. In each panel, the white regions within the MUSE field of view are due to the fact that these regions are located in the Seyfert or LINER regions according to the BPT diagram, where the SFR and log(O/H) can not be well determined based on emission lines.   
  Bottom three panels: The maps of $\Delta\log$sSFR, $\Delta$log(O/H)-\citetalias{Dopita-16} and $\Delta$log(O/H)-\citetalias{Pilyugin-16} for this galaxy. The $\Delta\log$sSFR, $\Delta$log(O/H)-\citetalias{Dopita-16} and $\Delta$log(O/H)-\citetalias{Pilyugin-16} of each individual spaxel are defined to the deviations from the red lines in the corresponding top panels. }

  \label{fig:6}
\end{figure*}


In Section \ref{sec:2}, we investigated the behavior of the SFR$(t)$, the cold gas mass $M_{\rm gas}(t)$ and the gas-phase metallicity $Z(t)$ in the gas regulator system in response to variations in the inflow rate $\Phi(t)$ and star-formation efficiency SFE$(t)$, and how this response depends on the (assumed constant) wind mass-loading factor $\lambda$, and metallicity of the inflowing gas $Z_{\rm 0}$.  Specifically, we found a {\it negative} correlation between $\Delta \log$SFR and $\Delta \log$Z (i.e. $\log ({\rm SFR}/\langle {\rm SFR}\rangle$) vs. $\log (Z/\langle {Z}\rangle)$) when driving the gas-regulator with time-varying inflow rate, and a {\it positive} correlation between $\Delta \log$SFR and $\Delta \log Z$ when driving with time-varying SFE$(t)$. Therefore, one can in principle identify the driving mechanism of star formation activity by looking at the sign of the correlation between SFR and gas-phase metallicity in observational data.  

However, as pointed out above in Section \ref{sec:2.4}, one should look at the correlations of the relative values of SFR and $Z$ (i.e. the residuals $\Delta \log$SFR and $\Delta \log Z$ in Figure \ref{fig:1} and Figure \ref{fig:2}), rather than the absolute values, in order to take out the effects of different $\langle {\rm SFR}\rangle$ and $\langle Z\rangle$ for different galaxies or of different regions within them, e.g. the overall mass-metallicity or mass-sSFR relations or radial gradients in metallicity or sSFR within galaxies.  

In this section, we will therefore construct radial profiles of sSFR and log(O/H) for the MAD galaxies, and use these to construct localized $\Delta \log$sSFR and $\Delta \log$(O/H) data points from the observations.

We here clarify why the sSFR is better than the SFR to characterize variations in the SFR within and between galaxies.  Ideally, we would wish, in order to compare observations with the model in Section \ref{sec:2}, to follow a given galaxy, or a region within a galaxy, as it changes temporally.   This is of course not possible.   Instead, we must compare different galaxies in the population, or different regions at the same radius within a galaxy, and, invoking ergodicity, assume that these reflect the temporal variations that we are interested in.   For each observational point, we therefore need the best estimate of the average (long-term) state of that location.   It is an observational fact that, both from galaxy to galaxy and within galaxies, the range of sSFR is smaller than the range of SFR.  For this reason, the average sSFR will be better defined than the average SFR.
Related to this, normalizing by the local $\Sigma_*$, even for spaxels at the same galactic radius, also removes the effect of azimuthal variations in the density of gas to the extent that both gas and stars vary in the same way.


We emphasize again that, in the model predictions, the $\Delta\log$SFR is replaceable by $\Delta\log$sSFR, and the $\Delta\log$Z is replaceable by $\Delta\log$(O/H), as mentioned in the end of Section \ref{sec:2}. 

Figure \ref{fig:6} shows an example of the radial profiles and 2-dimensional maps of sSFR and 12+log(O/H) for one individual MAD galaxy, NGC 1483. This typical galaxy is not specially selected in anyway, and is shown for illustration purposes as being representative of the general sample. 
The top three panels of Figure \ref{fig:6} show the sSFR$(r)$, and the 12+log(O/H) as estimated by {\tt N2S2H$\alpha$} and {\tt Scal} as a function of radius for individual spaxels. The log(O/H) based on {\tt N2S2H$\alpha$} diagnostic is denoted as log(O/H)-\citetalias{Dopita-16}, and the log(O/H) based on {\tt Scal} approach is denoted as log(O/H)-\citetalias{Pilyugin-16}. The radius used here is the de-projected radius scaled by the effective radius of the galaxy. In computing this de-projected radius we use the disk inclination based on the measured minor-to-major axis ratio and the position angle taken from the S4G photometry \citep{Salo-15}, assuming an infinitely thin disk. In each of the top panels, the blue line is a running median of 201 spaxels. 

As shown, for NGC 1483 the distribution of the sSFR at a given radius is quite strongly asymmetric, over nearly the whole range of galactic radius. While the sSFR of most spaxels are close to the median profile (or lightly less), a small fraction of spaxels have sSFR that is enhanced by up to an order of magnitude relative to the median profile. This is due to the fact that star formation activity is not uniform across the disk, but happens mostly in spiral arms or other star-formation regions.  The regions with strong of enhanced sSFR can clearly be seen in the sSFR map in the middle-left panel of Figure \ref{fig:6}. In addition, the sSFR profile shows a positive radial gradient, which is consistent with the inside-out growth expected in disk galaxies \citep[e.g.][]{Perez-13, Li-15, Ibarra-Medel-16, Lilly-16, Goddard-17, Rowlands-18, Wang-18a}.

The impression of strong asymmetry is reduced in 12+$\log({\rm O/H})$, for both metallicity indicators. In addition, the overall 12+$\log({\rm O/H})$ profiles for both indicators have a negative radial gradient, consistent with the previous studies of disk galaxies \citep[e.g.][]{Pilkington-12, Belfiore-17, Erroz-Ferrer-19}. 
This feature also can be seen in the maps of 12+$\log ({\rm O/H})$, shown in the middle row of Figure \ref{fig:6}.
The measurements of $\log ({\rm O/H})$ based on {\tt N2S2H$\alpha$} and {\tt Scal} are qualitatively consistent. However, for a given dataset, $\log ({\rm O/H})$-\citetalias{Dopita-16} is usually larger than $\log ({\rm O/H})$-\citetalias{Pilyugin-16}, and the range of $\log ({\rm O/H})$-\citetalias{Dopita-16} is nearly twice as that of $\log ({\rm O/H})$-\citetalias{Pilyugin-16}. 
We note that the particular galaxy is typical of the sample, and most of the SF disk galaxies show similar features in sSFR and $\log ({\rm O/H})$ as shown for this galaxy.

As pointed out in Section \ref{sec:2.4}, the gas-regulator predicts that the average SFR will only depend on the average inflow rate $\Phi_{\rm 0}$ and mass-loading factor $\lambda$, and the average $Z$ is determined by the effective yield, defined from the wind mass-loading as $y(1-R+\lambda)^{-1}$, and the metallicity of the inflowing gas $Z_{\rm 0}$. However, the $\lambda$ (and possibly also the $Z_{\rm 0}$) may well change radially within individual galaxies, because $\lambda$ should be directly related to the gravitational potential well. If so, the fitted average $12+\log({\rm O/H})$ profile should therefore reflect the radial dependence of the wind mass-loading and/or $Z_{\rm 0}$ (see further discussion in Section \ref{sec:4.2.2}).
We are not interested in this effect if it is caused by time-invariant factors ($\lambda$ and $Z_0$).  Rather we are interested in the spaxel-by-spaxel variations that are superposed on this smooth radial profile, because these are presumably caused by shorter term temporal variations.

Therefore, we focus on the values of individual spaxels relative to these underlying trends, 
i.e. on the $\Delta \log$sSFR and $\Delta \log$(O/H) residuals when the underlying sSFR$(r)$ and $\log ({\rm O/H})(r)$ profiles are subtracted from each spaxel (compare to the relation of $\Delta \log$SFR vs. $\Delta \log Z$ in Figure \ref{fig:1} and \ref{fig:2}). 
In this way, we are effectively not allowing a variation of $\lambda$ and $Z_0$ between spaxels at a given radius for an individual galaxy, but allowing these quantities to vary radially, and removing the effect of this from our analysis.


To achieve this, we first perform a linear fit to the sSFR$(r)$ and 12+$\log ({\rm O/H})(r)$ profiles based on all the individual spaxels. These fits are shown as red lines in the top panels of Figure \ref{fig:6}. As shown, for both sSFR or 12+$\log ({\rm O/H})$, the linear fit is quite a good representation of the median profile, although it is not perfect. We then define the $\Delta \log$sSFR and $\Delta \log ({\rm O/H})$ for each individual spaxel as the deviation of each spaxel from the fitted profile of sSFR$(r)$ or 12+$\log ({\rm O/H})(r)$ respectively.  In this way, we can eliminate the overall radial dependences of sSFR and $\log ({\rm O/H})$, as well as global differences between galaxies, such as the overall sSFR or effects due to the mass-metallicity relation.   As shown in Section \ref{sec:2}, in the gas-regulator framework, these changes in $\langle {\rm SFR} \rangle$ or $\langle Z \rangle$ will reflect differences (radially within galaxies or from galaxy to galaxy) in the overall inflow rate, mass-loading factor and $Z_{\rm 0}$. 

The bottom three panels of Figure \ref{fig:6} show the maps of $\Delta \log$sSFR and $\Delta \log ({\rm O/H})$ that are obtained for NGC 1483 after removing the radial gradients in $\log$sSFR and $\log ({\rm O/H})$. It is immediately apparent that regions with enhanced SFR, indicated by red bumps, nearly always show enhanced metallicity for both of the two metallicity indicators. This is consistent with the previous analysis of \cite{Kreckel-19} and  \cite{Erroz-Ferrer-19}, at the similar spatial resolution of $\sim$100 pc.  It should be noted that the color scale used for $\Delta \log ({\rm O/H})$-\citetalias{Dopita-16} has twice the range of that used for $\Delta \log ({\rm O/H})$-\citetalias{Pilyugin-16}. 

\begin{figure*}
  \begin{center}
    \epsfig{figure=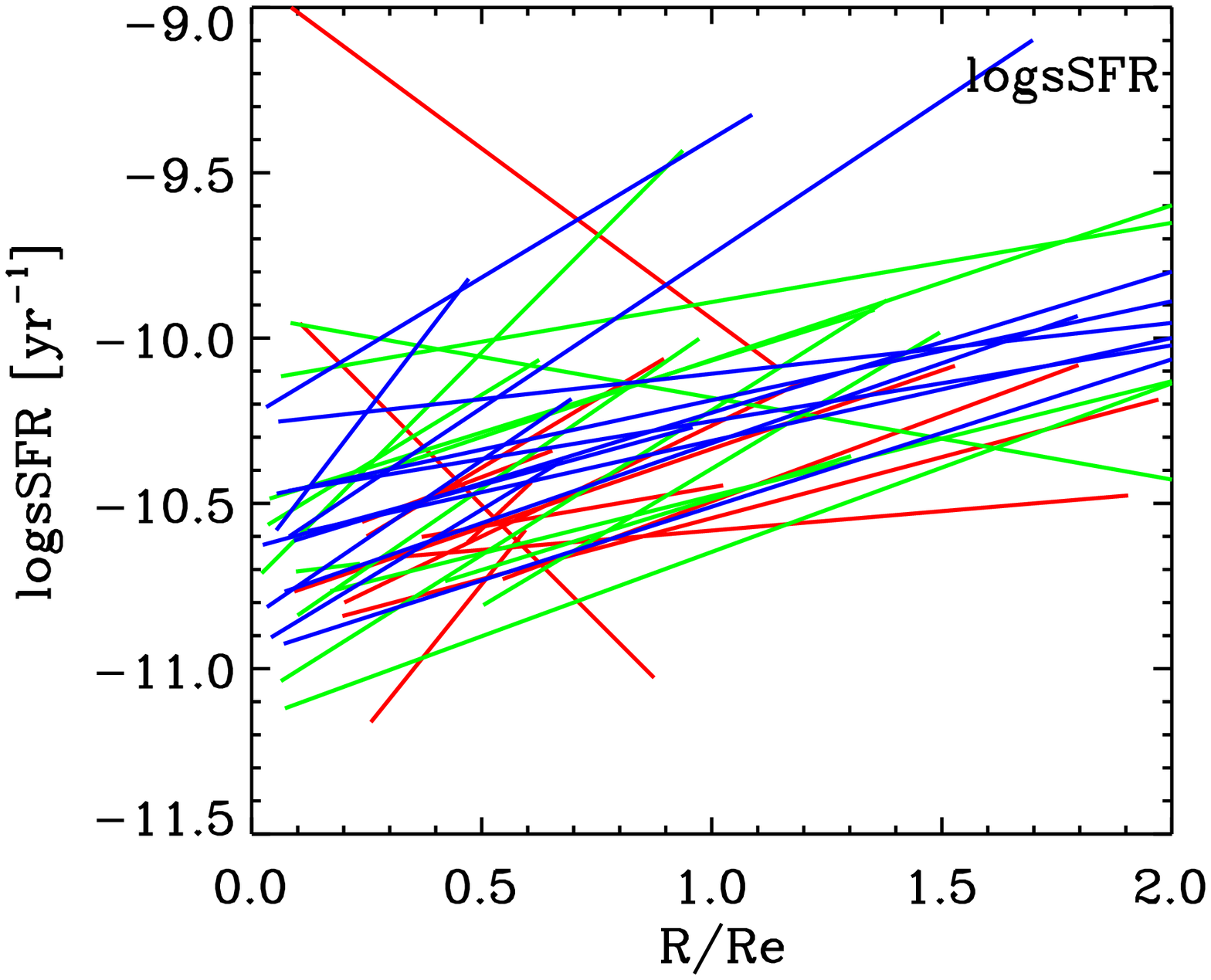,clip=true,width=0.33\textwidth}
    \epsfig{figure=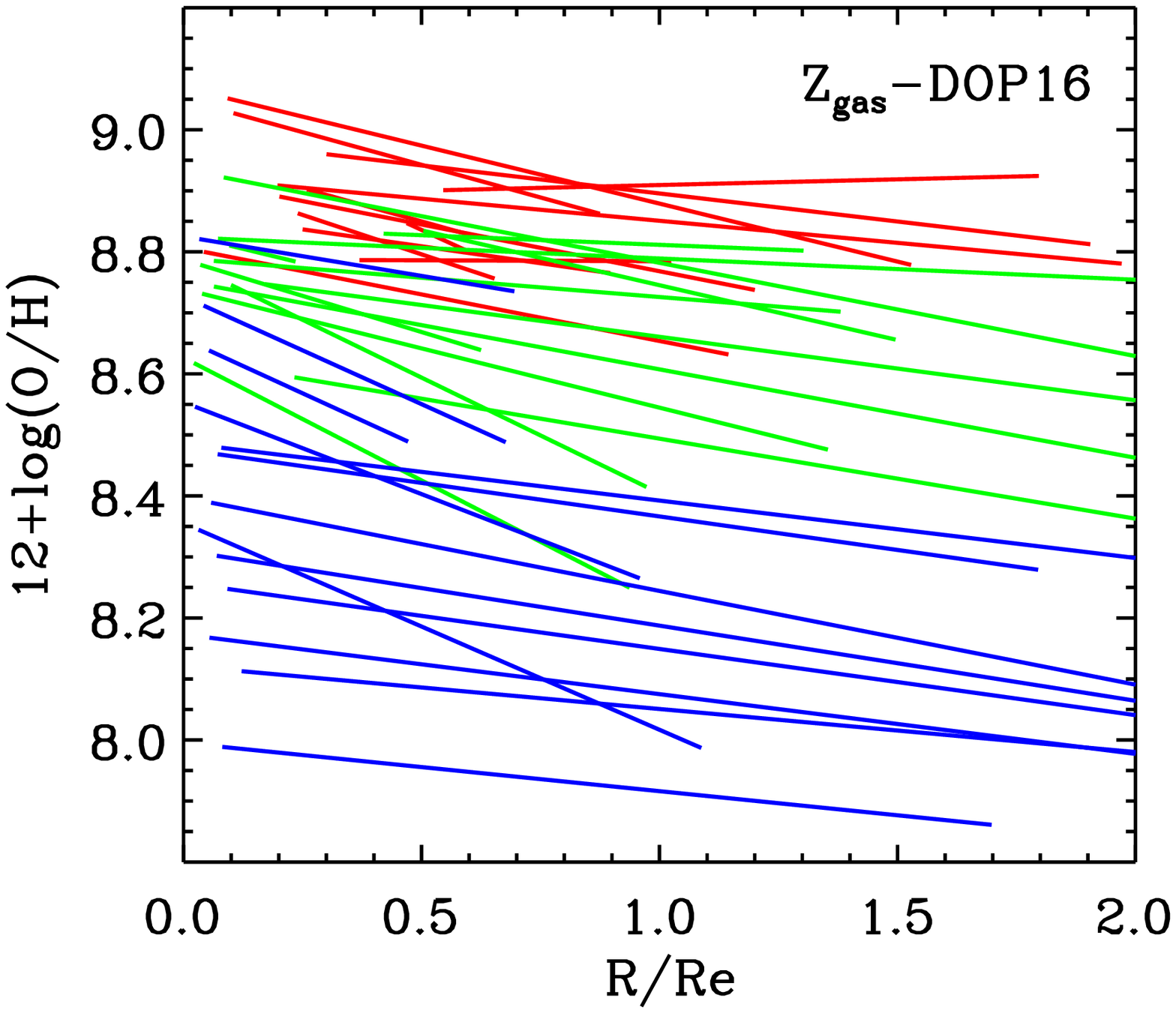,clip=true,width=0.33\textwidth}
    \epsfig{figure=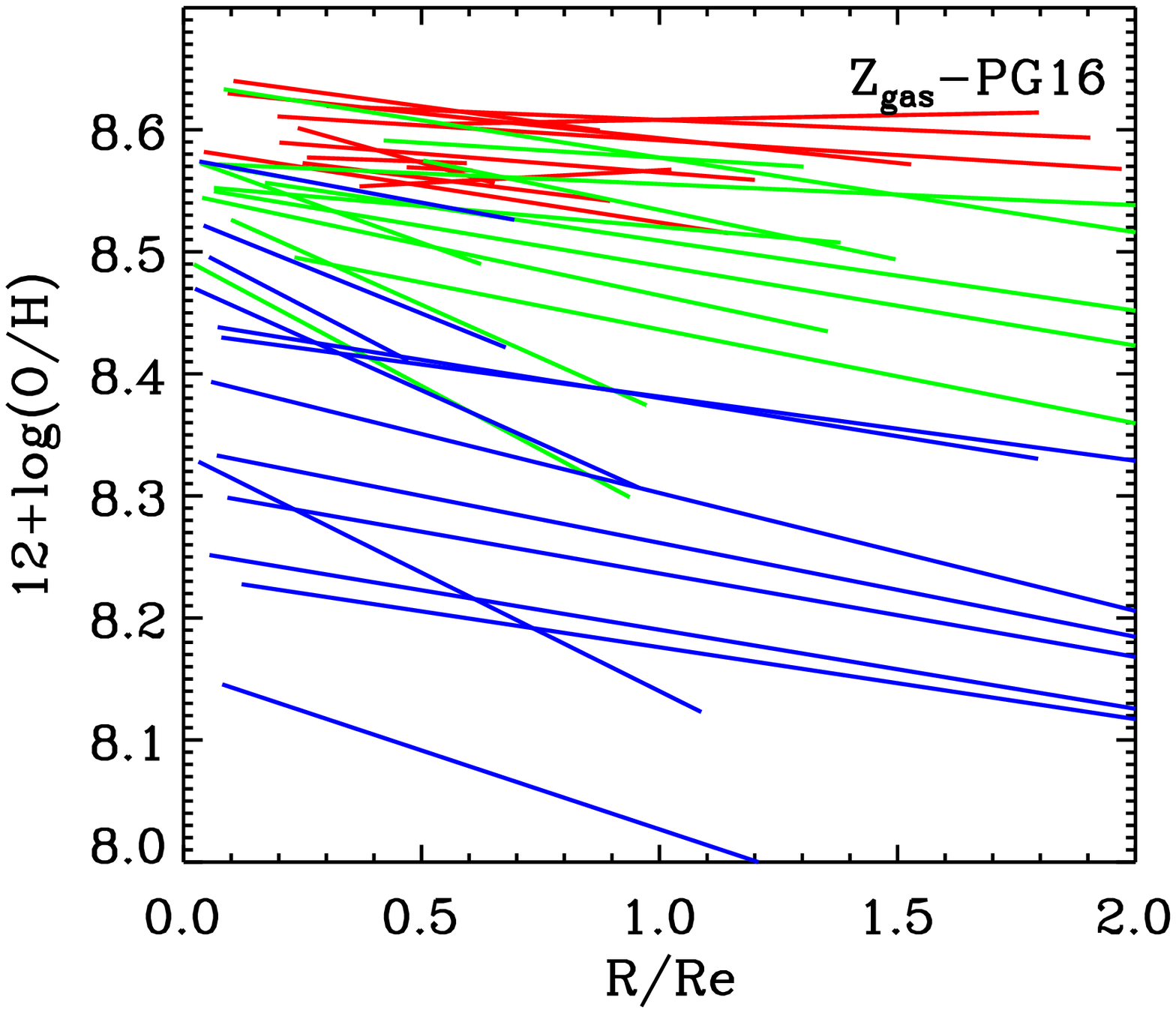,clip=true,width=0.33\textwidth}
    \end{center}
  \caption{ The linearly fitted profiles of sSFR (left panel),  12+log(O/H)-\citetalias{Dopita-16} (middle panel), and 12+log(O/H)-\citetalias{Pilyugin-16} (right panel) for all MAD galaxies.  In each panel, the MAD galaxies are separated into three color-coded stellar mass bins: \lgmstar$<10.0$ (blue), $10.0<$\lgmstar$<10.8$ (green), and $10.8<$\lgmstar\ (red). 
   }
  \label{fig:7}
\end{figure*}

Figure \ref{fig:7} shows the fitted linear profiles of sSFR, 12+$\log ({\rm O/H})$-\citetalias{Dopita-16} and 12+$\log ({\rm O/H})$-\citetalias{Pilyugin-16} for all the 38 MAD galaxies. In displaying these profiles, we separate galaxies into three mass bins: \lgmstar$<10.0$ (blue lines), $10.0<$\lgmstar$<10.8$ (green lines), and $10.8<$\lgmstar\ (red lines).  As shown, almost all of the MAD galaxies have positive radial gradients in sSFR with only four exceptions. It can be seen that there is no strong dependence of the sSFR profile on the stellar mass of galaxies.  While the profiles in 12+$\log ({\rm O/H})$ have similar slopes, the overall values show a very strong dependence on global stellar mass, reflecting the MZR.

\begin{figure*}
  \begin{center}
    \epsfig{figure=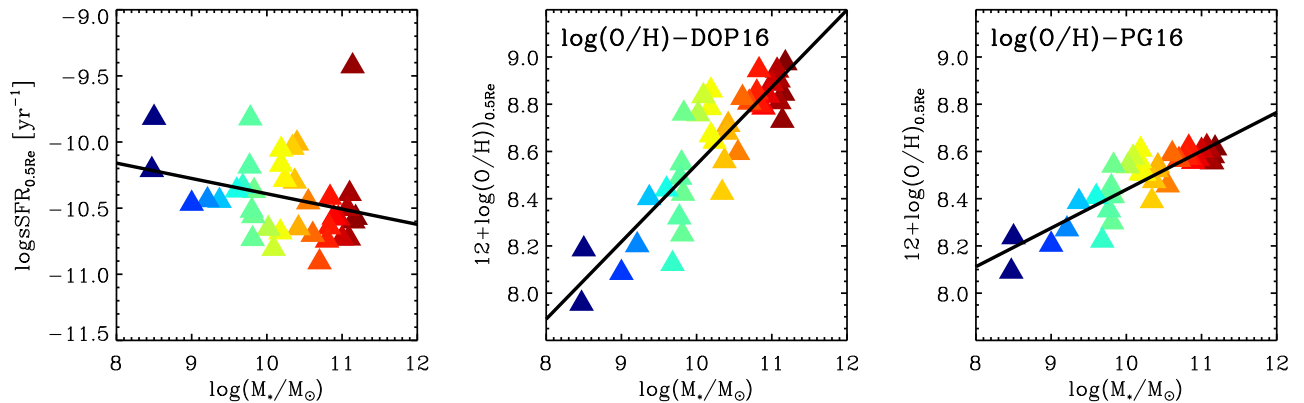,clip=true,width=0.95\textwidth}
    \end{center}
  \caption{ The sSFR (left panel), 12+log(O/H)-\citetalias{Dopita-16} (middle panel) and 12+log(O/H)-\citetalias{Pilyugin-16} (right panel) at 0.5\re\ as a function of the stellar mass for the MAD galaxies. In each panel, the data points are color-coded with the stellar mass, and the black solid line
  is the linear fit to the data points. }
   \label{fig:8}
\end{figure*}

The definition of $\Delta\log$sSFR and $\Delta \log ({\rm O/H})$ for the individual spaxels, based on Figure \ref{fig:7},  enables us to investigate the correlations of $\Delta\log$sSFR vs. $\Delta \log ({\rm O/H})$ for small-scale regions ($\sim$100 pc) within individual MAD galaxies. However, one can well imagine that the physical processes driving the small-scale star formation, may be very different from those driving the star formation on galactic scales and this motivates define analogous quantities, $\Delta\log$sSFR and $\Delta \log ({\rm O/H})$ to reflect the {it global} properties of galaxies with the MAD population. For this purpose, we choose the fitted values of sSFR and 12+$\log ({\rm O/H})$ at 0.5\re\ (see the red lines in the top panels of Figure \ref{fig:6}) as a representative of the global properties of individual galaxies\footnote{We realize that the sSFR and 12+$\log ({\rm O/H})$ at one specific galactic radius can not perfectly reflect the global sSFR and gas-phase metallicities, because both sSFR and 12+$\log ({\rm O/H})$ show radial gradients. In Section \ref{sec:5.1}, we will treat the sSFR and 12+$\log ({\rm O/H})$ that are measured within 1.5\re\ as a representative of global quantities for MaNGA galaxies. }, because the spatial coverage for MAD galaxies generally extends to at least 0.5\re.   

Figure \ref{fig:8} shows the sSFR$_{\rm 0.5Re}$, 12+$\log ({\rm O/H})_{\rm 0.5Re}$-\citetalias{Dopita-16} and  12+$\log ({\rm O/H})_{\rm 0.5Re}$-\citetalias{Pilyugin-16} as a function of the overall stellar mass of the galaxies. The overall stellar mass is obtained by broad-band spectral energy distribution fitting \citep{Erroz-Ferrer-19}.
The sSFR$_{\rm 0.5Re}$ decreases slightly with stellar mass, and 12+$\log ({\rm O/H})_{\rm 0.5Re}$ increases significantly with stellar mass, for both metallicity indicators. Both of these trends are of course well-established for SF galaxies in the literature.  As pointed out above, in the framework of gas-regulator model, the dependence of SFR and $Z$ on stellar mass is due to the stellar mass-dependence of the inflow rate, $\lambda$ and $Z_{\rm 0}$ \citep[see Equation \ref{eq:21} and Equation \ref{eq:22} and discussion in][]{Lilly-13}. To eliminate the mass dependence, we perform a linear fit for each of these two relation, as shown with the black lines in Figure \ref{fig:8}. In a similar way, for each individual MAD galaxy, we can then define the $\Delta \log$sSFR$_{\rm 0.5Re}$ or $\Delta \log ({\rm O/H})_{\rm 0.5Re}$ to be the deviation of an individual galaxy from the linearly fitted relation.  This is useful to study the driving mechanisms of star formation from galaxy to galaxy within the population. 

In Appendix \ref{sec:B}, we repeat the our basic analysis by defining the $\Delta\log$sSFR and $\Delta \log ({\rm O/H})$ for individual spaxels using average relations with $\Sigma_*$ rather than with galactic radius, and find that the basic results remain the same. 

\subsection{Correlations between $\Delta\log$sSFR and $\Delta\log$(O/H) on different spatial scales} \label{sec:4.2}

\begin{figure*}
  \begin{center}
    \epsfig{figure=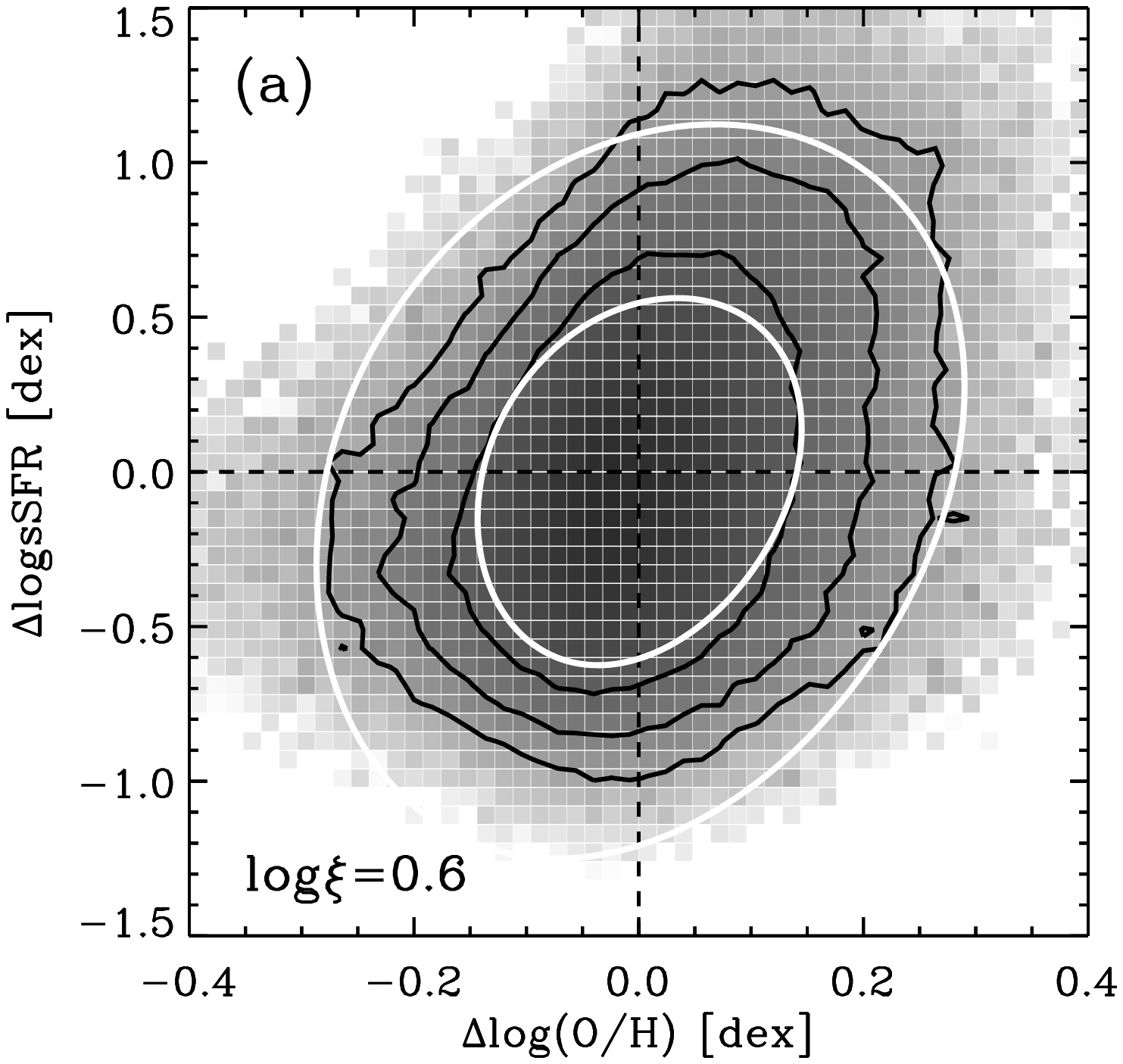,clip=true,width=0.33\textwidth}
    \epsfig{figure=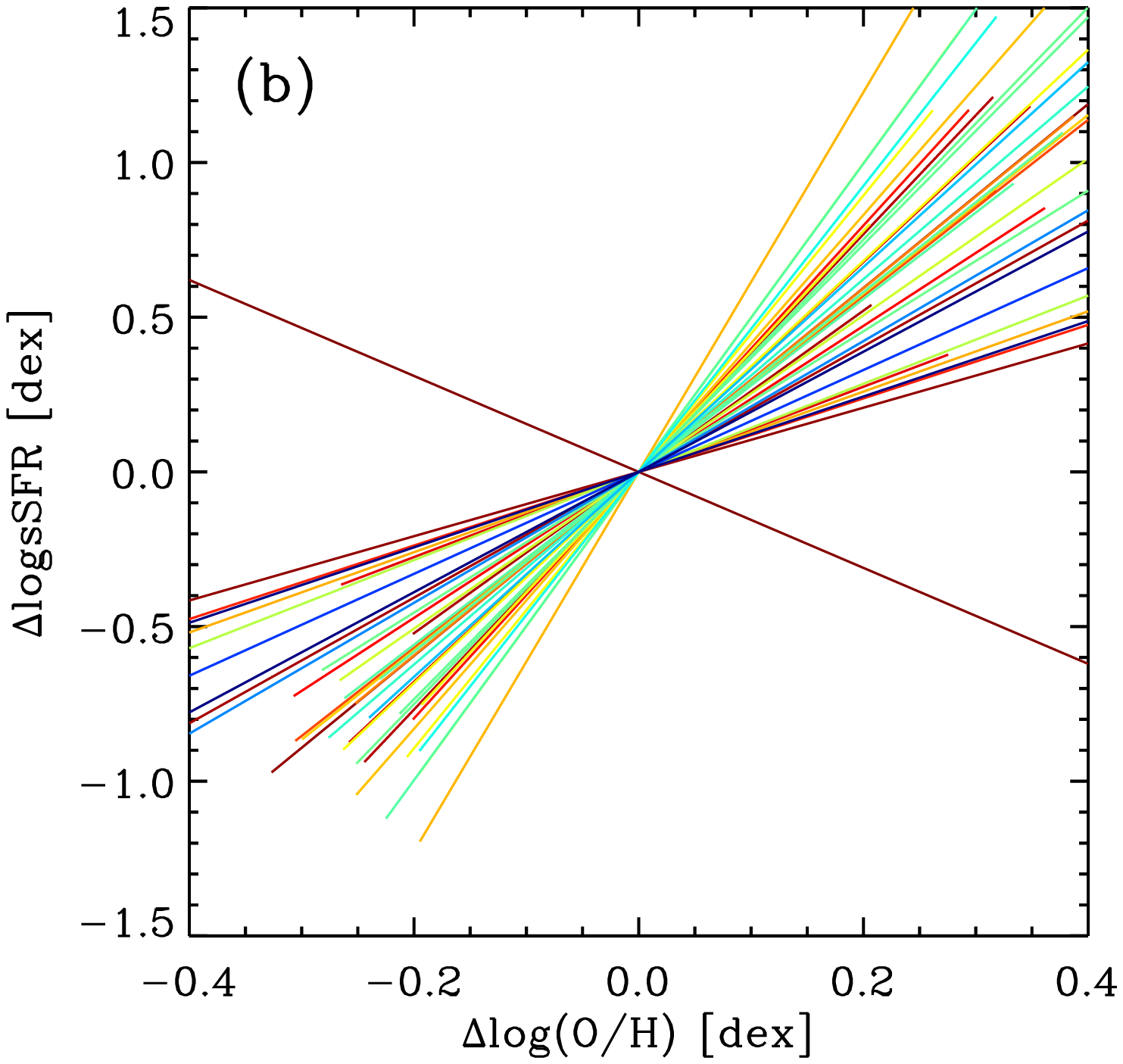,clip=true,width=0.33\textwidth}

    \epsfig{figure=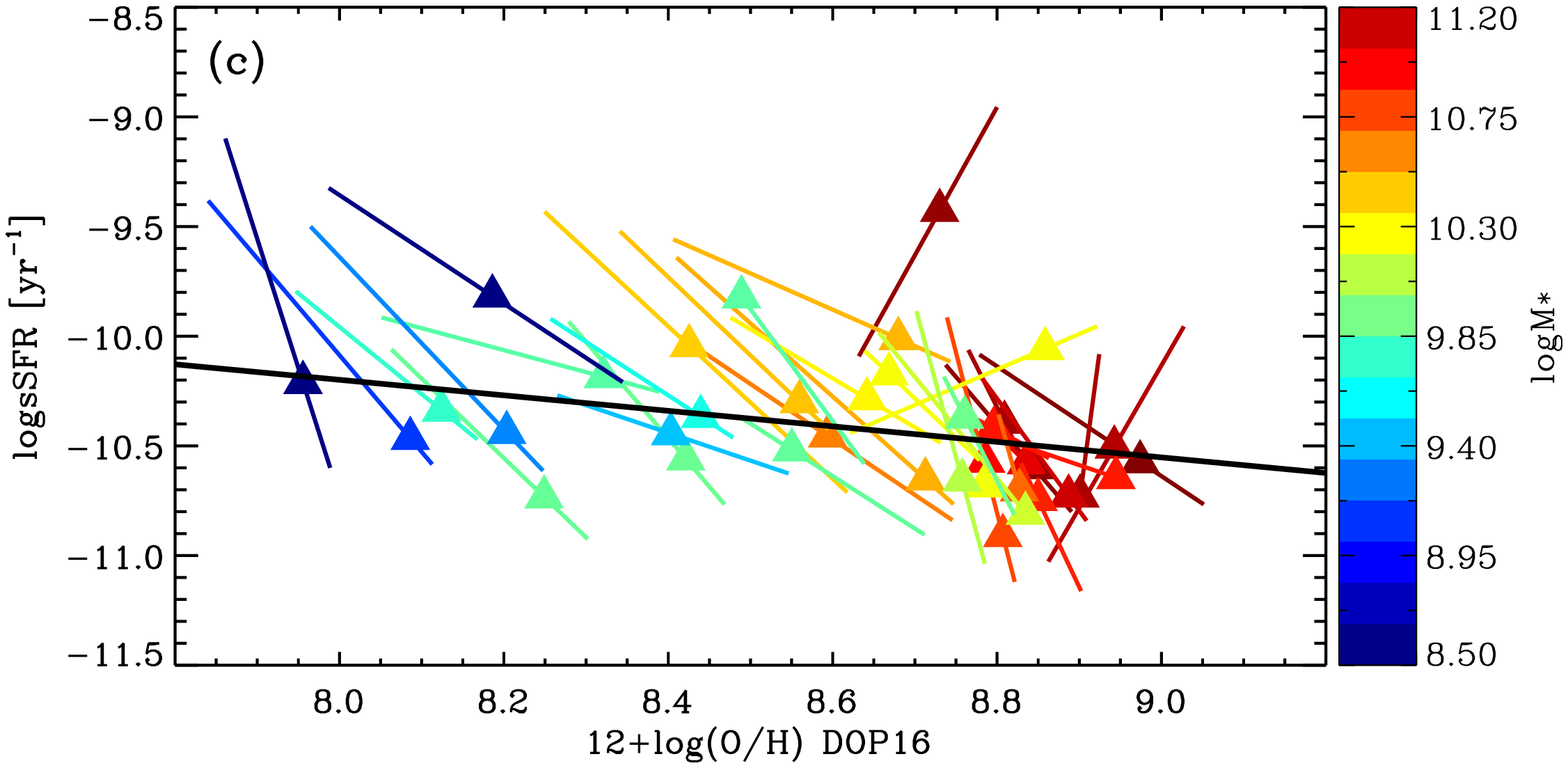,clip=true,width=0.663\textwidth}
    
    \epsfig{figure=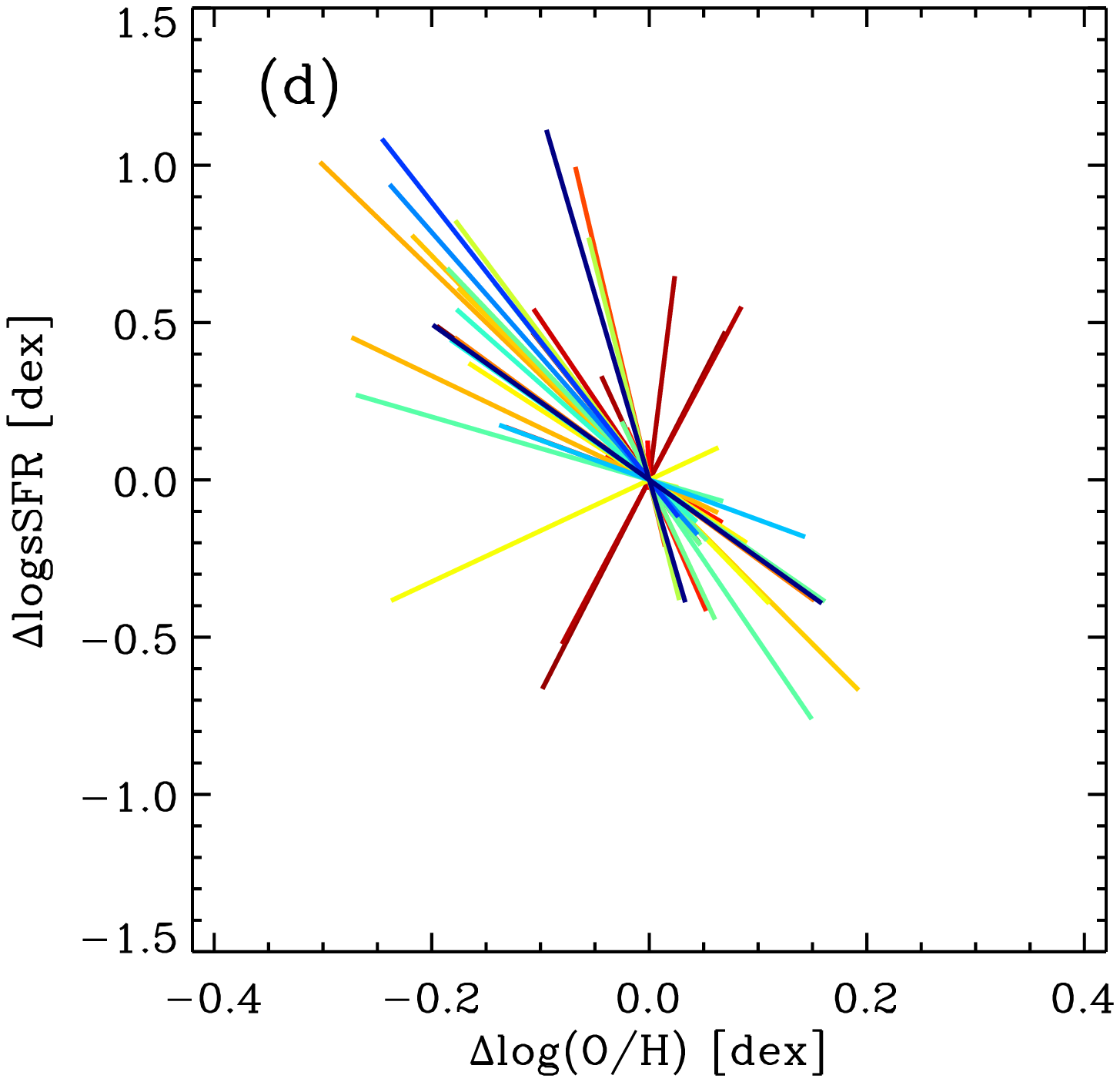,clip=true,width=0.33\textwidth}
    \epsfig{figure=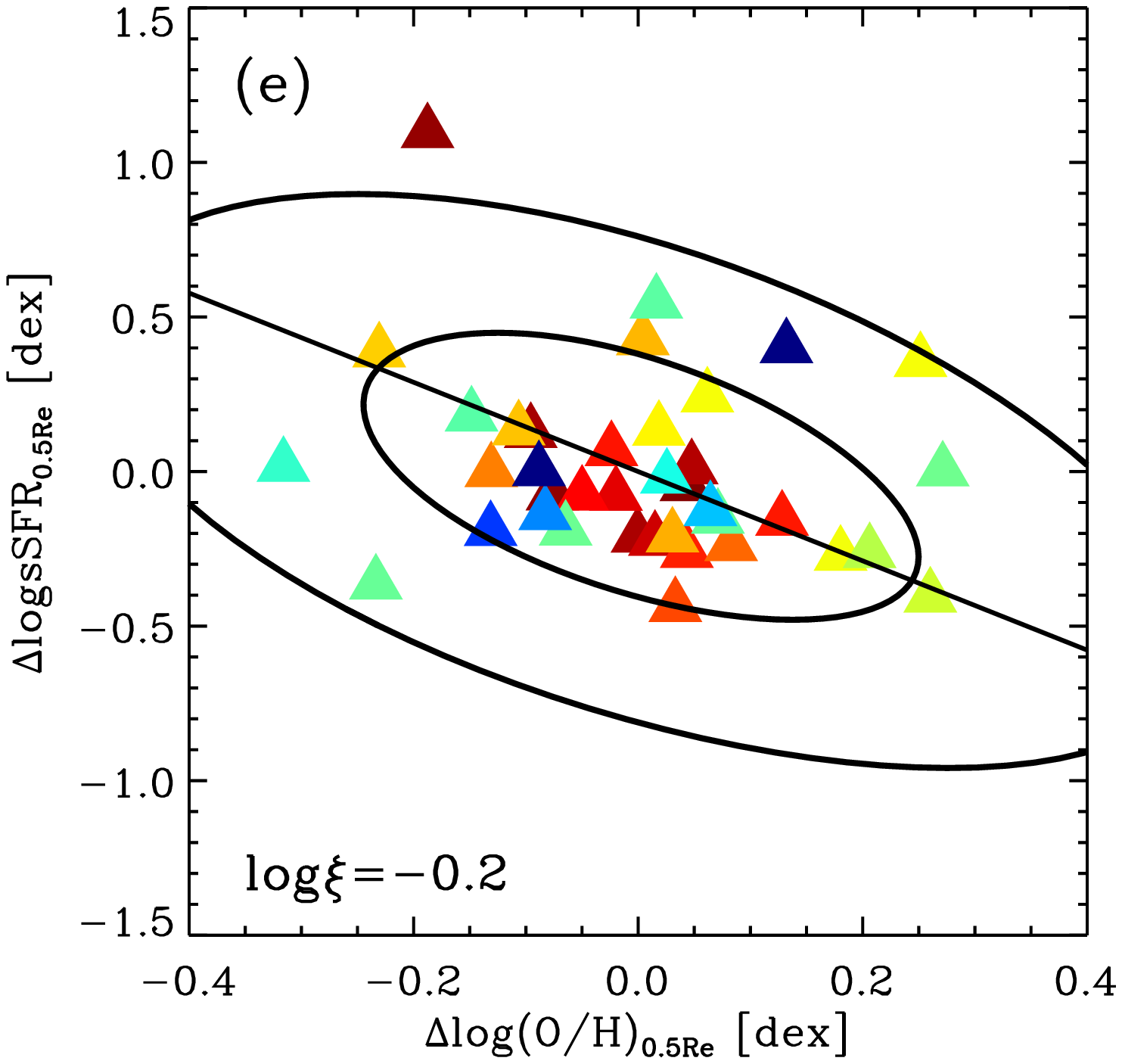,clip=true,width=0.33\textwidth}
    \end{center}
  \caption{ (a), The $\Delta\log$sSFR-$\Delta\log$(O/H) diagram for all the usable spaxels of MAD galaxies. The grayscale shows the number density of spaxels in logarithmic space.  The black contours show equal number densities of spaxels decreasing 0.5 dex from the inside outwards. For comparison, model predictions ($\log \xi$=0.6 and $Z_{\rm 0}=0$) from the top-middle panel of Figure \ref{fig:2} are shown as white ellipses, scaled to match the scatter  of $\Delta\log$sSFR as described in the text. 
 (b), Bisector fits for the individual spaxels on the $\Delta\log$sSFR-$\Delta\log$(O/H) diagram. The lines correspond to the individual MAD galaxies color-coded by the overall stellar mass. The length of the line is determined by the range of $\Delta\log$sSFR for each individual galaxy. (c), The sSFR-metallicity relation of the fitted profiles of sSFR and log(O/H) (shown in Figure \ref{fig:7}) for MAD galaxies, color-coded by the stellar mass. The triangles shows the values of sSFR and log(O/H) at 0.5\re\ for each individual galaxies.  The black solid line in this panel shows the sSFR-log(O/H) relation for the fitted relation of sSFR$_{\rm 0.5Re}$-$M_*$ and log(O/H)$_{\rm 0.5Re}$-$M_*$ relations, which are determined in Figure \ref{fig:8}. (d), The $\Delta\log$sSFR-$\Delta\log$(O/H) relation of the fitted profiles of sSFR and log(O/H) for individual MAD galaxies. The colored lines in panel (d) are taken from panel (c) but shifting the lines with the sSFR and log(O/H) at 0.5\re\ (indicated by the triangles) to be zero. (e), The $\Delta\log$sSFR$_{\rm 0.5Re}$-$\Delta$log(O/H)$_{\rm 0.5Re}$ diagram for the 38 MAD galaxies, with the color-coding of stellar mass. In panel (e), the black line is the bisector fits of the data points. The model predictions (for $\log \xi=-0.2$ and $Z_{\rm 0}=0$) from the top-middle panel of Figure \ref{fig:1} are overlaid as black ellipses, scaled to match the scatter of the $\Delta \log$sSFR of MaNGA galaxies (see Figure \ref{fig:12}). 
 In all five panels, the scale of x-axis (or y-axis) are the same in displaying, so that the readers can directly compare the slope of the lines in all five panels. We note that, in all the panels, the gas-phase metallicity is log(O/H)-\citetalias{Dopita-16}.   
  }
  \label{fig:9}
\end{figure*}

\begin{figure*}
  \begin{center}
    \epsfig{figure=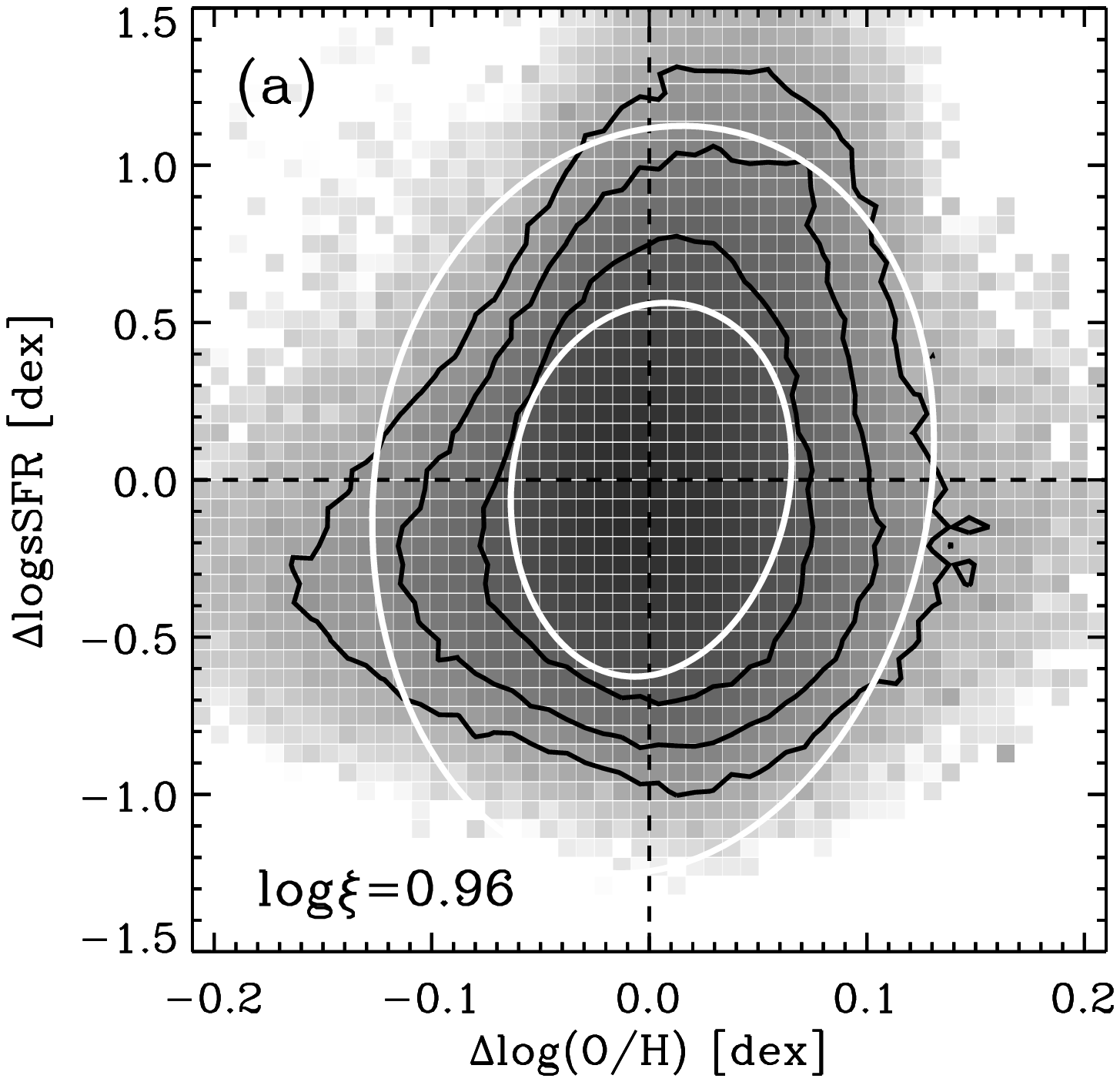,clip=true,width=0.33\textwidth}
    \epsfig{figure=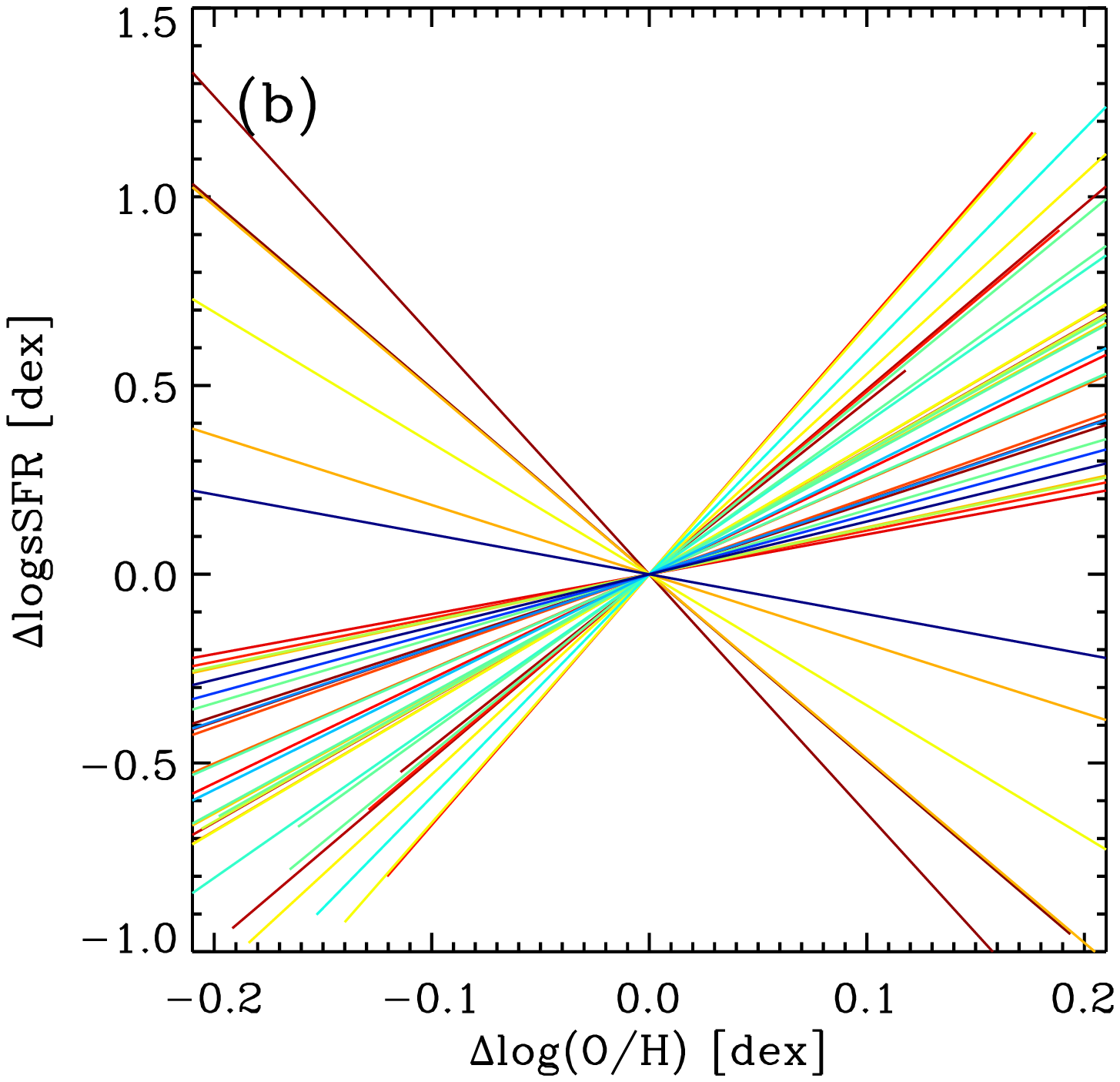,clip=true,width=0.33\textwidth}

    \epsfig{figure=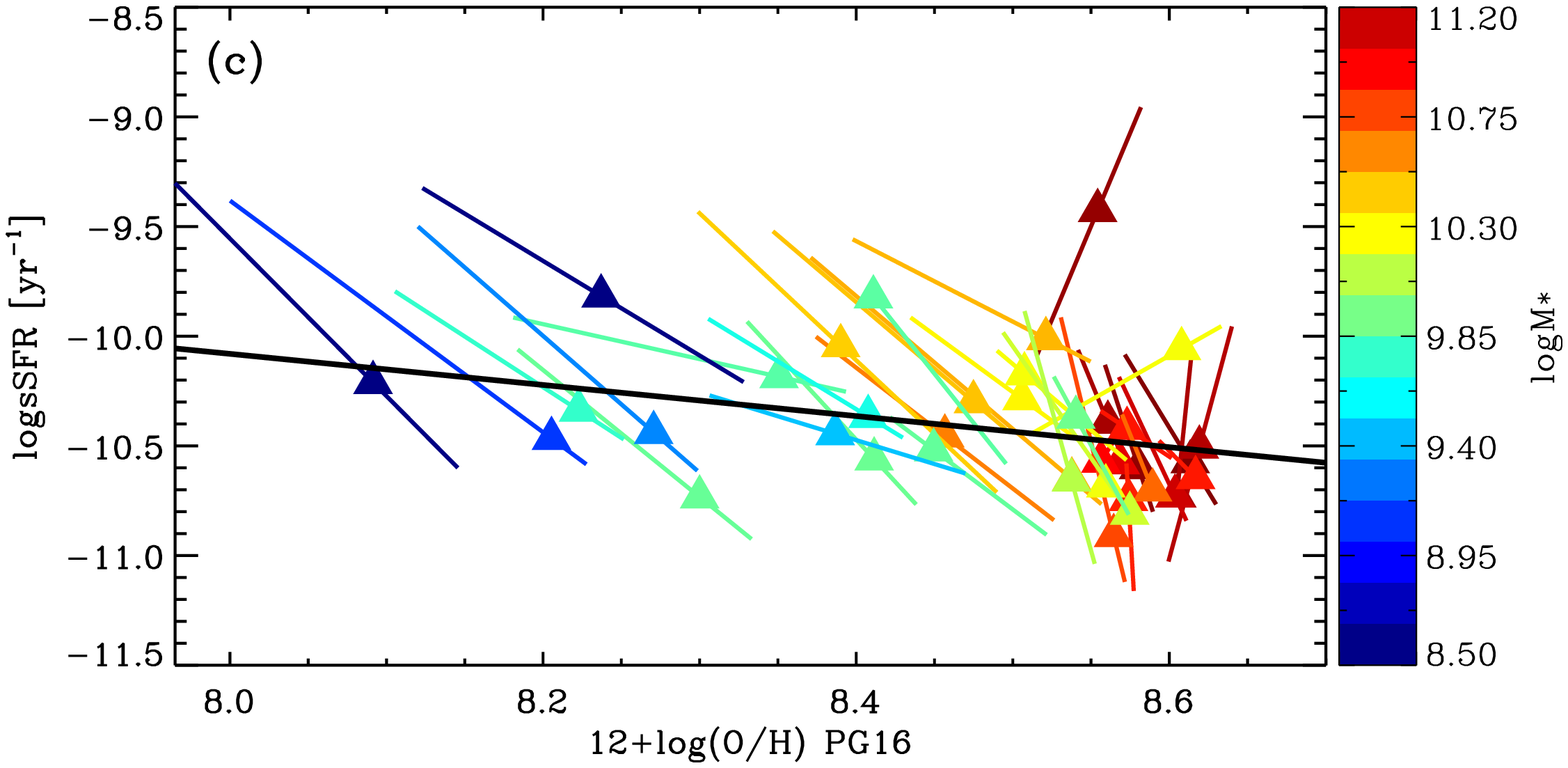,clip=true,width=0.663\textwidth}
    
    \epsfig{figure=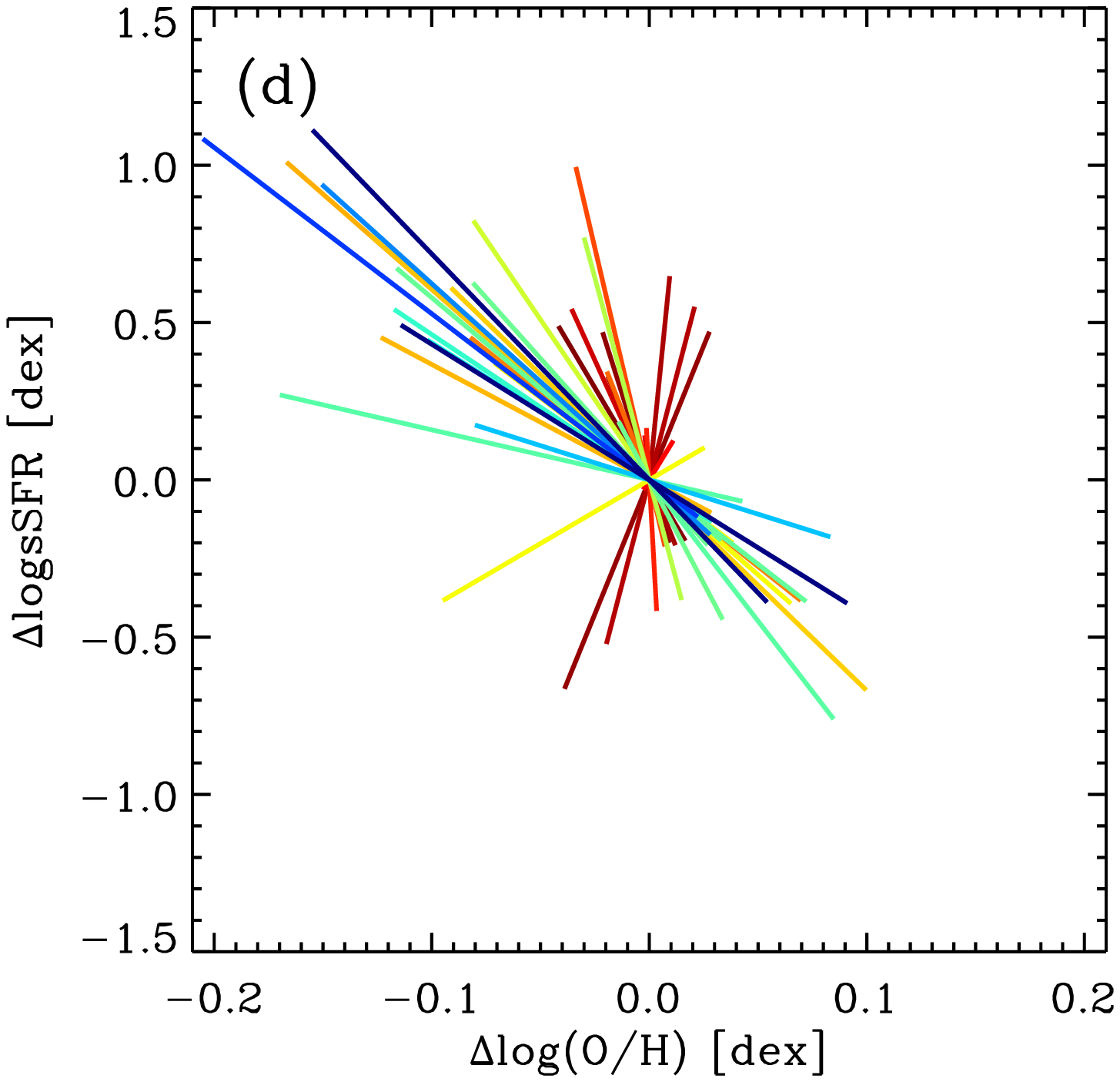,clip=true,width=0.33\textwidth}
    \epsfig{figure=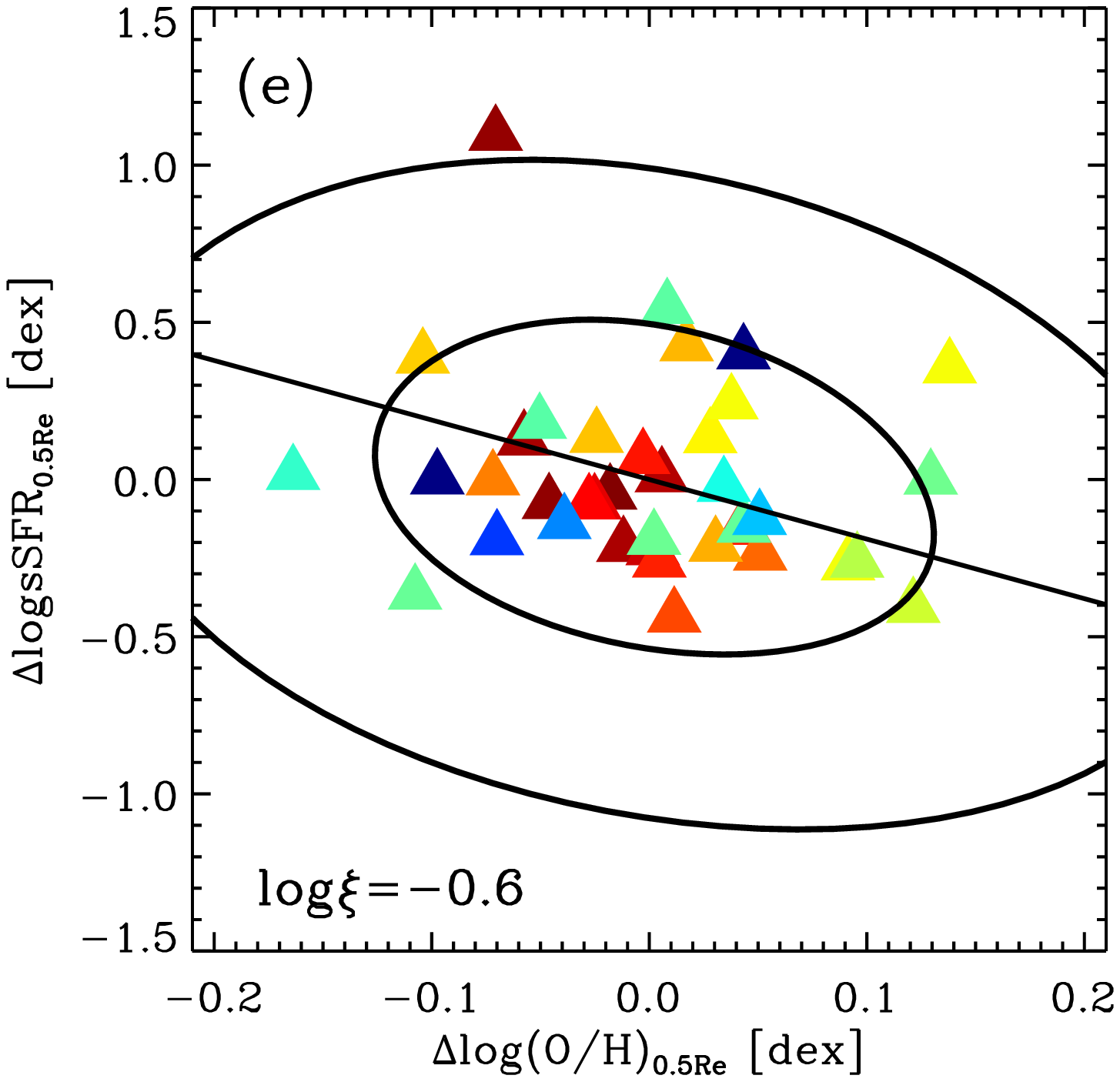,clip=true,width=0.33\textwidth}
    \end{center}
  \caption{The same as Figure \ref{fig:9}, but using the gas-phase metallicity based on the empirical formula from \citetalias{Pilyugin-16}. }
  \label{fig:10}
\end{figure*}


To compare with our theoretical expectations of the gas-regulator model constructed in Section \ref{sec:4.1}, we have defined in the previous section $\Delta\log$sSFR and $\Delta\log$(O/H) both on the 100-pc scale of individual spaxels in MAD galaxies, and for the much larger galactic scale of MAD galaxies as a whole (defined at 0.5\re). In this section, we will explore the correlations between $\Delta\log$sSFR and $\Delta\log$(O/H) on these scales, and interpret the results in terms of the predictions of the gas-regulator model.

  
\subsubsection{100-pc scales} \label{sec:4.2.1}

We first look at the $\Delta\log$sSFR-$\Delta\log$(O/H) relation on scales of $\sim$100 pc. The panel (a) of Figure \ref{fig:9} shows the distribution of all valid MAD spaxels on the $\Delta\log$sSFR vs. $\Delta\log$(O/H)-\citetalias{Dopita-16} diagram. The grayscale shows the number density of spaxels in logarithmic space.  In constructing panel (a), we give each MAD galaxy the same weight, by weighting the contribution of spaxels from each galaxy in panel (a) of Figure \ref{fig:9}. In other words, for a given MAD galaxy, we weight the individual spaxels of that galaxy by $N_{\rm spaxel}^{-1}$, where $N_{\rm spaxel}$ is the total number of valid spaxels for that particular galaxy. This ensures that the sum of the weights of all the spaxels in each galaxy is the same, and therefore that each galaxy is equally represented in the figure.

As a whole, we find a significant positive correlation between $\Delta\log$sSFR and $\Delta\log$(O/H)-\citetalias{Dopita-16} for all the individual spaxels in the 38 MAD galaxies.  Furthermore, we also investigate the correlation of $\Delta\log$sSFR vs. $\Delta\log$(O/H) for each individual MAD galaxy. The panel (b) of Figure \ref{fig:9} shows bisector fits of $\Delta\log$sSFR vs. $\Delta\log$(O/H)-\citetalias{Dopita-16} relation of individual spaxels for each of the 38 MAD galaxies. Here we adopt a bisector fitting \citep{Isobe-90}, because there is no reason for us to prefer regression of $\Delta\log$(O/H) on $\Delta\log$sSFR or $\Delta\log$sSFR on $\Delta\log$(O/H). As can be seen, consistent with the result of panel (a), the $\Delta\log$sSFR and $\Delta\log$(O/H)-\citetalias{Dopita-16} show positive correlation for 37 of the 38 MAD galaxies with only one exception. The exception is NGC 4030, the most massive galaxy in MAD sample. 
Inspection of the color-coding suggests that this result is not dependent on the mass of the galaxy.

The same analysis is repeated for the log(O/H)-\citetalias{Pilyugin-16}, in the panels (a) and (b) of Figure \ref{fig:10}. Overall, we find that $\Delta\log$sSFR and $\Delta$log(O/H)-\citetalias{Pilyugin-16} still show a positive correlation, although not as significant as in panel (a) of Figure \ref{fig:9}. Consistent with this, panel (b) shows that 32 of the MAD galaxies show positive correlations of $\Delta\log$sSFR vs. $\Delta\log$(O/H)-\citetalias{Pilyugin-16}, but now 6 MAD galaxies show negative correlations. We note that when using {\tt O3N2}, the result is qualitatively different from that when using {\tt Scal} (or {\tt N2S2H$\alpha$}) for galaxies with stellar mass below $\sim10^{10.5}$\msolar. This is likely due to the fact that the {\tt O3N2} indicator has a larger uncertainty ($\sim$0.18 dex) than {\tt Scal} ($\sim$0.05 dex) in determining the metallicity (see more discussion in Section \ref{sec:3.3.2} and \ref{sec:6.4}). 

For comparison purposes, we also show the model predictions of $\Delta \log$SFR vs. $\Delta \log$Z on the panels (a) of Figures \ref{fig:9} and \ref{fig:10}.  The two white roughly elliptical shapes are based on the model predictions shown in the top-middle panel of Figure \ref{fig:2}.
As discussed in Section \ref{sec:2}, the correlation of $\Delta \log$Z vs. $\Delta \log$SFR depends on both $\xi$ and $Z_0/y_{\rm eff}$. For simplicity, we only consider here the model predictions for $Z_0=0$.  We then choose the $\log \xi$ (i.e. the ellipse in Figure \ref{fig:2}) that has the same ratio of the dispersion in $\Delta \log$Z (x-axis) and $\Delta \log$SFR (y-axis) as the observed data points in panels (a), and then scale this model ellipse so as to match the 1$\sigma$ and 2$\sigma$ dispersions of the observed quantities (see further discussion in Section \ref{sec:4.3}).
Given the simplicity of the models in Section \ref{sec:2}, and especially given the 
different ranges of metallicity returned by the two metallicity indicators, we do not believe that a quantitative fitting would be meaningful. Nominally, we get $\log \xi=0.6$ for log(O/H)-\citetalias{Dopita-16}, and $\log \xi=0.96$ for log(O/H)-\citetalias{Pilyugin-16}. 
Within the limitations of model and metallicity estimators, this qualitative matching approach suggests that the prediction of time-varying SFE in the gas-regulator model is in good agreement with the data in predicting a positive correlation between $\Delta \log$sSFR vs. $\Delta \log$(O/H) at 100 pc scale. 

We emphasize that given the highly idealised nature of the model, any precise comparison of the model with the data would not be very useful.  Rather, the model was intended to establish two things:  the {\it sign} of the correlation between changes in the SFR and Z (positive or negative), and the approximate {\it relative} variation of these two quantities. Throughout the paper, the comparisons between the model and the data should be treated in this spirit.

\begin{figure*}
  \begin{center}
    \epsfig{figure=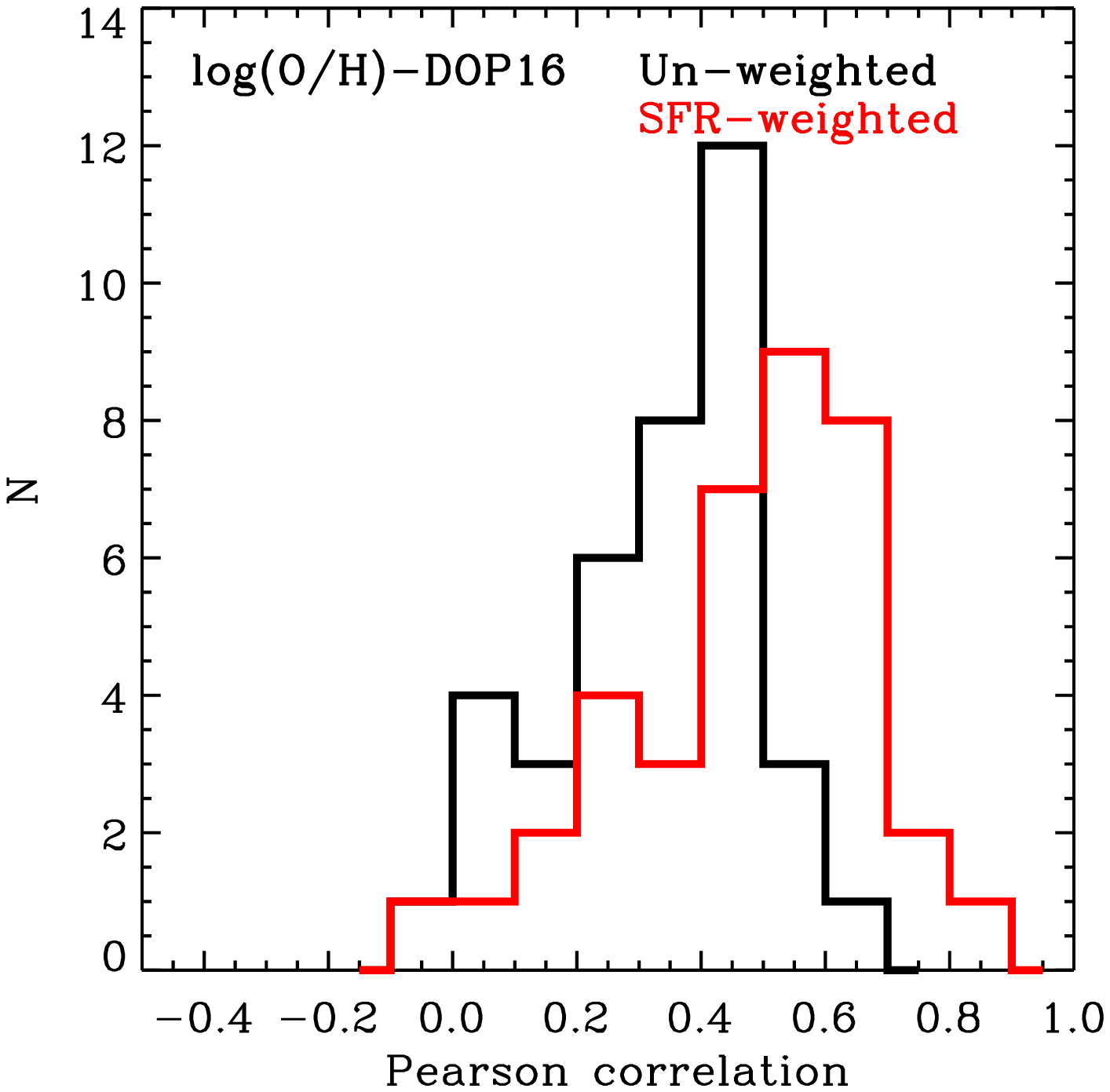,clip=true,width=0.4\textwidth}
    \epsfig{figure=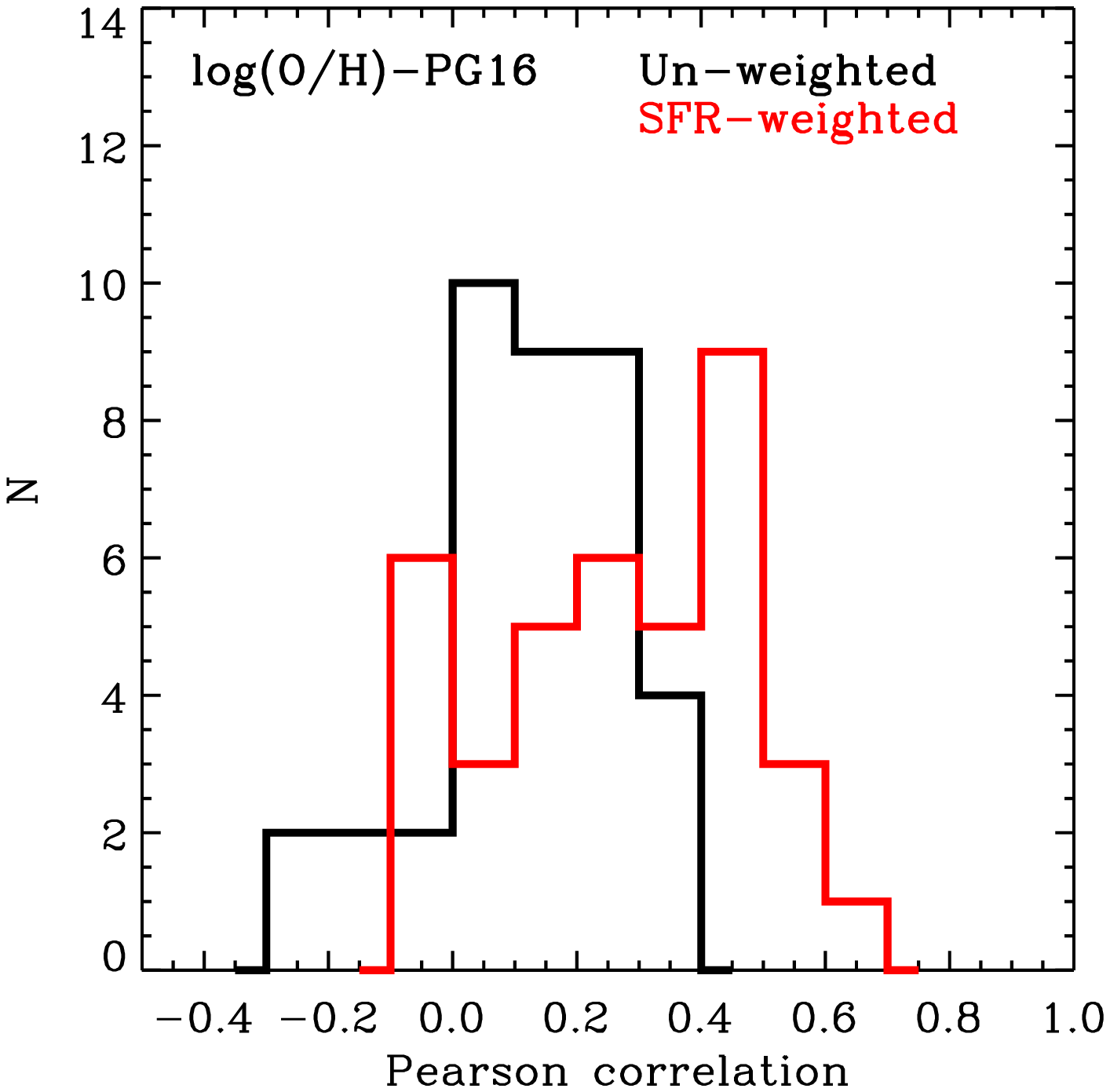,clip=true,width=0.4\textwidth}
    \end{center}
  \caption{ Left panel: The distribution of the Pearson correlation coefficient for the  $\Delta\log$sSFR-$\Delta\log$(O/H)-\citetalias{Dopita-16} relation of the individual spaxels for 38 MAD galaxies.  We measure two kinds of Pearson correlation coefficient: one is without weighting the spaxels (black histogram), and the other is after weighting the spaxels by their SFR (red histogram). 
  Right panel: The same as the left panel, but using \zgas-\citetalias{Pilyugin-16}. }
  \label{fig:11}
\end{figure*}

To try to quantify the significance of this positive correlation, we calculate the distribution of Pearson correlation coefficients for the 38 MAD galaxies in Figure \ref{fig:11}.  For each individual MAD galaxy, the coefficient is calculated in two ways, both by weighting the individual spaxels equally and by weighting them by the SFR of the spaxels. There are two reasons for doing the SFR-weighting.  First it ensures that the inner parts of galaxies, which have fewer but brighter pixels, are not swamped by the outer parts.  Secondly, it is clear in Figure \ref{fig:2} that while the sinusoidal variation in SFE produces symmetric changes in SFR and metallicity, the (possibly more realistic) step-function changes produce asymmetric changes, in which the positive correlation is most clearly established by the regions of highest SFR.  
As shown, for both the approaches of computing correlation coefficient, and for both the metallicity indicators, we find that most of the coefficients are positive with only a few less than zero. This is consistent with the results in panels (a) and (b) of the Figure \ref{fig:9} and Figure \ref{fig:10}. The correlation of $\Delta \log$sSFR vs. $\Delta \log$(O/H) becomes more significant when weighting the spaxels with their SFR. This is due to the fact that regions with strongly enhanced star formation, always show enhanced gas-phase metallicity (also see panel (a) of Figure \ref{fig:9} and \ref{fig:10}).  Both the positive correlation of $\Delta \log$sSFR vs. $\Delta \log$(O/H) and the increased significance of the correlation for regions of enhanced star formation are also visible to the eye in Figure \ref{fig:6}. This may reflect the point made in the context of Figure \ref{fig:2} in Section \ref{sec:2.3}, that the positive correlations caused by step function changes to the SFE are most clearly seen when the SFR is highest. 

We have also examined the significance of the correlation based on the p-value\footnote{https://en.wikipedia.org/wiki/P-value}, which is the probability of obtaining the real observed result under the assumption of no correlation. A smaller p-value means that the correlation is more significant. For both metallicity indicators, we find the p-values are very close to zero ($p<0.001$) for all except one or two (for different metallicity indicators) of the 38 MAD galaxies. This test shows that the positive correlation between $\Delta \log$sSFR and $\Delta \log$(O/H) is present also within individual galaxies as well as in the ensemble analysis. 


In the gas-regulator framework, we showed that the $\Delta\log$SFR (or $\Delta\log$sSFR) and $\Delta\log Z$ will be positively correlated when the gas-regulator system is driven by a time-varying SFE (see Figure \ref{fig:2}).  Comparing the model predictions with the panel (a) of Figure \ref{fig:9} or Figure \ref{fig:10}, we can conclude that at $\sim$100 pc scales within galaxies, the variation of SFR (and $Z$) is due to a time-varying SFE experienced by a particular packet of gas. This conclusion implies an assumption that different {\it spatial} regions of galaxies, around an annulus of given galactic radius, constitute different temporal phases of a gas packet, as modelled in Figure \ref{fig:2}. This assumption is not unreasonable, and is supported by other strong observational evidence in favour of a time-varying SFE at $\sim$100 pc scale.

\cite{Leroy-13} and \cite{Kreckel-18} have found that the dispersion of the SFE based on molecular gas measurements increases significantly towards smaller scales \citep[see also][]{Kruijssen-14}. Specifically, the scatter of the SFE is $\sim$0.4 dex at $\sim100$ pc \citep{Kreckel-18}. Consistent with this, \cite{Kruijssen-19} and \cite{Chevance-20} have shown that molecular gas mass and star formation rates are spatially de-correlated on the scales of individual GMCs in nearby disk galaxies, contrary to the tight correlation seen on kpc- and galactic-scales \citep[e.g.][]{Shi-11, Bigiel-08}. This de-correlation implies rapid temporal cycling between GMCs, star formation, likely involving feedback processes. By using this de-correlation,  \cite{Chevance-20} constrain the properties of GMC and claimed that the GMC lifetimes are typically 10–30 Myr, which consists of a long inert phase without 
massive star formation and a short phase (of less than 5 Myr) with a burst of star formation.  
These observational results indicate a strong spatial variation of SFE at $\sim$100 pc scales, simultaneously suggesting a strong temporal variation of SFE for a given packet of gas.  

In Section \ref{sec:2.3}, we constructed two models of time-varying SFE$(t)$ in the form of both a sinusoidal function and a periodic step-function. According to the above discussion, we find that the SFE$(t)$ at $\sim$100 pc scale may be better characterized by a step-function with a short duration of top-phase and a long duration of bottom-phase, rather than the sinusoidal model. Consistent with this, we find the model prediction with the step-function is in good qualitative agreement with the observational results. Specifically, the distribution of $\Delta\log$sSFR is strongly asymmetric with a very small fraction of spaxels showing strongly enhanced SFR, and with these regions also show strongly enhanced metallicity, as found in the panel (a) of both Figure \ref{fig:9} and \ref{fig:10}. These features can also be found in the model prediction shown in the bottom-middle panel of Figure \ref{fig:2}.  

It goes without saying that the models explored in Section \ref{sec:2} are simple heuristic models, which cannot be expected to explain all the details of the situation. Not least, timescales of star-formation of only 10-30 Myr \citep{Chevance-20} are comparable to the timescales for chemical enrichment, an effect neglected by our use of the instantaneous recycling approximation in Section \ref{sec:2}.  Nevertheless, we can conclude qualitatively that the variation of SFR and gas-phase metallicity on 100 pc scales is primarily due to a time-varying star formation efficiency experienced by the gas. 

\subsubsection{Sub-galactic scales} \label{sec:4.2.2}


The two Panels (c) on Figures \ref{fig:9} and \ref{fig:10} show the correlation between the average sSFR and the average 12+log(O/H), as at a given galactic radius, 
for all 38 MAD galaxies, color-coded by their stellar mass.  These are obtained by combining the linear fits of sSFR$(r)$ and 12+log(O/H)$(r)$ (see Figure \ref{fig:7} in Section \ref{sec:4.1}) to eliminate $r$ and thereby produce a linear sSFR-metallicity relation for each galaxy.  The triangular points on each line show the values of sSFR and 12+log(O/H) at a fiducial 0.5\re, chosen to be representative of the global quantities for the galaxies.  Since the metallicity profiles of MAD galaxies always show negative radial gradients, for a given line segment in these Panels, the central regions of the galaxies correspond to the higher log(O/H) end of each line segment, which generally also have lower sSFR. 

These individual lines in the Panels (c) therefore represent the radial variations of sSFR and 12+log(O/H) {\it within} a given galaxy, i.e. on {\it sub-galactic} scales. All azimuthal variations are eliminated and the radial variations are greatly smoothed out by the linear fits to sSFR and metallicity.   Shifting these lines to align the triangles would therefore produce a residual plot that would in principle be directly analogous to that in the Panels (b).   This is done in the Panels (d).

Comparing the lines in the panels (d) with those in the panels (b) of Figure \ref{fig:9} and \ref{fig:10}, it is clear that the correlation of $\Delta\log$sSFR and $\Delta\log$(O/H) on these larger ``sub-galactic'' scales show the {\it opposite} correlation to that seen on 100-pc scales. Almost all the individual galaxies show positive correlations on 100-pc scales in the Panels (b) but most, especially at low stellar masses, show a negative correlation between in the Panels (d).  It should be noted that the trend with stellar mass of the galaxy is much clearer than in the Panels (b).  

The interpretation of the (anti-)correlation of $\Delta\log$sSFR and $\Delta\log$(O/H) radially across the galaxy is less trivially interpreted than the positive correlation on 100-pc scales in Section \ref{sec:4.2.1} or, we will argue, on larger galactic scales in Section \ref{sec:4.2.3} below. 


An important difference between these radial ``sub-galactic'' variations and those on both smaller (Section \ref{sec:4.2.1}) and larger (Section \ref{sec:4.2.3} below) scales concerns the potential effects of the wind mass-loading term, $\lambda$ and/or possibly the metallicity of inflow gas, $Z_{\rm 0}$.  On 100-pc scales, we normalised each spaxel by the average properties of all the spaxels at the {\it same} galactic radius in the {\it same} galaxy. We might expect the $\lambda$ and $Z_{\rm 0}$ to be the same for all these pixels if they are determined by the location within the galactic potential well.  Likewise, on the larger scales when we consider the integrated properties of galaxies, we will normalise by the average properties of galaxies with the same integrated stellar mass, which again we may argue are likely have similar overall values of $\lambda$ and $Z_{\rm 0}$.  But, in the current subsection, in which we are looking at radial variations of sSFR and log(O/H) {\it within} galaxies, it is very likely that there will be a positive radial gradient in $\lambda(r)$ (and a possibly negative radial gradient in $Z_{\rm 0}$). From Equation \ref{eq:22}, in which the average $Z$ is determined by $Z_{\rm 0}+y_{\rm eff}$, it is easy to produce a negative gradient in $Z$ with a positive gradient in $\lambda$ and/or a negative gradient in $Z_{\rm 0}$.  
This explanation of the negative gradient of metallicity with radius does not take into account possible radial flows within the galaxy.
 Hydrodynamical simulations, including EAGLE \citep{Schaye-15} and IllustrisTNG \citep{Nelson-18}, show that the inflow is more substantial along the galaxy major axis, while outflow is strongest along the minor axis \citep{Peroux-20}. In parallel work (E. Wang et al, in preparation), we find that the negative metallicity gradient can be naturally produced assuming that a radial gas inflow is dominant within the disk. 

We will return to the radial profiles of (s)SFR and log(O/H) below in Section \ref{sec:5.2}, where we analyze the MaNGA sample, which is not only much larger but extends over a much larger range of radii range, albeit with much poorer spatial resolution.



\subsubsection{Galactic scales} \label{sec:4.2.3}


Finally, the two Panels (e) of Figures \ref{fig:9} and \ref{fig:10} show the correlations between the residuals of the overall sSFR and metallicity, i.e. $\Delta \log$sSFR$_{\rm 0.5Re}$ and $\Delta \log ({\rm O/H})_{\rm 0.5Re}$, for each MAD galaxy, once the overall trends of sSFR$_{\rm 0.5Re}$ and $\log({\rm O/H})_{\rm 0.5Re}$ with galactic stellar mass (shown in Figure \ref{fig:8}) are taken out.  Each triangle represents an individual MAD galaxy with the color-coding of its stellar mass.   

The Panels (e) on the two figures therefore show whether a given MAD galaxy is {\it overall} elevated or depressed in sSFR and log(O/H) relative to other galaxies of the same mass.  They are therefore completely analogous (albeit with a vastly different number of points) to the Panels (a) which showed whether individual spaxels within a MAD galaxy were elevated or depressed in these two quantities relative to the other spaxels at the same radial location within the same galaxy. As argued in the previous subsection, the effects of any systematic variations of wind mass-loading and metallicity of inflow gas with stellar mass should not be present in this diagram.

The black solid lines in the Panels (e) show a bisector fit for the data points.  For both metallicity indicators, a negative correlation between $\Delta\log$sSFR$_{\rm 0.5Re}$ and $\Delta \log({\rm O/H})_{\rm 0.5Re}$ on galactic scales can clearly be seen.  The Pearson correlation coefficients are $-$0.23 and $-$0.17 for the {\tt N2S2H$\alpha$} and {\tt Scal} indicators, respectively.  This negative correlation and fit is in clear contrast to the positive correlation and fits on 100-pc scales that are shown in the (directly analogous) Panels (a) and (b). 

Given the limit number of MAD galaxies that are available, one might be concerned about the statistical validity of the reversal of sign between the negative correlation in Panels (e). 
Therefore, we examined the p-values of the correlation in both Panels (e), which are 0.171 and 0.317 for $\log({\rm O/H})$-\citetalias{Dopita-16} and $\log({\rm O/H})$-\citetalias{Pilyugin-16}, respectively.  Also, in the next section of the paper, we perform a similar analysis on the much larger sample of 976 MaNGA galaxies, and find completely consistent results in that much larger sample. 

For comparison, we also present the model predictions from Section \ref{sec:2.2}, in which the gas-regulator is driven with a sinusoidal inflow rate. Similar to panels (a), we scaled the model prediction with an overall factor so as to match the scatter of $\Delta \log$sSFR for MaNGA galaxies, since the large number of MaNGA galaxies enables a reliable estimate of the scatter (see Figure \ref{fig:12}). We have $\log \xi=-0.2$ for log(O/H)-\citetalias{Dopita-16}, and $\log \xi=-0.6$ for log(O/H)-\citetalias{Pilyugin-16} at $Z_0=0$. 
As can be seen in panels (e) of Figure \ref{fig:9} and \ref{fig:10}, the model  of time-varying inflow rate is in good agreement with the observed diagram of $\Delta \log$sSFR vs.  $\Delta \log$(O/H) on galactic scales (again in the sense of reproducing the sign of the correlation and the relative amplitudes of the variations of these two quantities).  
We conclude that the inverse correlation between variations in the {\it overall} sSFR and $\log({\rm O/H})$ across the galaxy population at a given mass, are due to temporal variations in the inflow rate onto the galaxies.  This is in marked contrast to the situation on 100-pc scales, where we argued that the positive correlation between these quantities was the clear signature of temporal variations in the SFE as a given packet of gas enters and leaves regions of high SFE.

Finally, it should be noted in passing that the observed negative correlation between the overall $\Delta\log$sSFR and $\Delta\log$(O/H) in the Panels (e) is a straightforward manifestation of the existence of SFR as a second parameter in the mass-metallicity relation \citep[e.g.][]{Mannucci-10, Salim-14, Cresci-19, Curti-20}.  It has further been claimed that the $Z(M_*,{\rm SFR})$ relation is epoch-independent, i.e. that there is a so-called FMR \citep[e.g.][]{Richard-11, Nakajima-12, Huang-19}.  One of the successes of the gas-regulator model presented by \cite{Lilly-13} was to provide a natural analytic explanation for the presence of SFR as a second parameter and even to predict that the $Z(M_*,{\rm SFR})$ relation could well be more or less epoch independent.   The \cite{Lilly-13} analysis considered in the first instance a constant specific inflow rate. But a steady specific inflow implicitly produces an {\it increase} in the inflow rate.  If the specific inflow changes in such a way that the inflow rate is constant, then the sensitivity to the SFR vanishes (see \cite{Lilly-13} for discussion, also the Appendix to \citet{Onodera-16}).  

This emphasizes both that the anti-correlation in Panel (e) is not a new or controversial observational result, and also, in a very general sense, that this negative correlation between overall metallicity and star-formation rate on galactic scales is fundamentally driven by {\it changes} to the inflow rate, as discussed in this paper. 

\subsection{Quantitative interpretation of the dispersion of gas-phase metallicity and sSFR on 100-pc scales in MAD galaxies} \label{sec:4.3}

\begin{figure*}
  \begin{center}
    \epsfig{figure=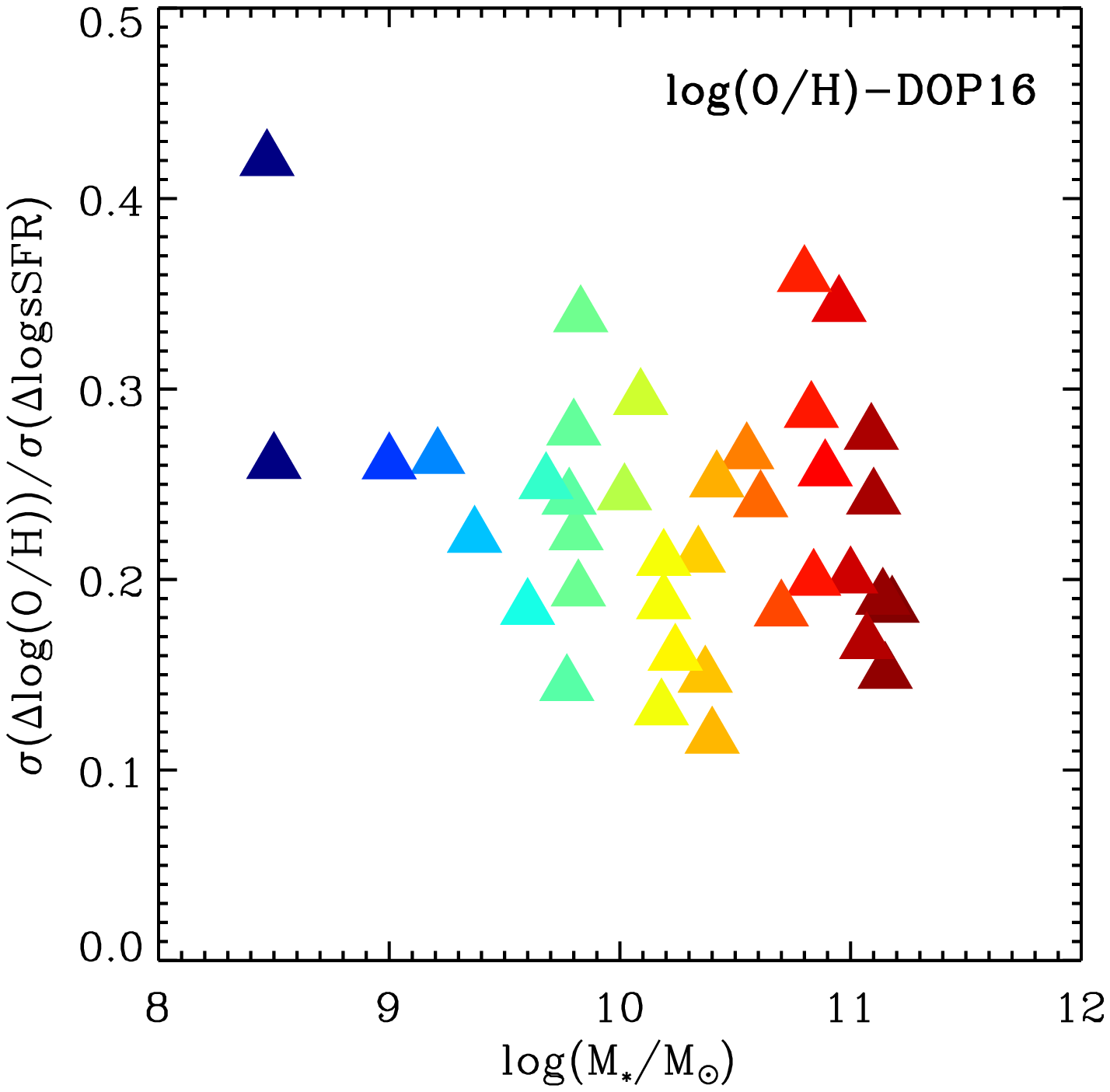,clip=true,width=0.45\textwidth}
    \epsfig{figure=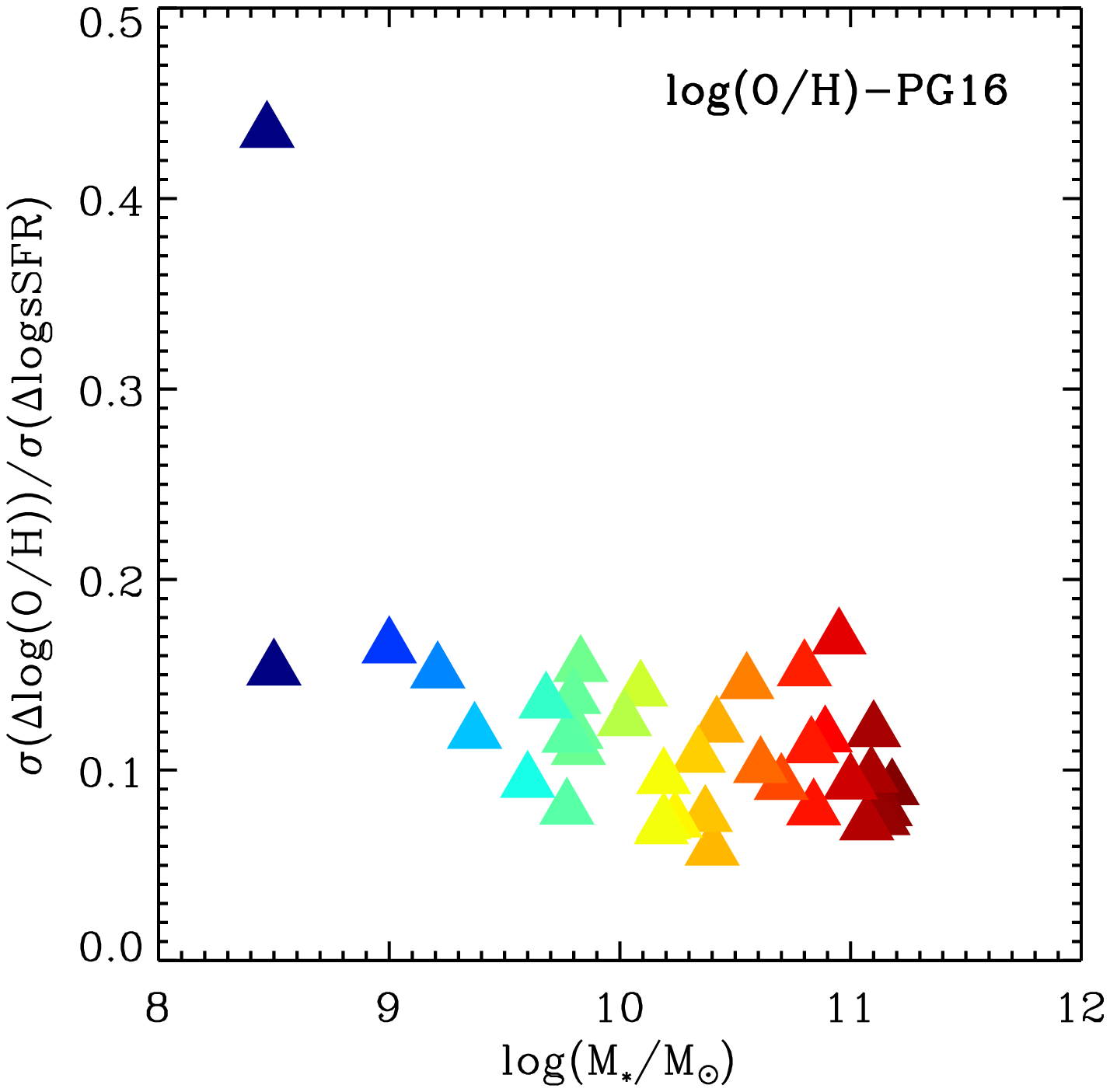,clip=true,width=0.45\textwidth}
    \end{center}
  \caption{ Left panel: The ratio of $\sigma$($\Delta\log$(O/H)) to  $\sigma$($\Delta\log$sSFR) as a function of stellar mass for the 38 MAD galaxies. For each MAD galaxies, the $\sigma$($\Delta\log$(O/H)) (or $\sigma$($\Delta\log$sSFR)) is the standard deviation of $\Delta\log$(O/H) (or $\Delta\log$sSFR) for the all the individual spaxels in this galaxy.  The data points are color-coded with the stellar masses. 
  Right panel: The same as the left panel, but using log(O/H)-\citetalias{Pilyugin-16}.
  }
  \label{fig:16}
\end{figure*}

We showed in Section \ref{sec:2} that the relative strength of variations in the SFR and metallicity of the gas-regulator should depend on the response timescale of the system, set by the gas depletion timescale, relative to the timescales of any driving variations in the inflow or in the SFE.
The Equation \ref{eq:20} describes the relative amplitude of these variations, characterized by $\sigma(\Delta\log({\rm O/H}))$/$\sigma(\Delta\log {\rm SFR})$ across the population, as a function of $\xi$ when driving the gas-regulator system with sinusoidal SFE$(t)$. According to Equation \ref{eq:20}, the  $\sigma(\Delta\log({\rm O/H}))$/$\sigma(\Delta\log {\rm SFR})$ should decreases with increasing $\xi$, i.e. decrease with increasing effective gas depletion time if we ignore the possible variation of the driving period of SFE$(t)$ (see Figure \ref{fig:5}).

We have therefore calculated the dispersions across the spaxels of the residual quantities $\Delta \log({\rm O/H})$ and $\Delta \log {\rm sSFR}$ that were plotted in the two Panels (a) in Figures \ref{fig:9} and \ref{fig:10} for the two metallicity indicators respectively. We calculate this dispersion for each of the MAD galaxies respectively.

Figure \ref{fig:16} shows the resulting ratios of $\sigma(\Delta\log({\rm O/H}))$ to $\sigma(\Delta\log {\rm sSFR})$ for each of the 38 MAD galaxies for the two metallicity indicators.   It is obvious that the $\sigma(\Delta\log({\rm O/H}))$ based on {\tt N2S2H$\alpha$} is overall greater than  that based on {\tt Scal}. This is not a noise issue, but is due to the fact that the range of log(O/H)-\citetalias{Dopita-16} is nearly twice the range of log(O/H)-\citetalias{Pilyugin-16} for a given dataset, as mentioned earlier in this paper. This systematic uncertainty hinders the quantitative interpretation of these dispersions, although trends established within a single estimator (i.e. within a single panel of Figure \ref{fig:16}) should have some validity. 

As can be seen, the $\sigma(\Delta\log({\rm O/H}))$/$\sigma(\Delta\log {\rm sSFR})$ based on {\tt N2S2H$\alpha$} is in the range of 0.12 to 0.42 with the median value of 0.24. The $\sigma(\Delta\log({\rm O/H}))$/$\sigma(\Delta\log {\rm sSFR})$ based on {\tt Scal} is about a half this value, mostly in the range of 0.06-0.17 with a median value of 0.11. We do not find a significant dependence of $\sigma(\Delta\log({\rm O/H}))$/$\sigma(\Delta\log {\rm sSFR})$ on stellar mass, except possibly for an increase at the lowest stellar masses below $10^9$\msolar.  The MAD galaxy with the largest $\sigma(\Delta\log({\rm O/H}))$/$\sigma(\Delta\log {\rm sSFR})$ is ESO499-G37.  A small fraction of spaxels in ESO499-G37 have $\log${\tt N2}$<-0.6$, resulting in very low metallicity in these regions (see Equation \ref{eq:27}), and producing a large dispersion in $\Delta\log({\rm O/H})$-\citetalias{Pilyugin-16}.


Although we suspect that a sinusoidally time-varying SFE is not likely to be realistic in the universe at $\sim$100 pc, Equation \ref{eq:20} (also see Figure \ref{fig:3}) does permit a rough order of magnitude estimate of whether these relative dispersions are reasonable and of the approximate timescales involved.

If we examine what values of $\xi$ in Equation \ref{eq:20} produce the observed $\sigma(\Delta\log({\rm O/H}))$/$\sigma(\Delta\log {\rm sSFR})$  for typical MAD galaxies, i.e. 0.24 and 0.12 for the two metallicity estimators,
then these are $\log \xi=$ 0.607 for {\tt N2S2H$\alpha$}, and $\log \xi=$ 0.956 for {\tt Scal} at $Z_{\rm 0}=0$.  If we take 1 Gyr as a reasonable estimate for the effective gas depletion timescale for the overall galaxy population (see \citetalias{Wang-19}), we get rough estimates of $T_{\rm p}=$ 1.5 Gyr for {\tt N2S2H$\alpha$}, and $T_{\rm p}=$ 0.7 Gyr for {\tt Scal} as the nominal period of a time-varying SFE.

Intriguingly, in the Milky Way, a periodic star formation history with a period of $\sim$0.5 Gyr has been suggested in the solar neighborhood, from analysis of the resolved stellar population \citep{Hernandez-00, de-la-Fuente-Marcos-04}.  As further pointed out by \cite{Egusa-09}, this periodic star formation history may be associated with the passage of the spiral potential in the density wave theory \citep{Lin-69}. Assuming that the potential has a two-armed pattern as suggested by \cite{Drimmel-00}, the pattern speed can be calculated as $\Omega_{\rm P}=\Omega(r=R_{\odot})-\pi/(0.5\ {\rm Gyr}) = 21$ km s$^{-1}$ kpc$^{-1}$. This is also consistent with the result from numerical simulations of the stellar and gaseous response to the spiral potential, presented by \cite{Martos-04}.  

It is quite suggestive that this same sort of timescale emerges in our own analysis of the metallicities in terms of a periodically varying SFE, and suggests that the periodic (or time-varying) SFE$(t)$ that is relevant on 100-pc scales may be explained by the passage of orbiting packets of gas through the spiral density wave. It should be noted in the context of the gas regulator that changes in the density of the gas due to {\it local} flows associated with the passage of a spiral density wave have no effect, except (quite possibly) to change the SFE because of the changed density of gas, and should not be considered as ``inflows'' or ``outflows", which are explicitly concerned with changes in the total (baryonic) mass of the gas regulator ``system" being considered. Both the metallicity and the specific star-formation rate, are formally {\it intrinsic} quantities that make no statement about the size of the system, i.e. whether the boundary of the reservoir is of fixed physical size or not. Put another way, neither the measurements of metallicity nor the sSFR will change (per se) due to a compression of the whole system of gas and stars (unless there are associated physical changes, e,g, in the SFE).

However, given the many steps and uncertainties in our analysis, we certainly caution against over-interpretation of our analysis.


\section{Analysis of MaNGA galaxies} \label{sec:5}


In the previous section of the paper, we have presented  results based on MAD galaxies. The high spatial resolution of MAD galaxies enables us to obtain a robust statistical analysis on the correlation of the $\Delta\log$sSFR-$\Delta$log(O/H) on 100 pc scales. However, the analysis at galactic scales needs to be verified because of the limited sample size of MAD galaxies. In this section, we therefore present a similar analysis as in Section \ref{sec:4} by using a well-defined set of SF MaNGA galaxies. The MaNGA sample used in this work includes 976 SF galaxies, which is $\sim$25 times larger than the MAD sample.  The spatial coverage of MaNGA sample is also greater than 1.5\re, which is larger than the MAD coverage as a whole. However, the spatial resolution of MaNGA galaxies (1-2 kpc) is much worse than that of MAD galaxies. Therefore, we only focus on the analysis on galactic and ``sub-galactic" scales for MaNGA galaxies in this section, rather than on individual spaxels. 





\subsection{Correlations of the integrated $\Delta\log$sSFR and $\Delta\log$(O/H) for MaNGA galaxies} \label{sec:5.1}

\begin{figure*}
  \begin{center}
    \epsfig{figure=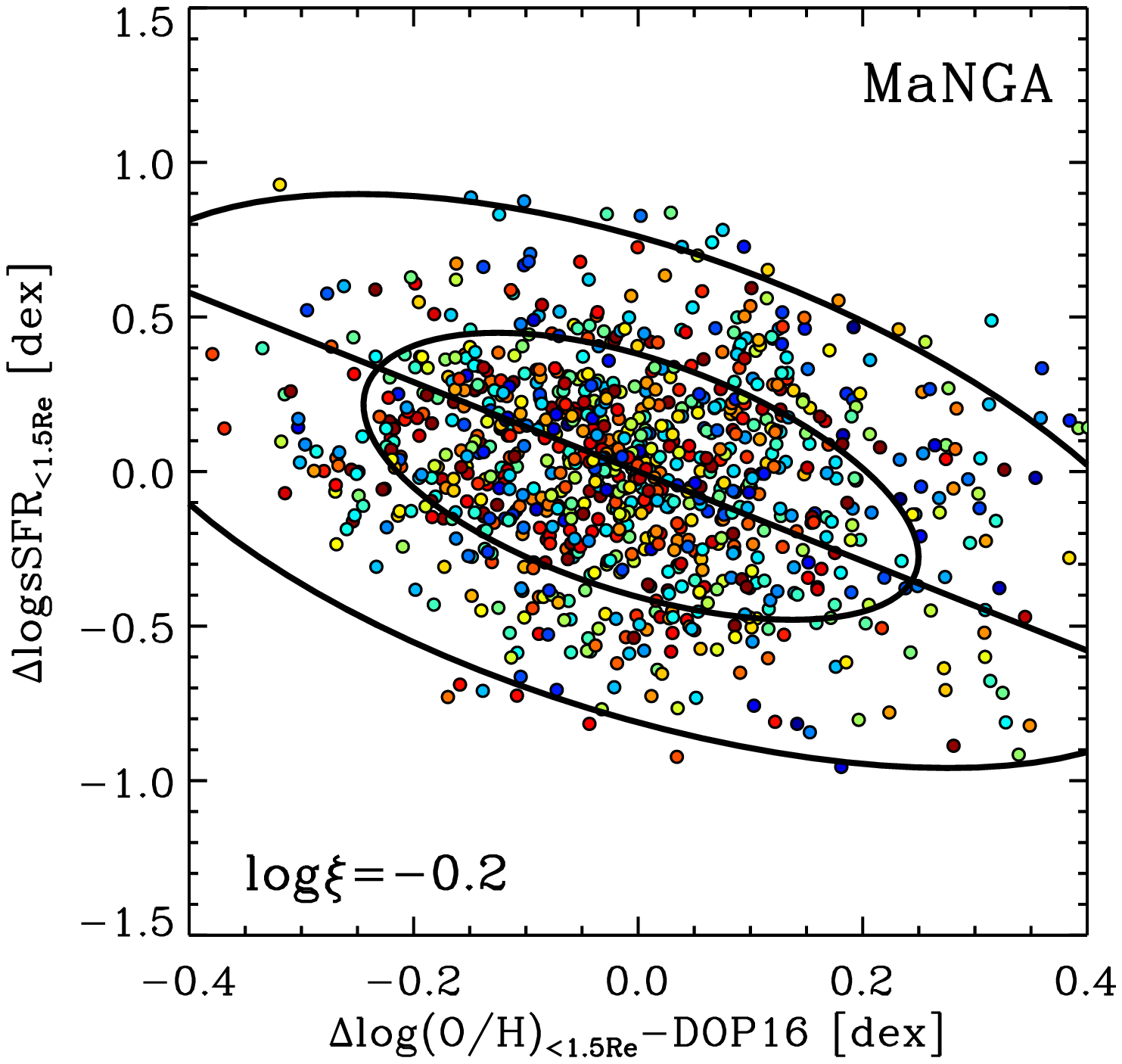,clip=true,width=0.43\textwidth}
    \epsfig{figure=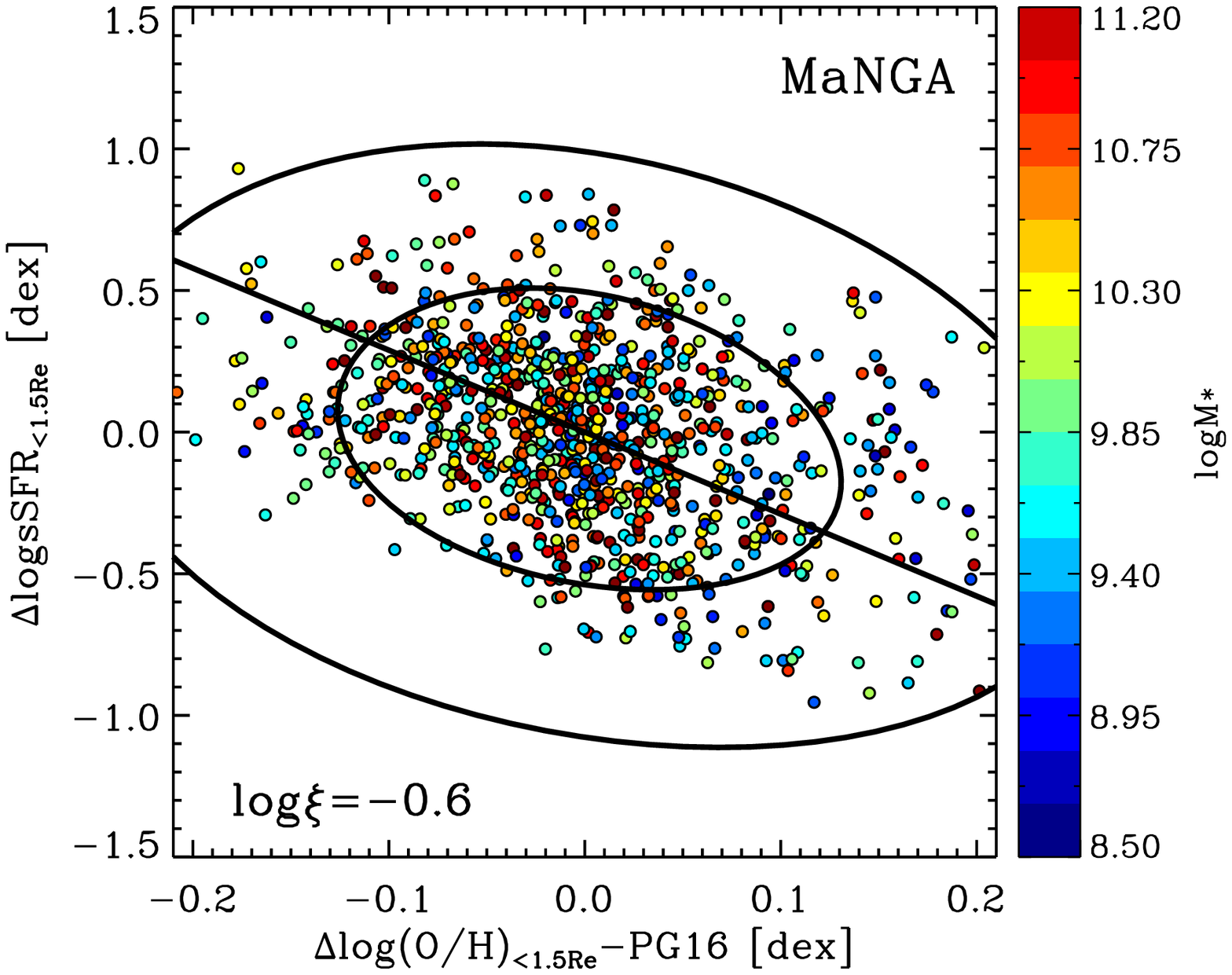,clip=true,width=0.516\textwidth}
    \end{center}
  \caption{ The $\Delta\log$sSFR$_{\rm <1.5Re}$-$\Delta\log$Z$_{\rm <1.5Re}$ diagram for MaNGA SF galaxies, color-coded by integrated stellar mass. Unlike in the panel (e) of Figure \ref{fig:9} and Figure \ref{fig:10}, we use the integrated sSFR (and metallicity) measured within 1.5\re\ to be representative of the global quantity, rather than using the values at one specific galactic radius. In both panels, the black lines are the bi-sector fittings of the data points. For comparison, the model predictions ($\log \xi=-0.2$ for log(O/H)-\citetalias{Dopita-16} and $\log \xi=-0.6$ for log(O/H)-\citetalias{Pilyugin-16}) are overlaid in black curves, taken from the top-middle panel of Figure \ref{fig:1}. The model predictions are scaled to match to the scatter of the $\Delta \log$sSFR$_{\rm <1.5Re}$ of MaNGA galaxies.  }
  \label{fig:12}
\end{figure*}
  
In Section \ref{sec:4.2.3} we examined the relation between the measure of the overall, i.e. ``galactic-scale",
$\Delta\log$sSFR-$\Delta\log$(O/H) for MAD galaxies, taking as measures of these quantities the values from the radial profiles at a fiducial radius of 0.5 \re, as shown for the 38 MAD galaxies in the two Panels (e) in Figures \ref{fig:9} and \ref{fig:10}. 
Since the MaNGA coverage usually extends further than 1.5\re\ for individual galaxies, we can now measure the integrated sSFR and metallicity within 1.5\re.  The metallicity within 1.5\re\ is computed as the H$\alpha$ luminosity weighted 12+log(O/H) for all the spaxels within 1.5\re.  This is probably more representative of the global quantities than the sSFR (or metallicity) that was measured in the MAD sample at one particular radius.   

As before, we first construct the sSFR$_{\rm <1.5Re}$ vs. mass 
and log(O/H)$_{\rm <1.5Re}$ vs. mass relations, and use these to normalize the measurements of individual galaxies.
The first is obtained with a linear fit of the sSFR$_{\rm <1.5Re}$ vs. stellar mass relation.  For the metallicity, we do a polynomial fit to the log(O/H)$_{\rm <1.5Re}$ vs. stellar mass relation, since it clearly flattens at the high-mass end.
For each individual MaNGA galaxy, we thereby define the $\Delta\log$sSFR$_{\rm <1.5Re}$ and $\Delta\log$(O/H)$_{\rm <1.5Re}$ to be the (logarithmic) deviation from these relations. 

The correlation between $\Delta\log$sSFR$_{\rm <1.5Re}$ and $\Delta\log$(O/H)$_{\rm <1.5Re}$ for the MaNGA sample is shown for the two metallicity indicators in the two panels of Figure \ref{fig:12}. As shown, enlarging the sample size by a factor of about 25, the negative correlation between $\Delta\log$sSFR and $\Delta\log$(O/H) is very clearly seen for both metallicity indicators.  The Pearson correlation coefficients for the MaNGA sample are $-$0.23 and $-$0.36 for {\tt N2S2H$\alpha$} and {\tt Scal} indicators, respectively.  The p-values are zero ($<10^{-10}$) for both metallicity indicators, substantially strengthening the results found in Panels (e) of Figure \ref{fig:9} and \ref{fig:10} for the much smaller MAD sample. 

For each metallicity indicator, the linear slopes obtained in MaNGA by bi-sector fitting are very similar to those seen in the MAD sample (Panels (e) of Figures \ref{fig:9} and \ref{fig:10}).  However, we note again that the slopes for the two metallicity indicators are significantly different,
due to the range of log(O/H)-\citetalias{Dopita-16} being nearly twice of that of log(O/H)-\citetalias{Pilyugin-16}.  As also noted above, this clear inverse correlation is a re-statement of the existence of SFR as a (negative) second parameter in the mass-metallicity relation. For comparison, we show the model predictions for a sinusoidally time-varying inflow rate as black curves scaled to the MaNGA data using the same procedure as described in Section \ref{sec:4.2.1}. These same curves were plotted in panels (e) of Figure \ref{fig:9} and \ref{fig:10} for comparison with the MAD data. As can be seen, a single model prediction is broadly consistent with both the observed MaNGA and MAD data points. 

\subsection{Radial profiles of sSFR and Oxygen abundance for MaNGA galaxies} \label{sec:5.2}



We now turn to look more closely at the radial variations of SFR and log(O/H) in galaxies and how these vary across the galaxy population, using again the MaNGA sample. This analysis may be regarded as an extension of the analysis of SFR profiles that was presented in \citetalias{Wang-19}.

In \citetalias{Wang-19}, we investigated the dependence of the normalized star-formation surface density profiles, $\Sigma_{\rm SFR}$(R/\re)  of star-forming MaNGA galaxies on the overall stellar mass and on $\Delta\log$sSFR$_{\rm <Re}$, the vertical deviation of the galaxy from the ``nominal'' SFMS\footnote{The ``nominal'' SFMS is the linear fitted relation of SFR($<$\re) versus $M_*$($<$\re) for the SF galaxies of MaNGA (see more details in section 2 of \citetalias{Wang-19}). }. We showed (figure 5 of that paper) that star-forming MaNGA galaxies that lie significantly above (or below) the overall SFMS show an elevation (or suppression) of SFR at all radii. In addition, we showed that whereas at low stellar masses this elevation (suppression) of star-formation is more or less uniform with radius, for the more massive galaxies the elevation (or suppression) of star-formation becomes more pronounced in the central regions of the galaxies.  As a direct consequence of this, the dispersion in the (normalized) $\Sigma_{\rm SFR}$ across the galaxy population, which we designated  $\sigma(\Delta\log\Sigma_{\rm SFR})$ was found to be correlated with the local surface mass density, which is equivalent to an inverse correlation with the gas depletion timescale in the extended Schmidt Law \citep{Shi-11}, since in that local surface density and the gas depletion timescale are directly related.  This result was shown in figure 9 of \citetalias{Wang-19}. 

\begin{figure*}
  \begin{center}
    \epsfig{figure=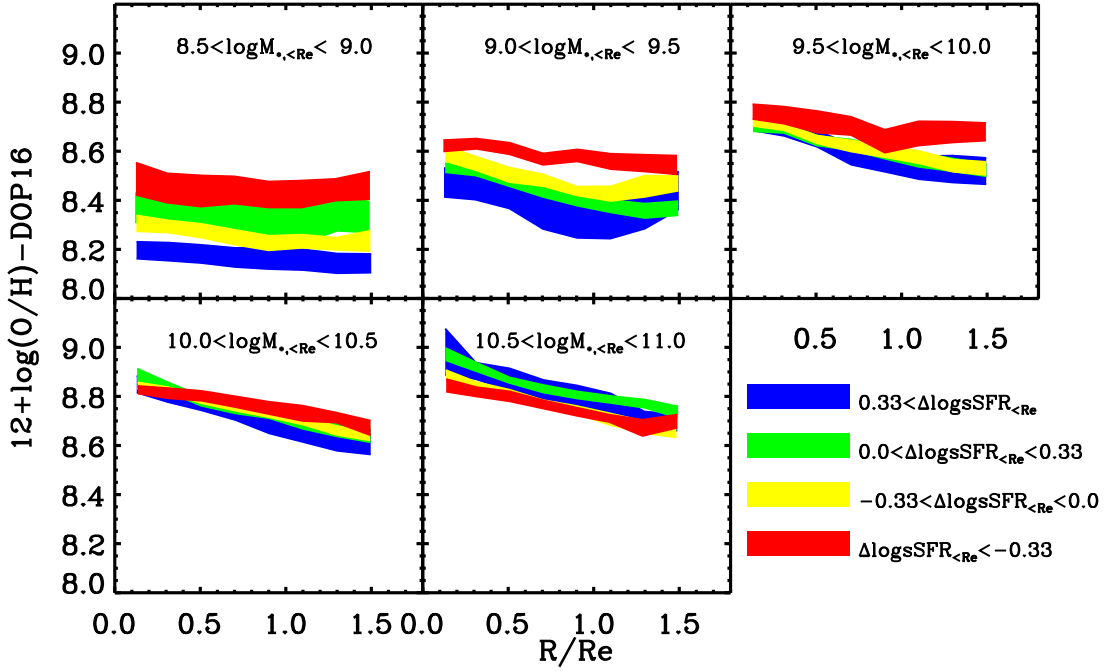,clip=true,width=0.84\textwidth}
    \end{center}
  \caption{The median gas-phase metallicity profile for MaNGA sample galaxies at given stellar mass and a given $\Delta\log$sSFR$_{\rm <Re}$.  
  The $\Delta\log$sSFR$_{\rm <Re}$ is defined to be the vertical deviation (i.e. in sSFR) from the ``nominal'' SFMS, i.e. sSFR$_{\rm <Re}$-$M_{\rm *,<Re}$ relation \citep{Wang-19}. The blue, green, yellow and red profiles are the median metallicity profiles of galaxies with different $\Delta\log$sSFR$_{\rm <Re}$.  The width of each profile is calculated using the boot-strap method from the sample in question. We note that the gas-phase metallicity profile here is generated based on log(O/H)-\citetalias{Dopita-16}.}
  \label{fig:14}
\end{figure*}

\begin{figure*}
  \begin{center}
    \epsfig{figure=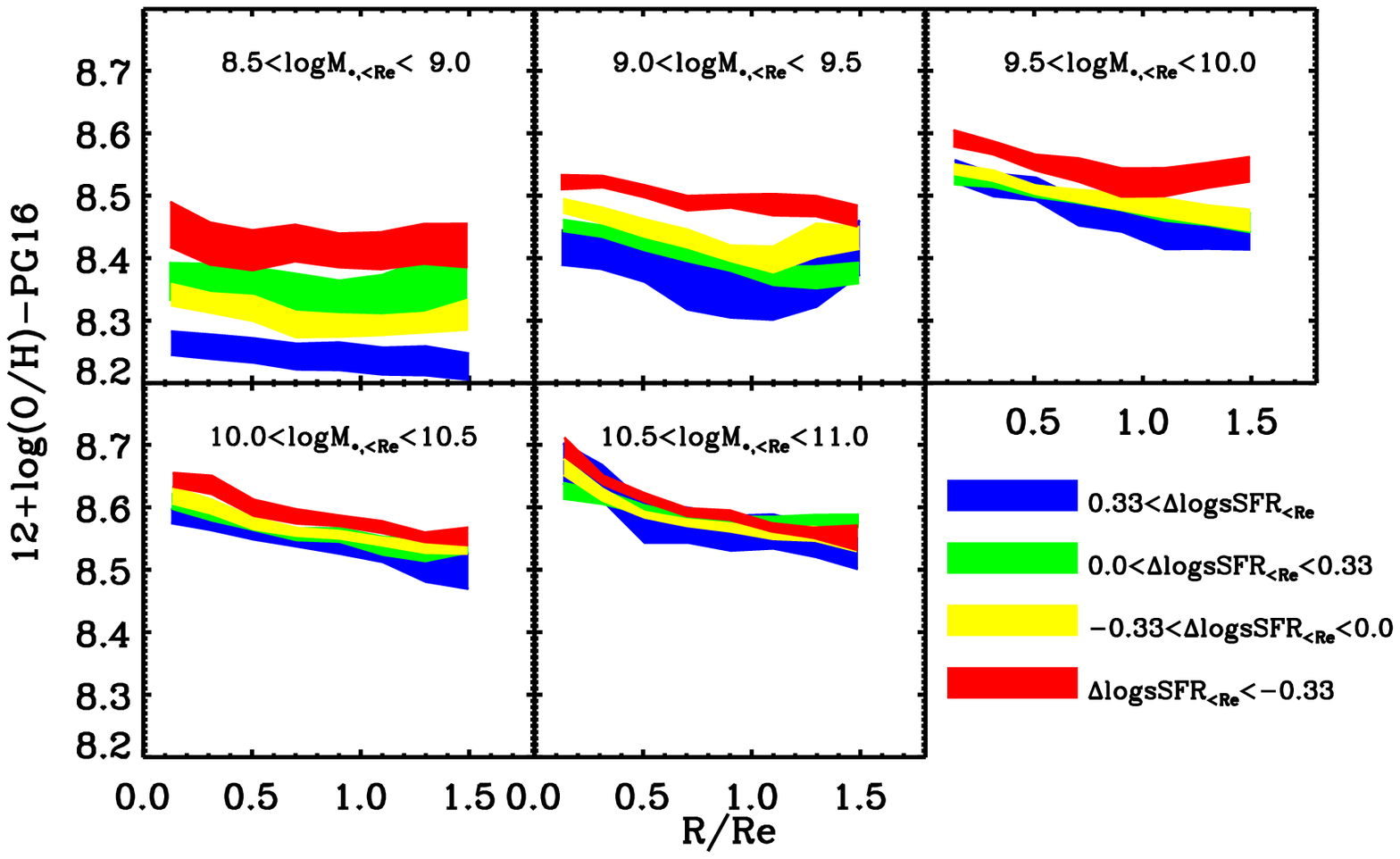,clip=true,width=0.84\textwidth}
    \end{center}
  \caption{ The same as Figure \ref{fig:14}, but using log(O/H)-\citetalias{Pilyugin-16}. }
  \label{fig:15}
\end{figure*}

In this subsection, we will carry out an analogous study of  
the metallicity profiles for the same MaNGA galaxies used in \citetalias{Wang-19}.  We will investigate the dependence of these metallicity profiles on stellar mass and the deviation from the SFMS, analogously to the analysis of $\Sigma_{\rm SFR}$ in that paper.  

For consistency with that earlier work, we use the $\Delta\log$sSFR$_{\rm <Re}$ to separate galaxies. 
Galaxies are classified into four sub-samples: 0.33 $<\Delta\log$sSFR$_{\rm <Re}$,  0.0 $<\Delta\log$sSFR$_{\rm <Re}<$ 0.33, $-$0.33 $<\Delta\log$sSFR$_{\rm <Re}<$ 0.0 and $\Delta\log$sSFR$_{\rm <Re}<-0.33$ and further split the sample into five bins of overall stellar mass, as in \citetalias{Wang-19}. For each individual galaxy, we first compute a radial profile of 12+log(O/H), determined as the median 12+log(O/H) of spaxels located within each radial bin.  
In each of these subsamples we then construct the median 12+log(O/H)($r$/\re) radial profiles, using the two metallicity estimators in turn, and estimating uncertainties by boot-strapping the sample galaxies.

Figure \ref{fig:14} shows the median 12+log(O/H)-\citetalias{Dopita-16} profiles for these sub-samples of galaxies. In each stellar mass bin, the metallicity profiles are displayed in the blue, green, yellow and red in descending order of their overall $\Delta\log$sSFR$_{\rm <Re}$. Figure \ref{fig:15} is the same as Figure \ref{fig:14} but for log(O/H)-\citetalias{Pilyugin-16}. 

The first-order result is clear: independent of which metallicity indicator is used, low mass galaxies that lie significantly above (or below) the SFMS in their overall sSFR, have systematically lower (or higher) log(O/H) over the whole galactic radii. This dependence of log(O/H) on $\Delta\log$sSFR however decreases (and even vanishes and possibly reverses) with increasing stellar mass. There is some evidence that it also decreases to the centers of the galaxies (see for example the higher mass bins in Figure \ref{fig:14}). 

The result of Figure \ref{fig:14} is broadly consistent with the result of Figure \ref{fig:15}, except for the highest mass bins.  In the highest mass bin of Figure \ref{fig:15}, a positive correlation between $\Delta\log$sSFR and $\Delta\log$(O/H) can be seen, which is clearly opposite to the result of other mass bins. This might be due to the failure of the {\tt N2S2H$\alpha$} indicator that the assumed N/O-O/H relation does not hold for the most massive SF galaxies.  

The overall negative correlation between the overall $\Delta\log$sSFR and log(O/H) shown by the sets of profiles in Figures \ref{fig:14} and \ref{fig:15} is another manifestation of the inverse correlations in Panels (e) of Figures \ref{fig:9} and \ref{fig:10} and in Figure \ref{fig:12}. We can then argue that these indicate that time-varying inflows are the primary drivers of variations of sSFR and log(O/H), across the galaxy population \citep[also see \citetalias{Wang-19};][]{Wang-20a}.  As noted above, this result is a manifestation of the general presence of SFR as a (negative) second parameter in the overall mass-metallicity relation \citep[e.g.][]{Mannucci-10}. The fact that we see the range of log(O/H) (at given R/\re) decreasing with stellar mass is also consistent with previous studies of the overall $Z(M_*,{\rm SFR})$ relation \citep[e.g.][]{Mannucci-10, Curti-20}.

In figure 5 of \citetalias{Wang-19}, the {\it dispersion} in the (normalized) $\Sigma_{\rm SFR}$ increases slightly with increasing stellar mass, and also increases towards the centers of galaxies for galaxies of a given stellar mass.  It is quite striking 
how the {\it dispersion} of $\log({\rm O/H})$ (at a given mass and radius) behaves in the {\it opposite} way to the dispersion in (normalized) $\Sigma_{\rm SFR}$ shown in our previous work.  Whereas the former decreases with increasing stellar mass (and possibly towards the centers of galaxies), the dispersion of $\Sigma_{\rm SFR}$ (or sSFR) increases slightly with mass, and increases towards the centers of galaxies. We will discuss this in detail in the next Section \ref{sec:5.3}. 

\subsection{Quantitative interpretation of the dispersion of gas-phase metallicity and sSFR in MaNGA galaxies} \label{sec:5.3}

\begin{figure*}
  \begin{center}
    \epsfig{figure=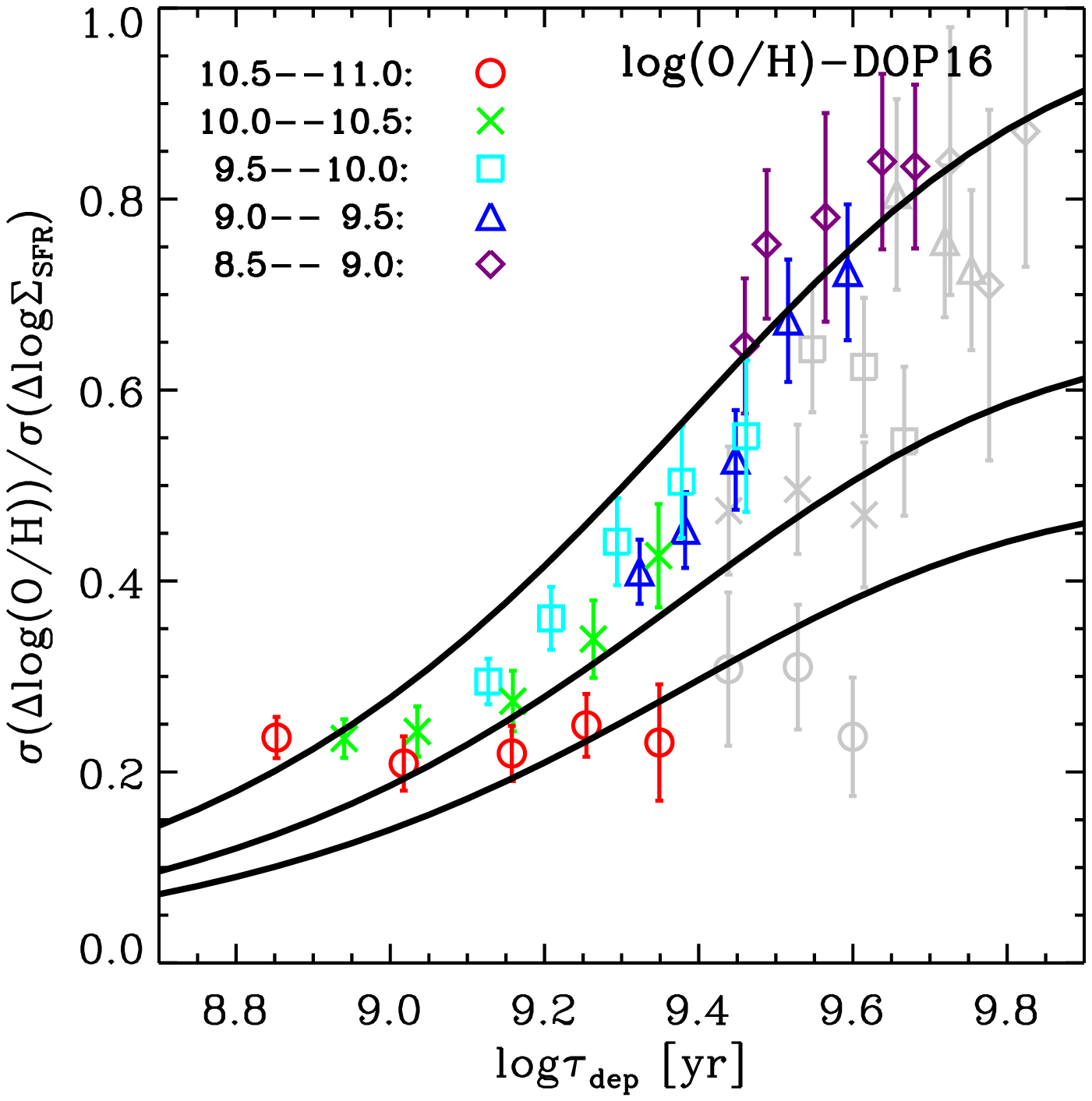,clip=true,width=0.45\textwidth}
    \epsfig{figure=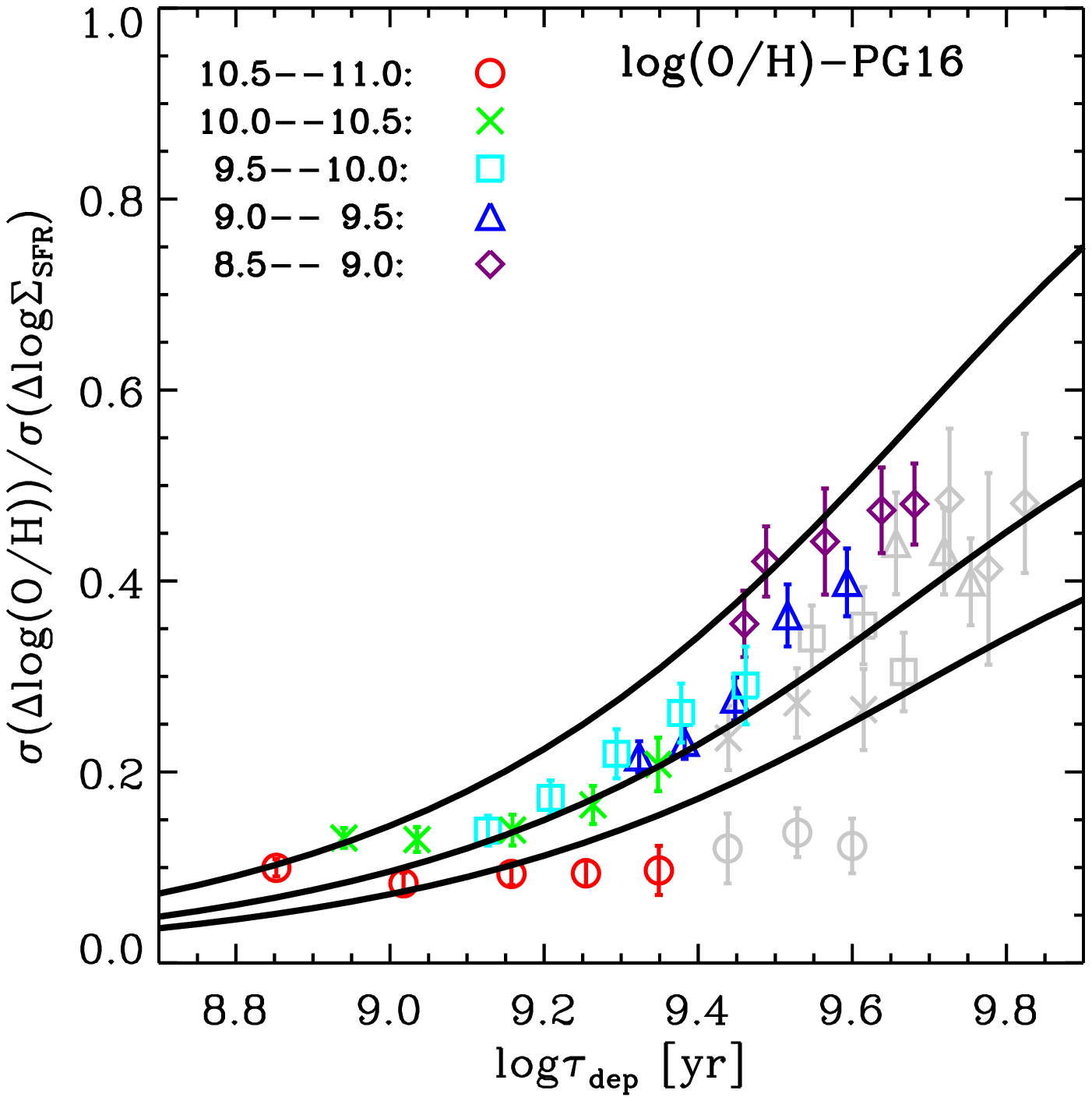,clip=true,width=0.45\textwidth}
    \end{center}
  \caption{Left panel: The ratio of $\sigma$($\Delta \log$(O/H))-\citetalias{Dopita-16} to $\sigma$($\Delta \log \Sigma_{\rm SFR}$) as a function of $\tau_{\rm dep}$ derived using the extended Schmidt law \citep{Shi-11}, based on the MaNGA galaxies. The different colors represent different stellar mass bins, as denoted in the top-left corner. Data points with the radius larger than \re, are indicated in gray. The uncertainties are measured by bootstrap method. 
  Right panel: The same as the left panel, but using log(O/H)-\citetalias{Pilyugin-16}. 
 In both panels, the black solid lines are the model predictions taken from the left panel of Figure \ref{fig:3} with $Z_0/y_{\rm eff}=$0.0, 0.5 and 1.0, but with horizontal shifts. 
  }
  \label{fig:17}
\end{figure*}

As discussed in Section \ref{sec:2}, the simple gas-regulator model predicts not only the sign of the correlation between $\Delta \log$SFR and $\Delta \log Z$, but also the variation of $\Delta \log$SFR and $\Delta \log Z$ as a function of $\xi$ (see Figure \ref{fig:3}) defined as the ratio of the driving period to the effective gas depletion timescale in Equation \ref{eq:6.1}. 
In this subsection, we will investigate more quantitatively the variation of $\Delta \log$SFR and $\Delta \log$(O/H) in MaNGA galaxies. This will inevitably run up against the systematic uncertainties arising from the significantly different outputs of our two chosen metallicity indicators. We can however try to minimize the effects of these by looking at relative effects across the galaxy population, expecting that systematic effects will thereby cancel out.

In \citetalias{Wang-19}, we constructed the $\Sigma_{\rm SFR}$ profiles for MaNGA galaxies at five different stellar mass bins and defined the parameter $\Delta \log \Sigma_{\rm SFR}$ as the deviation of a given galaxy from the median $\Sigma_{\rm SFR}$ profile at a given galactic radius and in a given stellar mass bin.   We then computed the dispersion of this quantity across the population within five stellar mass bins and within five radial bins of $r$/\re. It was found that the scatter of $\Delta \log \Sigma_{\rm SFR}$, which we denoted as $\sigma(\Delta \log \Sigma_{\rm SFR})$, increases significantly with stellar surface mass density, $\Sigma_*$.
We interpreted this trend in terms of a decreasing gas depletion time, linking the stellar surface mass density to the gas depletion timescale via the extended Schmidt law \citep{Shi-11}, i.e. $\tau_{\rm dep} \propto \Sigma_*^{-1/2}$.   The trend with the inferred gas depletion time was found to be consistent with the model predication of time-varying inflow rate driven scenario, provided that the driving period of the inflow was more or less the same for all galaxies. In this subsection, we look at the analogous variation of $\Delta \log ({\rm O/H})$ in the population, and compare the result with the model predictions.  

Similar to the definition of $\Delta \log \Sigma_{\rm SFR}$, we define the $\Delta \log$(O/H) to be the deviation from the median $\log$(O/H) at a given specific radius and a given stellar mass bin.  We then compute the dispersion of this quantity, $\sigma$($\Delta \log({\rm O/H})$), and compare this to the dispersion of the completely independent (normalised) star-formation surface density, $\sigma(\Delta \log \Sigma_{\rm SFR})$ used in \citetalias{Wang-19}. 

Figure \ref{fig:17} shows the ratio of $\sigma$($\Delta \log({\rm O/H})$) to $\sigma$($\Delta\log \Sigma_{\rm SFR}$) as a function of the inferred gas depletion time, for both metallicity indicators.  
The gas depletion time is again derived from the $\Sigma_*$ on the basis of the extended Schmidt law \citep{Shi-11}.  To be in line with \citetalias{Wang-19}, we here use the scatter of $\Delta\log\Sigma_{\rm SFR}$, rather than the scatter of $\Delta\log{\rm sSFR}$. We have checked that using the latter gives a consistent result to that in Figure \ref{fig:17}. 
In each panel of Figure \ref{fig:17}, different colors are used for different stellar mass bins.  The different data points of the same color represent the different radial bins, evenly spaced with a bin width of 0.2\re. At a fixed stellar mass bin, the $\tau_{\rm dep}$ monotonically increases with galactic radius as $\Sigma_*$ decreases. 
As in \citetalias{Wang-19}, data points at galactic radii larger than \re\ are indicated in gray, since these may be more easily affected by environmental effects \citep[also see \citetalias{Wang-19};][]{Wang-20a}. Readers are invited to consult figure 9 in \citetalias{Wang-19} and figures 15 and 17 in \cite{Wang-20a} for further insights into the variations of $\Sigma_{\rm SFR}$.

As a whole, the ratio $\sigma(\Delta \log (\rm O/H)$/$\sigma(\Delta \log \Sigma_{\rm SFR})$ increases quite sharply with the inferred gas depletion time for both the metallicity indicators. This is individually true for each the five stellar mass bins (where it reflects radial changes) except for the highest stellar mass bin for log(O/H)-\citetalias{Dopita-16} and it is true when comparing galaxies of different stellar masses.  These trends reflect quantitatively a combination of the effects that are seen in Figures \ref{fig:14} and \ref{fig:15} of this paper and figure 5 of \citetalias{Wang-19}. It should however be noted that the dispersions $\sigma$ are calculated using the individual galaxies whereas these figures plot the median values within the four bins of $\Delta \log$SFR, and so are not directly comparable. We have discussed the particular case of the highest mass bin for log(O/H)-\citetalias{Dopita-16} in Section \ref{sec:5.2}.

It is again clear that the dispersions $\sigma(\Delta \log ({\rm O/H})$ obtained using the different metallicity estimators differ by a factor of two, reflecting their two-fold difference in range within the sample. In the absence of a reliable reason to prefer one over the other, this makes any precise quantitative comparison with the predictions of the simple gas-regulator model (see Figure \ref{fig:3}) impossible.  However, three points may be made, independent of the choice of estimator.  

First, both metallicity estimators show about a factor of four {\it increase} in the ratio $\sigma(\Delta \log ({}\rm O/H)$/$\sigma(\Delta \log \Sigma_{\rm SFR})$ as the inferred gas depletion timescale increases by an order of magnitude from $\log \tau_{\rm dep}$ $\sim$ 8.8 to 9.8.  This is as expected for variations in inflow rate (around $\log \xi \sim 0$) and quite opposite to the trend expected for variations in SFE, which would have predicted a  decrease in this ratio as the gas depletion timescale increases. 

By horizontally shifting the model prediction in the left panel of Figure \ref{fig:3} (according to Equation \ref{eq:15}) so as to match the result of Figure \ref{fig:17} (see the black lines),  
we get $(1+\lambda)T_{\rm p}/2\pi \sim 10^{9.5}$ yr for log(O/H)-\citetalias{Dopita-16}, and $\sim 10^{9.8}$ yr for log(O/H)-\citetalias{Pilyugin-16}.  These are equivalent to driving periods of the inflow of $T_{\rm p}$ of a few to several Gyr.  This is broadly consistent with the independent argument presented in \cite{Wang-20a} that the variation of $\Delta\log$sSFR is mainly produced by a variation of inflow rate on relatively long timescales.  The driving period of changes in the inflow appears to be considerably longer than the period of the temporal variations in SFE discussed in Section \ref{sec:4.3}. 

Second, it can be seen that at a given gas depletion time (stellar surface mass density), more massive galaxies tend to have a {\it lower} value of $\sigma(\Delta \log({\rm O/H})$/$\sigma(\Delta \log \Sigma_{\rm SFR})$.  This could possibly reflect the quite plausible expectation that more massive galaxies might well have higher $Z_0/y_{\rm eff}$ than less massive galaxies due to either a lower wind-mass loading $\lambda$ or a higher inflow metallicity $Z_0$, leading to a reduction in $\sigma(\Delta \log({\rm O/H}))$/$\sigma(\Delta \log \Sigma_{\rm SFR})$, as shown in Figure \ref{fig:3}.

Finally, it is noticeable that, even with the larger range of the log(O/H)-\citetalias{Dopita-16}, the ratio of $\sigma(\Delta \log ({\rm O/H})$/$\sigma(\Delta \log \Sigma_{\rm SFR})$ never exceeds unity, the maximum value permitted by the gas-regulator model, as shown in Figure \ref{fig:3}.

\section{Discussion} \label{sec:6}

\subsection{The scale effect of $\Delta\log${\rm sSFR}-$\Delta\log$(O/H) relation} \label{sec:6.1}

In this work, we find that on $\sim$100 pc scales within individual galaxies, the local $\Delta \log$(O/H) appears to positively correlated with the local $\Delta \log$sSFR, while when looking into the integrated quantities of the same galaxies across the galaxy population, the $\Delta \log ({\rm O/H})$ is found to be negatively correlated with $\Delta\log$sSFR.  These results are quite consistent with previous findings, as discussed in Section \ref{sec:introduction}.  
Specifically, based on the $\sim$1000 SAMI galaxies,  \cite{Sanchez-19} found the $\Delta \log$sSFR and $\Delta \log$(O/H) (defined in a similar way as in the present work) show a negative correlation across the galaxy population for a wide range of metallicity indicators with the correlation coefficient between $-$0.32 to $-$0.14. At highly-resolved scale ($\sim$100 pc), many authors have found that regions with strongly enhanced star formation show enhanced gas-phase metallicity \citep[e.g.][]{Ho-18, Erroz-Ferrer-19, Kreckel-19}. Our results are consistent with these previous results. 


In the context of our simple gas-regulator framework, the opposite sign of the correlation between $\Delta \log$sSFR and $\Delta \log$(O/H) on 100-pc (GMC-scales) and on galactic-scales and large sub-galactic scales indicates that different physical processes regulate the star formation and chemical enhancement on these different scales. As a whole, a positive $\Delta \log$sSFR-$\Delta \log$(O/H) relation arises driving the gas-regulator with time-varying SFE$(t)$, and a negative $\Delta \log$sSFR-$\Delta \log$(O/H) relation is the result of driving it with a time-varying inflow rate. 
A time-varying SFE$(t)$ at $\sim$100 pc scale \citep[see][]{Kruijssen-19, Chevance-20}, and a time-varying inflow rate at galactic scale \citep[see][]{Wang-19, Wang-20b}, are also suggested by other recent works. 



In this work, we have not examined the intermediate scales of, say, 1 kpc. However, it is not difficult to infer that, as the scale is increased, the effect of time-varying SFE$(t)$ becomes weaker and that of time-varying inflow rate becomes stronger. This is likely the reason that at $\sim$1 kpc scale, the correlation between $\Delta \log$sSFR and $\Delta \log$(O/H) is weaker or even disappeared, as seen by previous works \citep{Moran-12, Barrera-Ballesteros-17, Berhane-Teklu-20}.    

\subsection{What determines the gas-phase metallicity?}  \label{sec:6.2}

As shown in Equation \ref{eq:22}, the metallicity of the steady state (i.e. constant inflow rate and constant SFE) is only determined by the metallicity of the inflow gas $Z_0$ and the effective yield including the effects of any wind, $y(1-R+\lambda)^{-1}$.  These two parameters are expected to be strongly correlated with the global stellar mass. The $Z_0$ may be expected to increase with stellar mass, because the circumgalactic medium of galaxies was enriched by the outflow of enriched material driven by star formation in the past. The wind-loading $\lambda$ is expected to decrease with stellar mass, because more massive galaxies have deeper gravitational potential wells.
This is probably the origin of the observed mass-metallicity relation. 
In addition, we emphasize that the $Z$ does not depend on the {\it absolute} value of SFE or inflow rate, but depends on the change (with time) of it, under the gas-regulator model frame. 



Based on the analysis of the present work (and also some previous works), the so-called ``fundamental metallicity relation'' is clearly not valid at sub-kpc scales. In the gas-regulator framework, we predict that the mass of the gas reservoir is always negatively correlated with metallicity (see Figure \ref{fig:1} and \ref{fig:2}), regardless of the driving mechanisms by time-varying SFE or time-varying inflow rate.  Observationally, \cite{Bothwell-13} and \cite{Bothwell-16} found that the cold gas mass appears is more fundamental in determining the MZR than SFR, for both atomic and molecular gas, consistent with this picture.  
In this sense, the mass of gas reservoir is a better secondary parameter than SFR in determining the metallicity across the GMC-scale to galactic scale.


The importance of the SFR is that, in studying the correlation between $\Delta\log$sSFR and $\Delta\log$(O/H) (as in this paper) we can distinguish the underlying physical processes that are driving variations in gas content, SFR and metallicity, even for individual galaxies. Specifically, even though the negative $\Delta\log$sSFR vs. $\Delta\log$(O/H) relation across the galaxy population indicates that time-varying inflow is the main driver of variations in SFR, it is completely possible that in some particular galaxies, existing reservoirs of cold gas are undergoing strong gravitational instability, leading to a temporary increase in their SFE. 

For instance, \cite{Jin-16} identified 10 SF galaxies with kinematically misaligned gas and stars from MaNGA survey. These galaxies have intense on-going star formation and high gas-phase metallicity in their central regions with respect to normal SF galaxies, which can be easily interpreted as evidence of a temporary increase in SFE due to gas-gas collision between the pre-existing gas and the misaligned inflowing gas. 

\subsection{Caveats}  \label{sec:6.4}

In Section \ref{sec:2.2} (or Section \ref{sec:2.3}), 
we always explore the effects of time-varying inflow rate (or SFE) while assuming the other to be time-invariant. However, we note that in the real universe, both inflow rate and SFE could vary with time simultaneously.  On galactic scales, the negative correlation between $\Delta\log$sSFR and $\Delta\log$(O/H) in the observations indicates that time-varying inflow rates are dominant. However, this does not means that the SFE is fully time-invariant at galactic scale for all SF galaxies. Actually, as also mentioned in Section \ref{sec:6.2}, the SFE may be temporally enhanced in some galaxies, due to some physical processes, such as a merger or interaction with close companions. Mergers and interactions may trigger gravitational instabilities in the cold gas, and further enhance the SFE.  At the small 100-pc scale, although the variation of SFE is dominant, we can probably ignore the possible variation of inflow rate for different regions within individual galaxies. In addition, we do not consider other feedback processes of star formation in the model, except for the outflow, which is assumed to be simply proportional to the SFR.  

The dispersal and mixing of enriched supernova ejecta with the interstellar medium are assumed for simplicity in this paper to be instantaneous. The mixing timescale is expected to be strongly dependent on physical scale \citep{Roy-95, deAvillez-02}. Specifically, on scales of 100 pc or less, the mixing timescale should be a few tens of Myr, of order ten times shorter than the few hundred Myr variation timescales of the SFE that we proposed in this paper. Therefore, the simplifying assumption of instantaneous mixing is unlikely to invalidate the conclusions drawn here.

In the model, we also assume that the yield $y$ of metals is uniform both within galaxies and across the galaxy population. The yield $y$ is closely related with the relative number of Type II supernova, and therefore to the IMF. In the real universe, the IMF may be different from galaxy to galaxy, or even different in different parts of the same galaxy. Indeed, by using a sensitive index of the IMF, $^{13}$CO/C$^{18}$O, \cite{Zhang-18} found that the IMF in dusty star-burst galaxies at redshift $\sim$2-3 may be more top-heavy with respect to \cite{Chabrier-03} IMF. The top-heavy IMFs would result in larger $y$ than the bottom-heavy IMF, which increases the complexity of understanding the metal enhancement process by star formation. 

In Section \ref{sec:4}, we presented the observational results obtained using {\tt N2S2H$\alpha$} and {\tt Scal} metallicity indicators. The two indicators produce broadly consistent results. As discussed in Section \ref{sec:3.3}, these two indicators, proposed by \citetalias{Dopita-16} and \citetalias{Pilyugin-16} respectively, offer significant improvements and advantages over the previous indicators, like {\tt N2} and {\tt O3N2}.  However, we note that when using {\tt O3N2}, the derived results differ in part from those presented here.  Specifically, a negative correlation between $\Delta\log$sSFR and $\Delta\log$(O/H) across the MaNGA galaxy population can be still seen with {\tt O3N2}, while the $\Delta\log$sSFR-$\Delta\log$(O/H) relation of individual spaxels for MAD galaxies is found to depend quite strongly on stellar mass. For galaxies with stellar mass above $\sim10^{10.5}$\msolar, the $\Delta\log$sSFR and $\Delta\log$(O/H) of spaxels show positive correlation, which is similar to the results based on {\tt N2S2H$\alpha$} and {\tt Scal}. But for galaxies with stellar massed below $\sim10^{10.5}$\msolar, the correlation between $\Delta\log$sSFR and $\Delta\log$(O/H) of spaxels becomes negative, different from the results of {\tt N2S2H$\alpha$} and {\tt Scal} shown in this paper. This may be due to the fact that both the {\tt N2S2H$\alpha$} and {\tt Scal} indicators break the degeneracy between metallicity and ionization parameter.  Although we prefer to use the {\tt N2S2H$\alpha$} and {\tt Scal} indicators, we mention this alteration of the results with using {\tt O3N2} here for those readers who may prefer that metallicity indicator. 

In the present work, we have only compared the observational results on low redshift galaxies with the model predictions. However, we note that our model prediction is also suitable for the high-redshift galaxies. One may expect to push the analysis in the current work to high-redshift in the near future, based on near-infrared spectroscopic galaxy surveys with the JWST. 

\section{Summary and Conclusions} \label{sec:7}
\label{sec:conclusion}

The present work consists mainly of two parts. One is the theoretical prediction of the correlation between SFR and gas-phase metallicity in the gas-regulator framework (see Section \ref{sec:2}).  The other is the study of this correlation directly from the observation and the comparison of the results with the model predictions (see Section \ref{sec:4} and \ref{sec:5}). We may summarize the results of these two parts in the following. 

The gas-regulator model is based on the interplay between inflow, outflow and star formation, assuming that the star formation is instantaneously determined by the mass of cold gas reservoir \citep[][\citetalias{Wang-19}]{Lilly-13}. According to the continuity equations for the mass of metals and of the gas reservoir, we build the two basic continuity equations, shown in Equation \ref{eq:2} and Equation \ref{eq:3}. There are in total five quantities that determine the solution of the equations, which are the (assumed here varying) inflow rate $\Phi(t)$ and SFE$(t)$, and the (assumed here constant) mass-loading factor $\lambda$, metallicity of inflow gas $Z_{\rm 0}$ and the yield $y$. Once these five quantities are input, the solution of SFR$(t)$, $M_{\rm gas}(t)$ and $Z(t)$ are unique. 
The model predictions are listed below. 

\begin{itemize}

\item When driving the gas-regulator system with a sinusoidal inflow rate and a time-invariant SFE, the resulting SFR$(t)$, $M_{\rm gas}(t)$ and $M_{\rm Z}(t)$ are also in the form of an exact sinusoidal function with time, but with some phase-delay to the inflow rate (see Equation \ref{eq:6} and \ref{eq:8}). The $\Delta\log$SFR and $\Delta \log Z$, defined as  $\log {\rm SFR}(t)/\langle {\rm SFR}\rangle$ and $\log Z(t)/\langle {Z}\rangle$, are found to be negatively correlated, and the ratio of $\sigma(\Delta \log{\rm SFR})$ to $\sigma(\Delta \log Z)$ increases with increasing $\xi$, defined in terms of the effective gas depletion timescale to be  $2\pi\tau_{\rm dep,eff}/T_{\rm p}$ (see Equation \ref{eq:15}).  If the gas-regulator is driven by a periodic step-function in the inflow rate, a similar negative correlation between $\Delta \log$SFR and $\Delta \log Z$ is produced. 

\item When driving the gas-regulator system with a sinusoidal SFE and time-invariant inflow rate, the resulting SFR$(t)$, $M_{\rm gas}(t)$ and $M_{\rm Z}(t)$ can be solved approximately by a sinusoidal function if the variation of SFE is small (see the approximate solution in Equation \ref{eq:17} and \ref{eq:18}). Opposite to the case of time-varying inflow rate, the $\Delta\log$SFR and $\Delta \log Z$ are now positively correlated, and the ratio of $\sigma(\Delta \log{\rm SFR})$ to $\sigma(\Delta \log Z)$ decreases with increasing $\xi$ (see Equation \ref{eq:20}). When driving the gas-regulator with a periodic SFE in the form of step-function, we find the positive correlation between $\Delta\log$SFR and $\Delta \log Z$ becomes less significant with respect to the case of sinusoidal SFE. However, one thing is clear: the states with highly enhanced SFR are always metal-enhanced with respect to the mean metallicity. 

\item  Regardless of whether the gas regulator is driven with time-varying inflow or time-varying SFE, the $\Delta \log M_{\rm gas}$ is always predicted to be negatively correlated with $\Delta \log Z$ (see Figure \ref{fig:1} and Figure \ref{fig:2}). 

\item The scatter of $\Delta\log$Z always decreases with increasing $Z_0/y_{\rm eff}$, where the $y_{\rm eff}$ is defined as the $y(1-R+\lambda)^{-1}$ (see Equation \ref{eq:15}, Equation \ref{eq:20} and Figure \ref{fig:3}). 

\item The mean SFR is determined by the mean inflow rate and mass-loading factor, and the mean metallicity is determined by the $Z_{\rm 0}+y_{\rm eff}$ (see Equation \ref{eq:21} and \ref{eq:22}). The resulting $Z$ does not depend on the SFE (or inflow rate) itself, but does depend on the temporal {\it changes} of it. 

\end{itemize}

The key point is that a time-varying inflow rate leads to the opposite correlation between $\Delta\log$SFR and $\Delta \log Z$ from that produced by a time-varying SFE. Therefore, studying the $\Delta \log$SFR-$\Delta \log Z$ relation in observational data on different spatial scales can in principle directly distinguish the driving mechanisms of the variation of star formation and gas-phase metallicity in galaxies.  In the model predictions, we note that the conclusions concerning the $\Delta \log$SFR-$\Delta \log Z$ relation are also valid for the $\Delta \log$sSFR-$\Delta \log$(O/H) relation, when the $\Delta \log$sSFR is defined as the displacement of logsSFR(t) from its smooth cosmic evolution. 

We then utilize the two-dimensional spectroscopic data of 38 SF galaxies from the MAD survey \citep{Erroz-Ferrer-19}, as well as a well-defined SF sample of $\sim$1000 galaxies from MaNGA survey (\citetalias{Wang-19}). The spatial resolution of MAD galaxies is $\sim$100 pc or less, while the spatial resolution of MaNGA galaxies is 1-2 kpc. The MAD sample enables us to study the $\Delta \log$sSFR-$\Delta \log({\rm O/H})$ relation down to 100-pc (GMC) scales, while the large sample size of MaNGA  enables us to statistically study the $\Delta\log$sSFR-$\Delta \log({\rm O/H})$ relation at galactic or large (radial) sub-galactic scales across the galaxy population. The SFR is measured based on the dust attenuation corrected H$\alpha$ luminosity \citep{Kennicutt-98}.  The two versions of gas-phase metallicity are measured by adopting two recently-developed indicators: {\tt N2S2H$\alpha$} (\citetalias{Dopita-16}) and {\tt Scal} (\citetalias{Pilyugin-16}), which represent improvements and advantages over the previously widely-used indicators. The results of these two metallicity indicators are very similar. Here we summarize the main observational results, which are valid for both these metallicity indicators. 
\begin{itemize}

\item Consistent with previous studies, we find that MAD galaxies generally show a positive sSFR profile, confirming an inside-out growth scenario. As a whole, the gas-phase metallicity increases strongly with stellar mass, and decreases with galactic radius within individual galaxies, as expected. 

\item At $\sim$100 pc scale in MAD galaxies, we find that $\Delta\log$sSFR and $\Delta \log$(O/H) are positively correlated. This positive correlation shows little or no dependence on the overall stellar mass of the galaxy. We note that the positive correlation of $\Delta\log$sSFR vs. $\Delta \log$(O/H) at 100 pc scale does not hold when using {\tt O3N2} metallicity indicator for galaxies with stellar mass below $\sim10^{10.5}$\msolar. The inconsistency is likely arising from the fact that the {\tt O3N2} shows much larger uncertainty than {\tt Scal} in determining the metallicity, as discussed in Section \ref{sec:3.3.2}.   

\item At galactic scale, we find in contrast that $\Delta \log$sSFR and $\Delta \log({\rm O/H})$ are negatively correlated across the galaxy population. This is true for both MAD and the larger MaNGA samples. The correlation between $\Delta\log$sSFR and $\Delta \log({\rm O/H})$ shows a strong dependence of global stellar mass and galactic radius. 

\item At the $\sim$100 pc scale, the ratio of $\sigma(\Delta \log({\rm O/H})$ to $\sigma(\Delta \log \Sigma_{\rm SFR})$ show almost no dependence on the global stellar mass. However, at galactic scale, the $\sigma(\Delta \log({\rm O/H}))$/$\sigma(\Delta \log \Sigma_{\rm SFR})$ increases with the inferred gas depletion time (inferred from the surface mass density using the extended Schmidt law). At fixed gas depletion time, the $\sigma(\Delta  \log({\rm O/H}))$/$\sigma(\Delta \log \Sigma_{\rm SFR})$ appears to be smaller for galaxies of higher stellar mass. 

\end{itemize}

We interpret the observational results in the frame of the gas-regulator model. The overall increase of metallicity with global stellar mass and the decrease of metallicity with galactic radius, can be well explained as the mass and radial dependence of the metallicity of inflow gas $Z_0$ and the mass-loading factor $\lambda$.  At 100-pc scales, the positive correlation between $\Delta \log$sSFR and $\Delta \log({\rm O/H})$ indicates that the time-varying SFE plays a dominant role in governing the star formation and metal enhancement. This is also consistent with the fact that the variation of SFE increases strongly towards smaller scale \citep{Kreckel-18, Chevance-20} and is likely caused by the passage of orbiting gas through regions of higher SFE, such as spiral arms.  At galactic or sub-galactic scales, the negative correlation across the galaxy population indicates that the time-varying inflow rate plays a dominant role. 

In addition, the variation of $\Delta \log$sSFR and $\Delta \log({\rm O/H})$ as a function of gas depletion time are in quite good agreement with the model predictions.  This strongly strengthens the conclusion that on galactic scales the star formation and metal-enhancement is primarily regulated by the time-varying inflow rate of gas from the surrounding medium.

We emphasize that the sign of the correlation between gas-phase metallicity and SFR is a powerful diagnostic of the driving mechanisms of star formation. Our study provides a new perspective in understanding the correlation between star formation rate, gas-phase metallicity and mass of cold gas reservoir, that is applicable from 100 pc-scales up to galactic scales, from individual galaxies up to the overall galaxy population, and at both low and high redshifts.  


\acknowledgments

Funding for the Sloan Digital Sky Survey IV has been provided by
the Alfred P. Sloan Foundation, the U.S. Department of Energy Office of
Science, and the Participating Institutions. SDSS-IV acknowledges
support and resources from the Center for High-Performance Computing at
the University of Utah. The SDSS web site is www.sdss.org.

SDSS-IV is managed by the Astrophysical Research Consortium for the
Participating Institutions of the SDSS Collaboration including the
Brazilian Participation Group, the Carnegie Institution for Science,
Carnegie Mellon University, the Chilean Participation Group, the French Participation Group, 
Harvard-Smithsonian Center for Astrophysics,
Instituto de Astrof\'isica de Canarias, The Johns Hopkins University,
Kavli Institute for the Physics and Mathematics of the Universe (IPMU) /
University of Tokyo, Lawrence Berkeley National Laboratory,
Leibniz Institut f\"ur Astrophysik Potsdam (AIP),
Max-Planck-Institut f\"ur Astronomie (MPIA Heidelberg),
Max-Planck-Institut f\"ur Astrophysik (MPA Garching),
Max-Planck-Institut f\"ur Extraterrestrische Physik (MPE),
National Astronomical Observatory of China, New Mexico State University,
New York University, University of Notre Dame,
Observat\'ario Nacional / MCTI, The Ohio State University,
Pennsylvania State University, Shanghai Astronomical Observatory,
United Kingdom Participation Group,
Universidad Nacional Aut\'onoma de M\'exico, University of Arizona,
University of Colorado Boulder, University of Oxford, University of Portsmouth,
University of Utah, University of Virginia, University of Washington, University of Wisconsin,
Vanderbilt University, and Yale University.


\bibliography{rewritebib.bib}

\appendix

\section{A. Detailed derivation of Equations 13 and 15.} \label{sec:A}

In Section \ref{sec:2.2}, we have presented the analytic solution of the continuity equations for SFR$(t)$, $M_{\rm gas}(t)$, and $Z_{\rm gas}(t)$ when driving the system with sinusoidal inflow rate and time-invariant SFE.  We have also presented the approximate analytic solution for $\sigma(\log {\rm SFR})$/$\sigma(\log \Phi)$, and $\sigma(\log Z_{\rm gas})$/$\sigma(\log {\rm SFR})$ in Equation \ref{eq:14} and Equation \ref{eq:15}.  Here we present the detailed derivation of them. In this process, we assume the variation of inflow rate is a small perturbation, i.e. $\Phi_{\rm t} \ll \Phi_{\rm 0}$. According to the Taylor expansion, the $\log(1+x)$ can be written as $x-x^2/2+x^3/3-x^4/4+ \cdots$. Therefore, the $\log {\rm SFR}(t)$ can be written as: 
\begin{equation} \label{eq:A1}
\begin{split}
    \log {\rm SFR}(t) = & \ \log [{\rm SFE}\cdot M_{\rm gas}(t)] \\ 
                = & \ \log ({\rm SFE} \cdot M_{\rm 0}) + \log[1+\frac{M_{\rm t}}{M_{\rm 0}}{\rm sin}(2\pi t/T_{\rm p} - \delta)] \\
                \approx & \  \log ({\rm SFE} \cdot M_{\rm 0}) + \frac{M_{\rm t}}{M_{\rm 0}}{\rm sin}(2\pi t/T_{\rm p} - \delta)
\end{split}
\end{equation}
Here we ignore the second-order terms or higher. Since $\sigma(X)=\textbf{Var}(X)^{1/2}=[\textbf{E}(X^2)-\textbf{E}(X)^2]^{1/2}$ with the \textbf{Var}($X$) denoting the variance of $X$ and $\textbf{E}(X)$ denoting the expected value of $X$,  we can rewrite the $\sigma(\log {\rm SFR})$ as: 
\begin{equation} \label{eq:A2}
\begin{split}
    \sigma(\log {\rm SFR}) \approx & \ \sigma(\frac{M_t}{M_0}{\rm sin}(2\pi t/T_p + \delta)) \\
           = & \  [\textbf{E}(\frac{M_t^2}{M_0^2}{\rm sin^2}(2\pi t/T_p - \delta))]^{1/2} \\ 
           = & \  \frac{1}{\sqrt{2}}\cdot \frac{1}{(1+\xi^2)^{1/2}} \frac{\Phi_t}{\Phi_0}. 
\end{split}
\end{equation}
In the same way, the $\sigma(\log \Phi)$ can be written as: 
\begin{equation} \label{eq:A3}
\begin{split}
    \sigma(\log \Phi) \approx  \ \frac{1}{\sqrt{2}}\cdot \frac{\Phi_t}{\Phi_0}.
\end{split}
\end{equation}
Therefore, we can obtain the approximate solution of $\sigma(\log {\rm SFR})$/$\sigma(\log \Phi)$ (Equation \ref{eq:14}): 
\begin{equation} \label{eq:A4}
\begin{split}
    \frac{\sigma(\log {\rm SFR})}{\sigma(\log \Phi)} \approx  \ \frac{1}{(1+\xi^2)^{1/2}}.
\end{split}
\end{equation}

According to Equation \ref{eq:5}, $\log Z(t)$ can be written as: 
\begin{equation} \label{eq:A5}
\begin{split}
    \log {Z}(t) = & \ \log [M_{\rm Z}(t)/M_{\rm gas}(t)] \\ 
                = & \ \log\frac{M_{\rm Z0}}{M_{\rm 0}} + \log[1+\frac{M_{\rm Zt}}{M_{\rm Z0}}{\rm sin}(2\pi t/T_{\rm p} - \beta)] - \log[1+\frac{M_{\rm t}}{M_{\rm 0}}{\rm sin}(2\pi t/T_{\rm p} - \delta)] \\
            \approx & \  \log\frac{M_{\rm Z0}}{M_{\rm 0}} + \frac{M_{\rm Zt}}{M_{\rm Z0}}{\rm sin}(2\pi t/T_{\rm p} - \beta) - \frac{M_{\rm t}}{M_{\rm 0}}{\rm sin}(2\pi t/T_{\rm p} - \delta)
\end{split}
\end{equation}
Therefore, $\sigma(\log Z)$ can be written as: 
\begin{equation} \label{eq:A6}
    \sigma(\log {Z}) 
           \approx \sigma(\frac{M_{\rm Zt}}{M_{\rm Z0}}{\rm sin}(2\pi t/T_{\rm p} - \beta) - \frac{M_{\rm t}}{M_{\rm 0}}{\rm sin}(2\pi t/T_{\rm p} - \delta))
\end{equation}
Since $\textbf{Var}(X-Y) = \textbf{Var}(X) + \textbf{Var}(Y) - 2\textbf{Cov}(X,Y)$ with the $\textbf{Cov}(X,Y)$ denoting the covariance of $X$ and $Y$, then we have 
\begin{equation} \label{eq:A7}
\begin{split}
    \sigma^2(\log {\rm Z}) 
           \approx  & \ \textbf{Var}(\frac{M_{\rm Zt}}{M_{\rm Z0}}{\rm sin}(2\pi t/T_{\rm p} - \beta) - \frac{M_{\rm t}}{M_{\rm 0}}{\rm sin}(2\pi t/T_{\rm p} - \delta)) \\ 
           = & \ \textbf{Var}(X) + \textbf{Var}(Y) - 2\textbf{Cov}(X,Y), 
\end{split}
\end{equation}
where we let $X=M_{\rm Zt}/M_{\rm Z0}\cdot {\rm sin}(2\pi t/T_{\rm p} - \beta)$ and $Y=M_{\rm t}/M_{\rm 0}\cdot {\rm sin}(2\pi t/T_{\rm p} - \delta)$. We then solve \textbf{Var}($X$), \textbf{Var}($Y$) and \textbf{Cov}($X,Y$) separately.  
Similar to Equation \ref{eq:A2}, we can write the  \textbf{Var}($X$) and \textbf{Var}($Y$) as: 
\begin{equation} \label{eq:A8}
\begin{split}
    \textbf{Var}(X) = & \ \textbf{E}[\frac{M_{\rm Zt}^2}{M_{\rm Z0}^2}{\rm sin^2}(2\pi t/T_{\rm p} - \beta)] \\
      = & \ \frac{1}{2}\cdot\frac{M_{\rm Zt}^2}{M_{\rm Z0}^2} \\
      = & \ \frac{1}{2}\cdot \frac{1+\eta^2}{(1+\xi^2)^2}\cdot \frac{\Phi_{\rm t}^2}{\Phi_{\rm 0}^2},
\end{split}
\end{equation}
where $\eta$ is defined in Equation \ref{eq:10}, and 
\begin{equation} \label{eq:A9}
\begin{split}
    \textbf{Var}(Y) = & \ \textbf{E}[\frac{M_{\rm t}^2}{M_{\rm 0}^2}{\rm sin^2}(2\pi t/T_{\rm p} - \delta)] \\
      = & \ \frac{1}{2}\cdot\frac{M_{\rm t}^2}{M_{\rm 0}^2} \\
      = & \ \frac{1}{2}\cdot \frac{1}{1+\xi^2}\cdot \frac{\Phi_{\rm t}^2}{\Phi_{\rm 0}^2}.
\end{split}
\end{equation}
The $\textbf{Cov}(X,Y)$ can be written as: 
\begin{equation} \label{eq:A10}
\begin{split}
    \textbf{Cov}(X,Y) = & \ \textbf{E}(XY) - \textbf{E}(X)\textbf{E}(Y) = \ \textbf{E}(XY) \\
      = & \ \frac{M_{\rm Zt}}{M_{\rm Z0}}\cdot \frac{M_{\rm t}}{M_{\rm 0}} \cdot \textbf{E}[{\rm sin}(2\pi t/T_{\rm p}-\beta)\cdot {\rm sin}(2\pi t/T_{\rm p} - \delta)] \\
      = & \ \frac{M_{\rm Zt}}{M_{\rm Z0}}\cdot \frac{M_{\rm t}}{M_{\rm 0}} \cdot \textbf{E}[{\rm cos^2}(2\pi t/T_{\rm p})\cdot {\rm sin}\beta{\rm sin}\delta + {\rm sin^2}(2\pi t/T_{\rm p}) \cdot {\rm cos}\beta {\rm cos}\delta] \\ 
      = & \ \frac{1}{2} \cdot \frac{M_{\rm Zt}}{M_{\rm Z0}}\cdot \frac{M_{\rm t}}{M_{\rm 0}} \cdot ({\rm sin}\beta{\rm sin}\delta + {\rm cos}\beta {\rm cos}\delta)
\end{split}
\end{equation}
Inserting the solutions of $M_{\rm t}/M_{\rm 0}$, $M_{\rm Zt}/M_{\rm Z0}$, $\beta$, and $\delta$ from Equation \ref{eq:6} and \ref{eq:8} into Equation \ref{eq:A10}, the \textbf{Cov}($X,Y$) then can be simplified as: 
\begin{equation} \label{eq:A11}
    \textbf{Cov}(X,Y)= \frac{1}{2}\frac{1+\eta\xi}{(1+\xi^2)^2}\cdot \frac{\Phi_{\rm t}^2}{\Phi_{\rm 0}^2}.
\end{equation}
Inserting Equation \ref{eq:A8}, \ref{eq:A9} and \ref{eq:A11} into Equation \ref{eq:A7}, we can obtain the approximate solution of $\sigma(\log Z)$ as: 
\begin{equation} \label{eq:A12}
\begin{split}
    \sigma(\log Z) \approx & \  \frac{1}{\sqrt{2}}\frac{\xi-\eta}{1+\xi^2}\cdot \frac{\Phi_{\rm t}}{\Phi_{\rm 0}} \\
     = & \ \frac{1}{\sqrt{2}} \frac{\xi}{1+\xi^2}\cdot \frac{1}{1+Z_{\rm 0}/y_{\rm eff}} \cdot \frac{\Phi_{\rm t}}{\Phi_{\rm 0}}
\end{split}
\end{equation}
We therefore have the solution of $\sigma(\log Z)$/$\sigma(\log {\rm SFR})$ (Equation \ref{eq:15}): 
\begin{equation} \label{eq:A13}
\frac{\sigma(\log Z)}{\sigma(\log {\rm SFR})} \approx \frac{\xi}{(1+\xi^2)^{1/2}} \cdot \frac{1}{1+Z_0/y_{\rm eff}}.
\end{equation}

In the same way as before, we can also obtain the approximate solutions of  $\sigma(\log {\rm SFR})$/$\sigma(\log {\rm SFE})$ and $\sigma(\log Z)$/$\sigma(\log {\rm SFR})$ when driving the gas-regulator with sinusoidal SFE and time-invariant inflow rate (see Equation \ref{eq:19} and Equation \ref{eq:20}). Here we do not present the details of derivation.  Although we assume the variation of inflow rate (or SFE) is a small perturbation in the derivation, we note that the analytic solutions are not bad approximations even for the cases that the variation of inflow rate (or SFE) is quite significant, as confirmed by the numerical solutions (see Appendix \ref{sec:C}).

\section{B. The diagram of $\Delta \log$SFR vs. $\Delta \log Z$ when the variation amplitude of inflow rate or SFE is quite significant } \label{sec:C}

\begin{figure*}
  \begin{center}
    \epsfig{figure=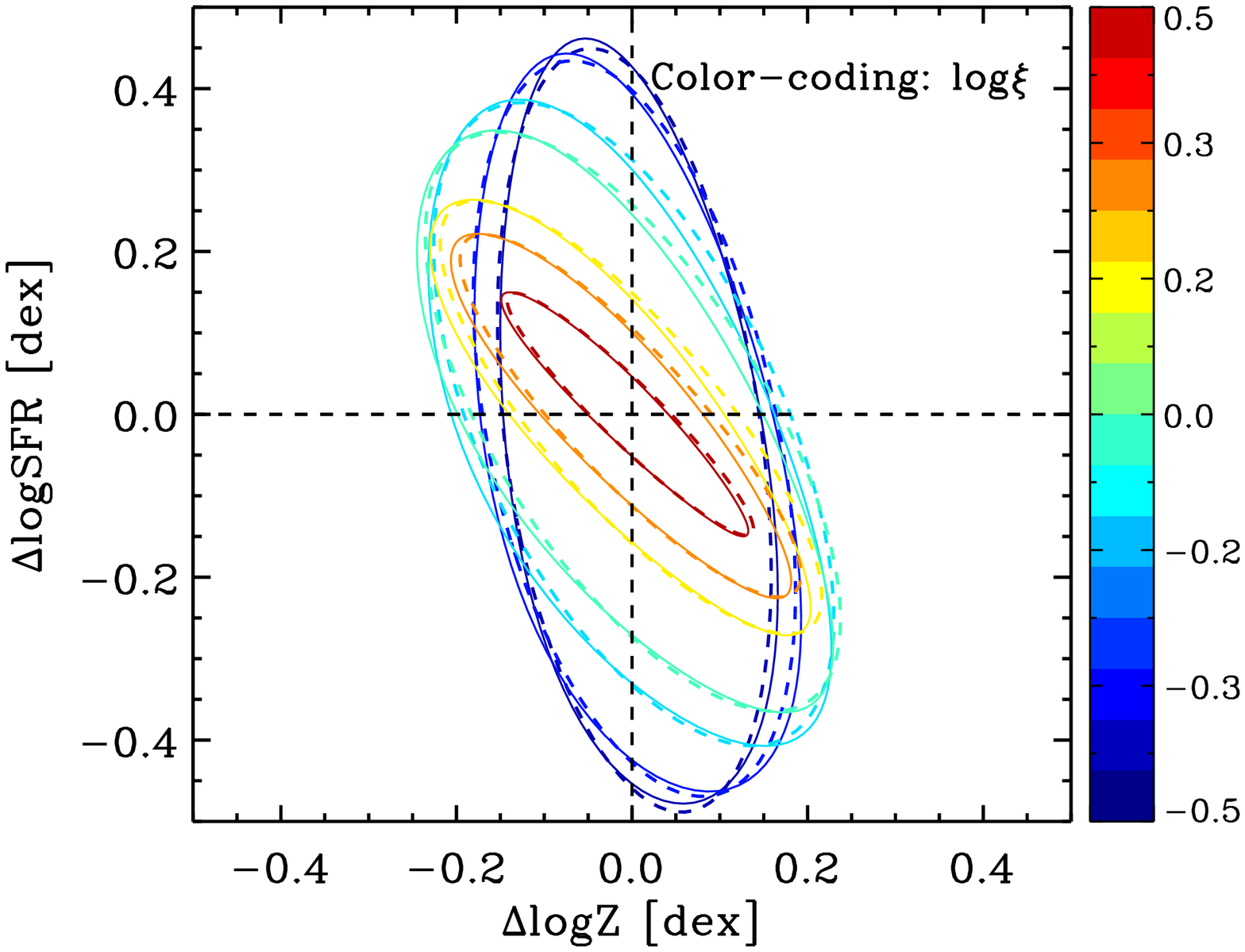,clip=true,width=0.45\textwidth}
    \epsfig{figure=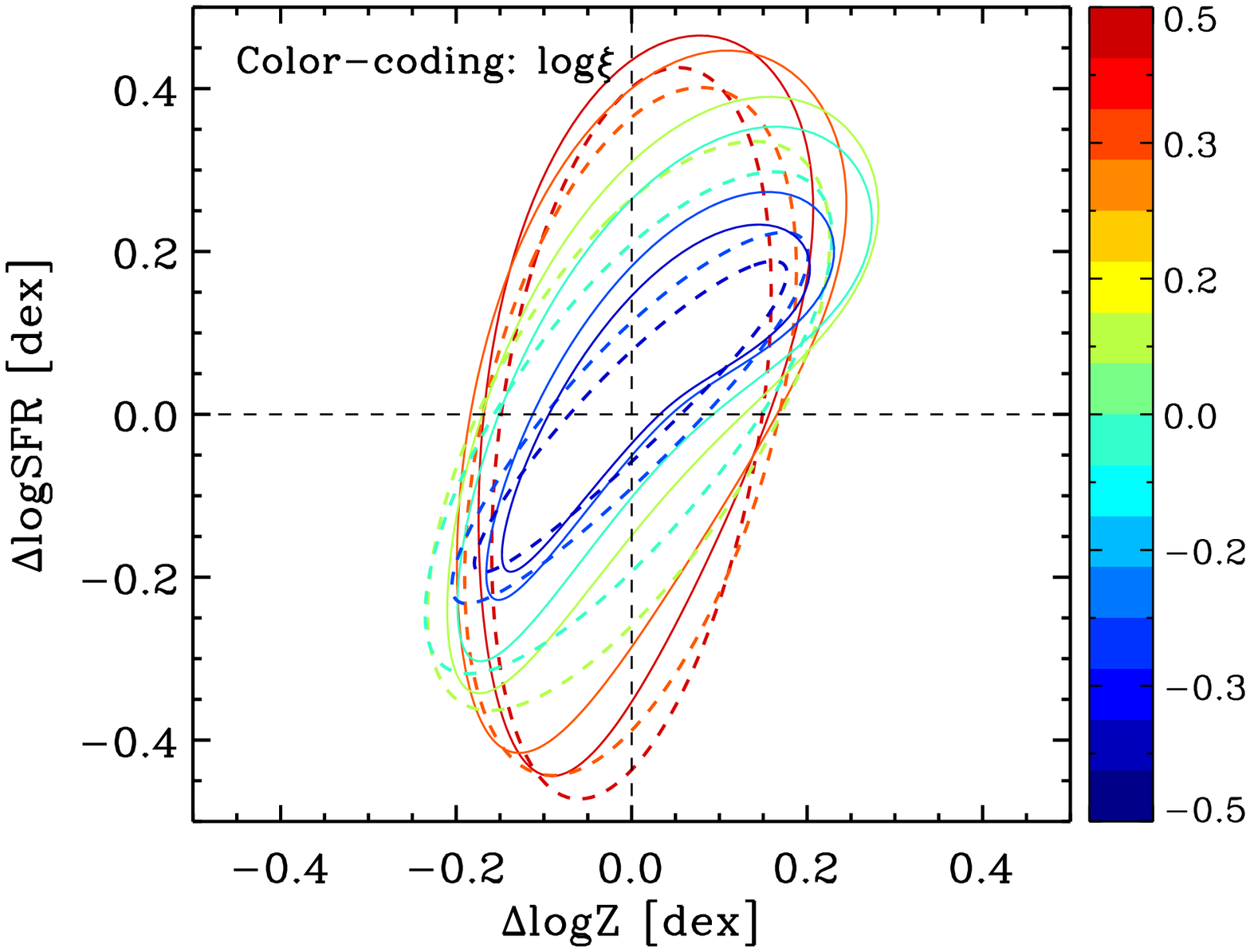,clip=true,width=0.45\textwidth}
    \end{center}
  \caption{Left panel: The correlation of $\Delta \log$SFR vs. $\Delta \log Z$ (the solid curves) at a set of $\log\xi$ when driving the gas-regulator with the sinusoidal inflow rate in logarithmic space.  Right panel:  The same as left panel by driving the gas-regulator with sinusoidal SFE in logarithmic space.  The variation amplitude for inflow rate or SFE is 0.5 dex. For comparison, we also present the correlation of  $\Delta \log$SFR vs. $\Delta \log Z$ (the dashed curves) taken from the top-middle panel of Figure \ref{fig:1} or \ref{fig:2}, but scaled to match the scatter of $\Delta \log$SFR here.
 }
  \label{fig:20}
\end{figure*}

In Section \ref{sec:2.2} and \ref{sec:2.3}, we have investigated the correlation of $\Delta \log$SFR vs. $\Delta \log Z$ when driving the gas-regulator with the sinusoidal inflow rate and SFE in linear space. However, in the top-middle panel of Figure \ref{fig:1} and \ref{fig:2}, we only presented the case for small variations, for illustration purpose. Here, we examine the cases for a very significant variation of inflow rate or SFE.  When the inflow rate varies linearly, the $\Phi_{\rm t}$ is at most equal to $\Phi_0$, since the inflow rate can not be negative. This limits the settings of the amplitude of inflow rate (or SFE). Therefore, in this test, we drive the gas-regulator model with sinusoidal $\Phi(t)$ (or SFE) in logarithmic space, i.e. $\Phi(t) = \Phi_0\cdot 10^{Am\cdot {\rm sin}(2\pi t/T_{\rm p})}$, where the amplitude of variation $Am$ is set to be 0.5 dex. We solve the Equation \ref{eq:2} and \ref{eq:3} numerically with all other settings to be the same of those in Figure \ref{fig:1} (or Figure \ref{fig:2}).

Figure \ref{fig:20} shows the numerical solutions (solid lines) for the correlation between $\Delta \log$SFR and $\Delta \log Z$ for large amplitude variations inflow rate (left panel) and SFE (right panel), compared with scaled versions of the small-amplitude results (dashed lines). The correlation of  $\Delta \log$SFR vs. $\Delta \log Z$ in the left panel hardly varies as the amplitude increases by a factor of $\sim$10. There are larger distortions for the locus of $\Delta \log$SFR vs. $\Delta \log Z$ (right panel), but the correlations are still broadly comparable over this same wide range of amplitude. This test indicates that the model predictions of $\Delta \log$SFR vs. $\Delta \log Z$ in Figure \ref{fig:1} and \ref{fig:2}, are also applicable when the variation of inflow rate or SFE is quite large. 

\section{C. Defining the $\Delta\log$sSFR and $\Delta\log$(O/H) of the spaxels using $\Sigma_*$ for MAD galaxies} \label{sec:B}

\begin{figure*}
  \begin{center}
    \epsfig{figure=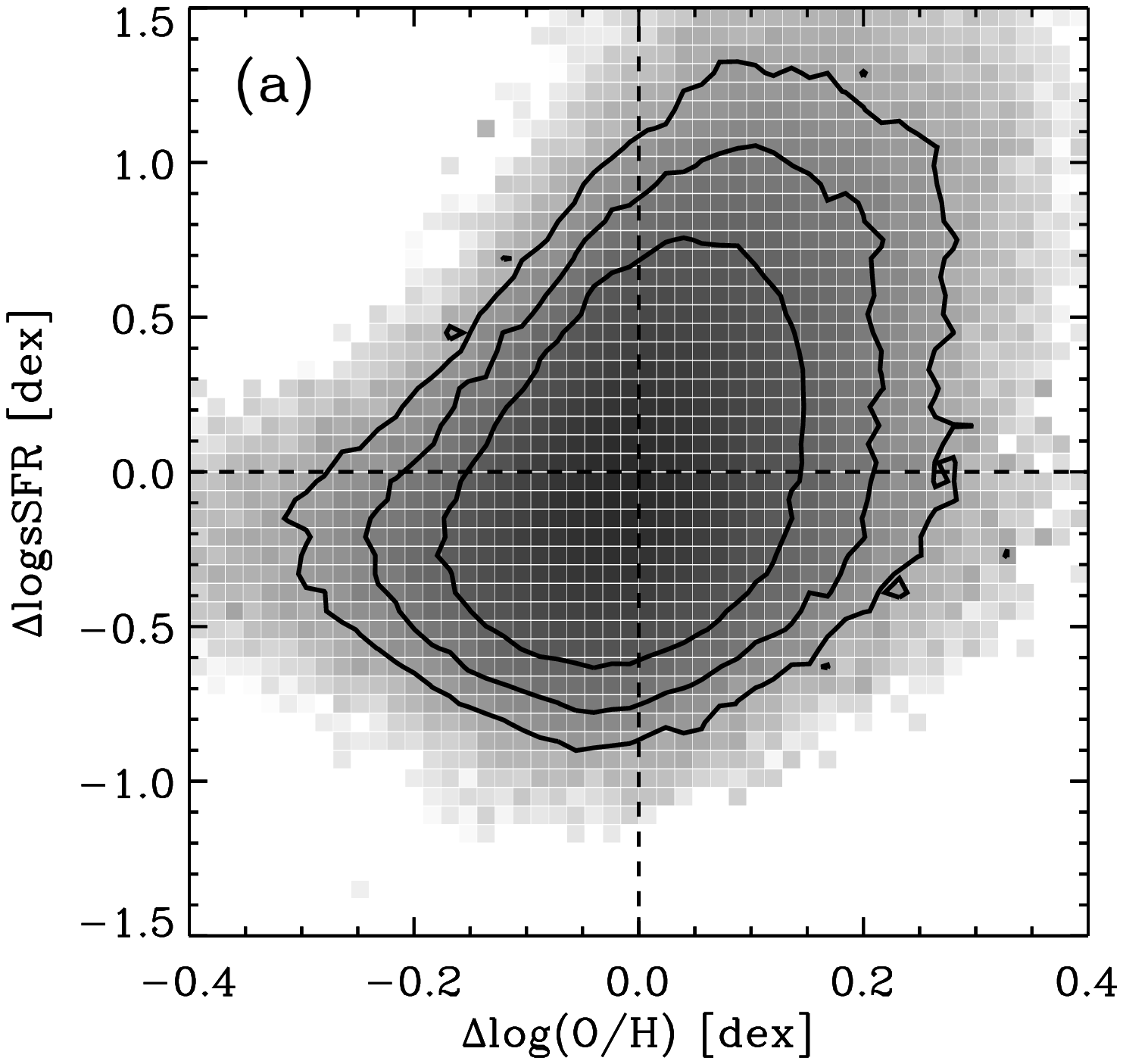,clip=true,width=0.33\textwidth}
    \epsfig{figure=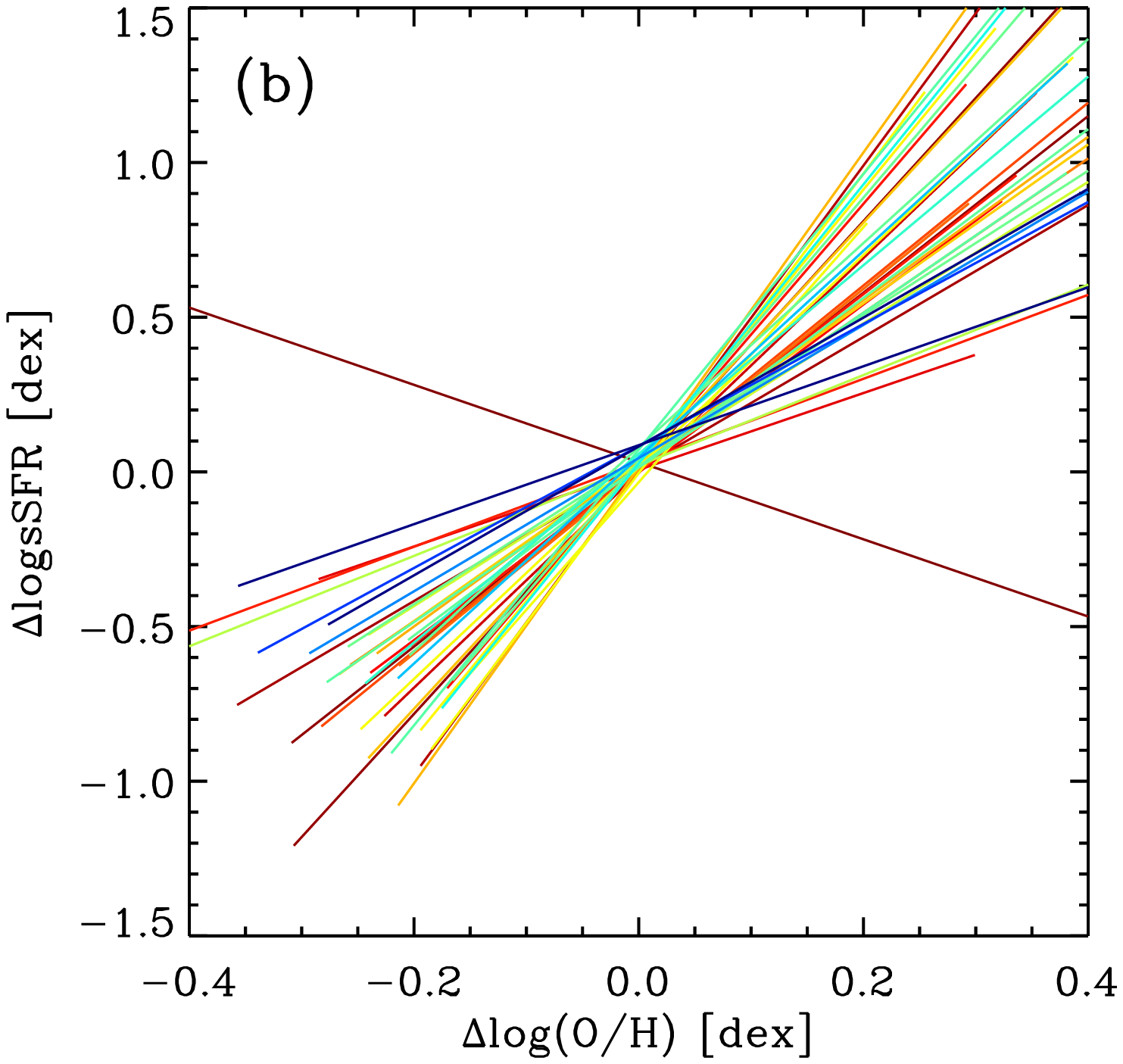,clip=true,width=0.33\textwidth}

    \epsfig{figure=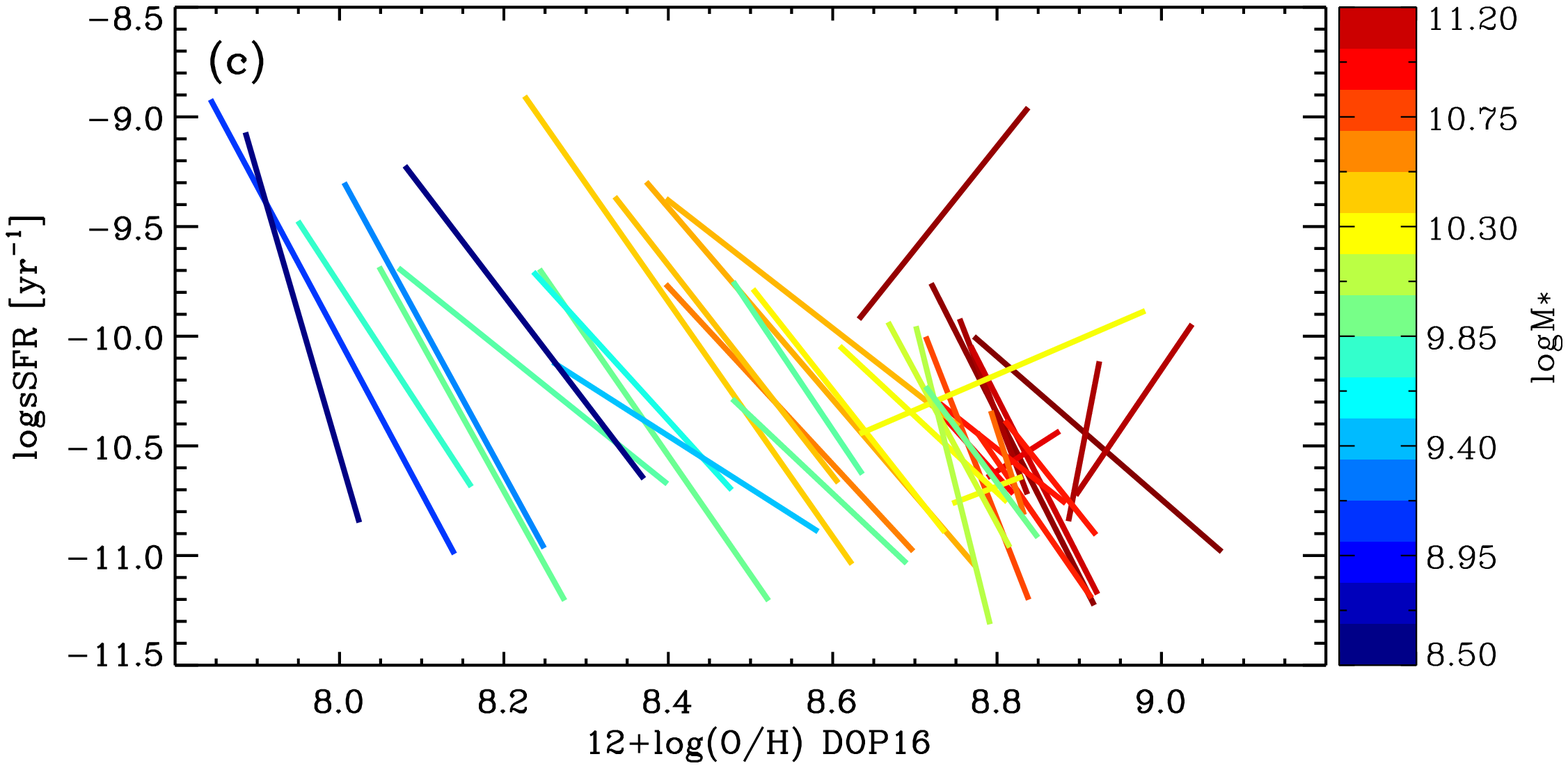,clip=true,width=0.663\textwidth}
    \end{center}
  \caption{This figure is the same as the panel (a), (b) and (c) of Figure \ref{fig:9}, but the definition of $\Delta\log$(O/H)-\citetalias{Dopita-16} and $\Delta\log$sSFR is based on the stellar surface density, rather than the galactic radius.  }
  \label{fig:18}
\end{figure*}

\begin{figure*}
  \begin{center}
    \epsfig{figure=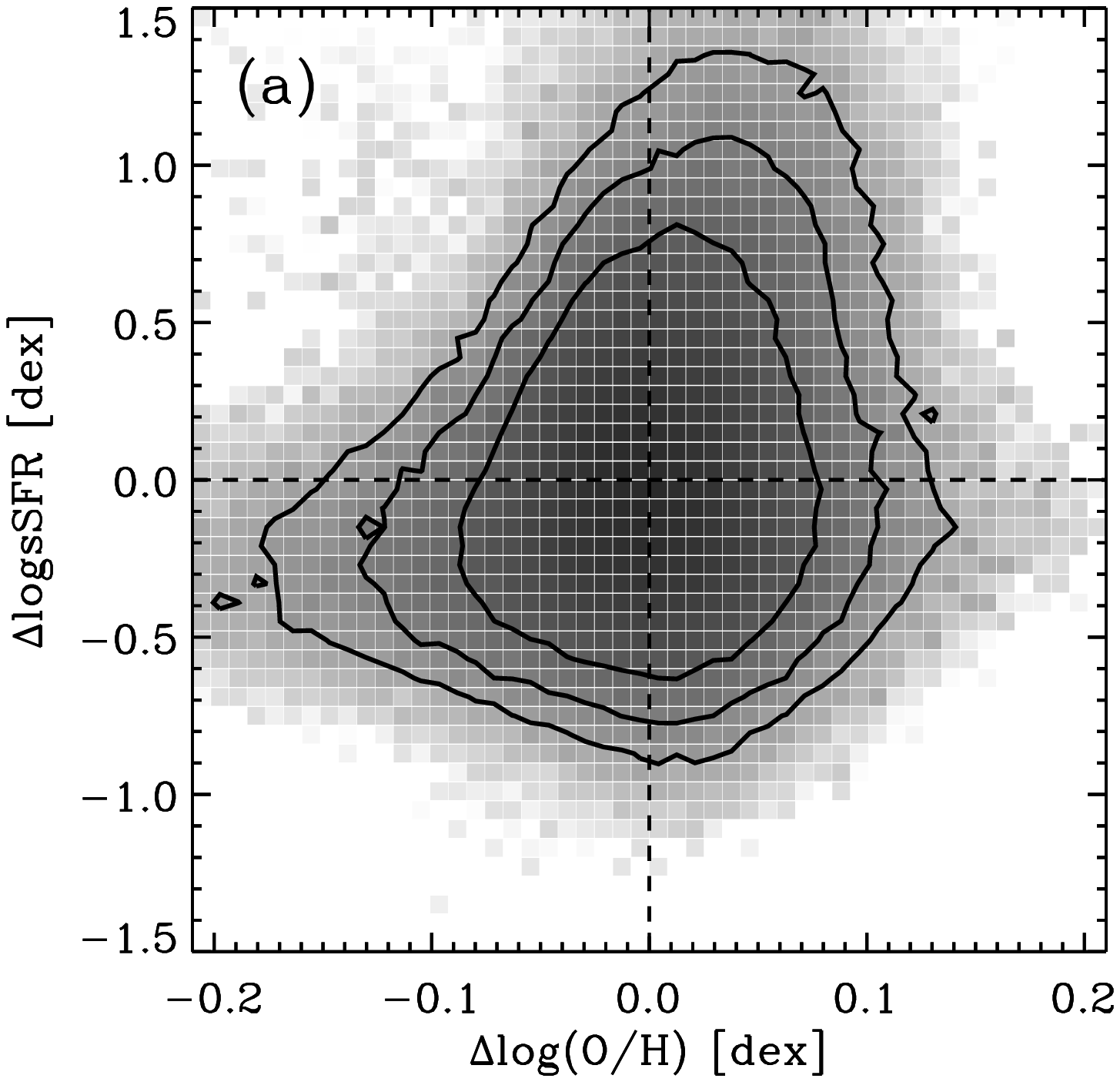,clip=true,width=0.33\textwidth}
    \epsfig{figure=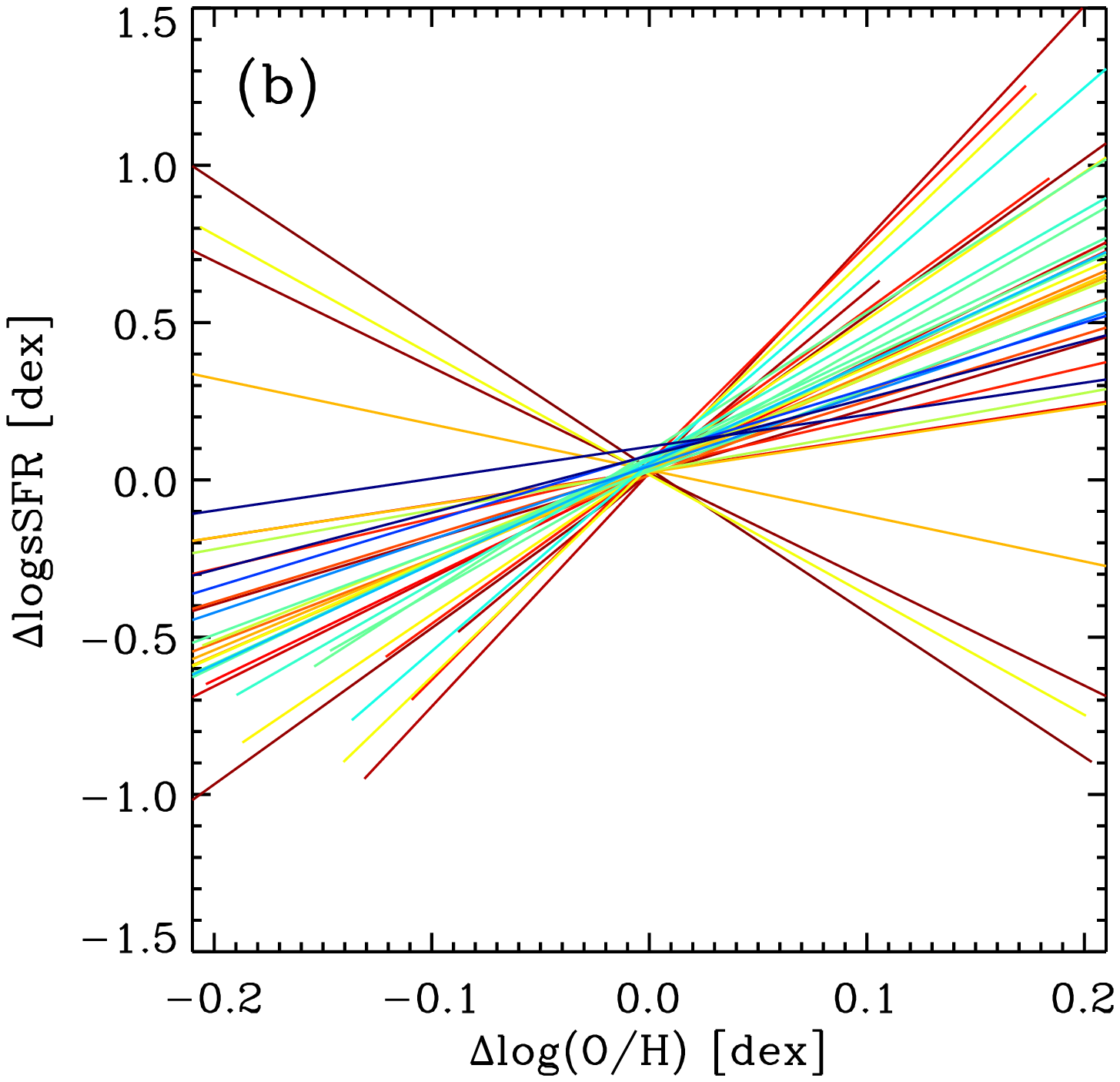,clip=true,width=0.33\textwidth}

    \epsfig{figure=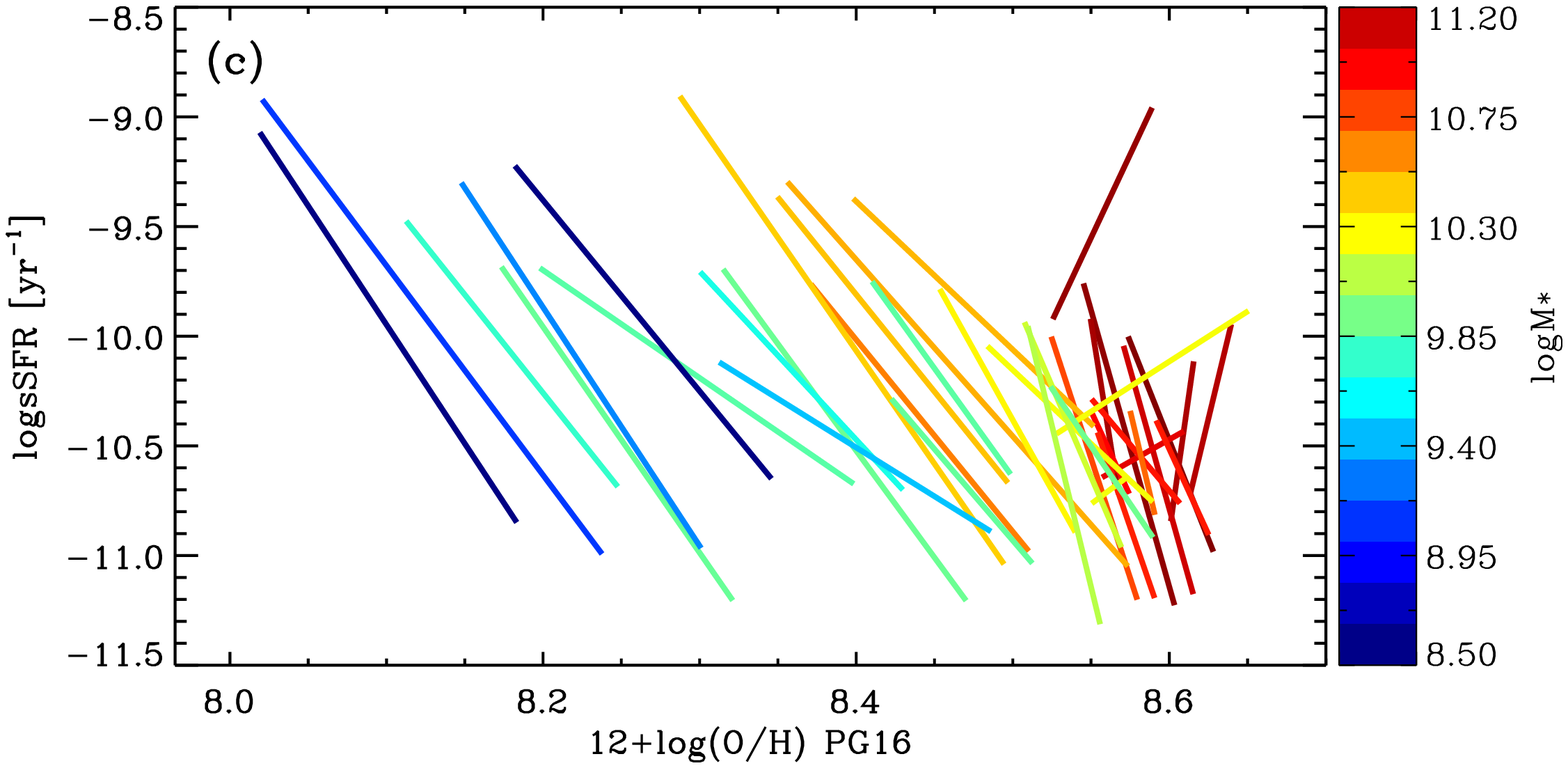,clip=true,width=0.663\textwidth}
    \end{center}
  \caption{The same as Figure \ref{fig:18}, but using log(O/H)-\citetalias{Pilyugin-16}. }
  \label{fig:19}
\end{figure*}

In Section \ref{sec:4.1}, we defined the $\Delta\log$sSFR and $\Delta\log$(O/H) of individual spaxels based on the sSFR and 12+log(O/H) profiles of each MAD galaxies.  However, some authors have argued that the resolved sSFR-$\Sigma_*$ relation is more fundamental than the global SFR-$M_*$ relation \citep[e.g.][]{Barrera-Ballesteros-17, Liu-18, Erroz-Ferrer-19}.  Here we re-examine our basic result by using the sSFR-$\Sigma_*$ relation to define the $\Delta\log$sSFR and $\Delta\log$(O/H) of individual spaxels. Specifically, we first define a sSFR-$\Sigma_*$ (or log(O/H)-$\Sigma_*$) relation for each individual MAD galaxy by linearly fitting all the valid spaxels. Then, for each spaxel of a given MAD galaxy, the $\Delta\log$sSFR (or $\Delta\log$(O/H)) is the vertical deviation from the fitted sSFR-$\Sigma_*$ (or log(O/H)-$\Sigma_*$) relation of this galaxy.  

Figure \ref{fig:18} (or Figure \ref{fig:19}) is the same as the panel (a), (b) and (c) of Figure \ref{fig:9} (or Figure \ref{fig:10}), but defining the $\Delta\log$(O/H) and $\Delta\log$sSFR with the stellar surface density rather than the galactic radius.  According to panel (c) of Figure \ref{fig:18}, for most of individual MAD galaxies, the sSFR decreases with increasing $\Sigma_*$, consistent with a positive radial gradient of sSFR.  The log(O/H) appears to increase significantly with global stellar mass between the galaxy population, and within individual MAD galaxies, the metallicity increases with increasing $\Sigma_*$, consistent with a negative radial gradients of log(O/H).  As discussed in Section \ref{sec:4.2}, this dependence of log(O/H) on stellar mass or stellar surface density can be interpreted as the mass dependence of the metallicity of inflow gas $Z_{\rm 0}$ and the mass-loading factor $\lambda$, under the gas-regulator frame.

According to panel (a) and (b) of Figure \ref{fig:18} (or Figure \ref{fig:19}), a clear positive correlation between $\Delta\log$sSFR and $\Delta\log$(O/H) can be seen.  This is true even for individual MAD galaxies (with one or four exceptions for {\tt N2S2H$\alpha$} or {\tt Scal} indicators).  We argue that our result is robust, regardless of whether we define the $\Delta\log$sSFR and $\Delta\log$(O/H) with $\Sigma_*$ or galactic radius. As pointed out in Section \ref{sec:4.2}, this result supports that at $\sim$100 pc scale, the star formation and metal-enhancement is primarily driven by the time-varying SFE.

\label{lastpage}
\end{document}